\documentclass[11pt,oneside]{book}

\usepackage{graphicx,fullpage,natbib,url,multirow,setspace,amsmath}
\usepackage[ms]{thesisfrontmatter}
\bibpunct{(}{)}{;}{a}{}{,}

\newcommand\arcsec{\mbox{$^{\prime\prime}$}}

\degreetitle{Modeling the Infrared Emission from Cygnus A}
\degreeauthor{George C. Privon}
\prevdegreeA{B.S. Rochester Institute of Technology, 2006}
\prevdegreeB{ }
\prevdegreeC{ }
\degreedate{23 May 2009}
\advisor{Dr. Stefi Baum}
\memberA{Dr. David Axon}
\memberB{Dr. Joel Kastner}
\memberC{Dr. Andrew Robinson}

\begin{document}


\begin{abstract}
The Spitzer Space Telescope provides a unique view of the Universe at infrared wavelengths. Improved sensitivity and angular resolution over previous missions enable detailed studies of astrophysical objects, both in imaging and spectroscopic modes. Spitzer observations of active galactic nuclei can help shed light on the physical conditions of the central regions of these active glalaxies.

The nearby radio galaxy Cygnus A is one of the most luminous radio sources in the local Universe. In addition to the high radio power, it is also very luminous in the infrared. New Spitzer spectroscopy and photometry of Cygnus A is combined with data from the literature at radio and sub-mm wavelengths. The resulting complication is modeled with a combination of: a synchrotron emitting jet, a burst of star formation, and emission from an AGN torus. 

The infrared emission in Cyngus A shows contributions from all three processes and the models are able to reproduce the observed emission over almost 5 dex in frequency. The bolometric AGN luminosity is found to be $\sim10^{45}$ erg s$^{-1}$, with a clumpy torus size of $\sim7$ pc. Evidence is seen for a break in the synchrotron spectrum in the mid-infrared. The relevant component of the infrared emission suggests Cygnus A has a star formation rate of $\sim20$ M$_{\odot}$ yr$^{-1}$. Even in the absence of the AGN, it would still be a luminous infrared source.
\end{abstract}

\begin{acknowledgements}

I am deeply indebited to my advisors Dr. Stefi Baum and Dr. Chris O'Dea for their helpful guidance and instruction, both for this thesis and previous research. I would also like to thank my committee members Dr. David Axon, Dr. Andrew Robinson, and Dr. Joel Kastner for their helpful insight. Dr. Jake Noel-Storr and Dr. Jack Gallimore were both invaluable resources and I am indebited to them for their assistance.

I am grateful for the support provided by other faculty and staff of the Center for Imaging Science at the Rochester Institute of Technology, including: Marilyn Lockwood, Sue Chan and Joe Pow. 

David Whelan and Dr. Mark Whittle have made themselves available for many helpful discussions, for which I am grateful. I also wish to thank the Department of Astronomy at the University of Virginia for their support as I worked to finish this research.

Finally, I would like to thank my parents (Chris \& Keron Privon) and brothers (Peter \& Chris II) for their love, support and encouragement. I would not have been able to do this without them.

This work was supported in part by the Research Computing group at the Rochester Institute of Technology. This research has made use of the NASA/IPAC Extragalactic Database (NED) which is operated by the Jet Propulsion Laboratory, California Institute of Technology, under contract with the National Aeronautics and Space Administration. 
\end{acknowledgements}

\tableofcontents

%

\chapter{Introduction}
\label{chap:introduction}

The advent of astronomical observations in non-optical wavelength regimes led to the discovery of bright celestial sources at radio wavelengths. Many of these sources of non-thermal radio emission have since been hypothesized to be powered by the release of gravitational energy from an accreting supermassive black hole at or near the center of a galaxy. These ``active galactic nuclei'' (AGN) represent some of the most energetic phenomenon in the Universe, and may play an important role in the formation and evolution of galaxies and galaxy clusters \citep[e.g.][]{Sanders88,Antonuccio-Delogu08,Nesvadba08}.

There are many observational types of active galaxies, with drastically different morphologies and powers. Radio-loud AGN  are classified as those sources whose radio flux exceeds the optical flux by at least a factor of $\sim10$. The first radio-loud AGN to be discovered were associated with point sources at optical wavelengths. These ``quasi-stellar objects'' (quasars) are some of the most intrinsically luminous objects in the Universe. 

Frequently these radio-loud AGN exhibit extended structure in the radio wavelengths, and are classified as radio galaxies. The host galaxies for these AGN are generally large elliptical galaxies \citep{McCarthy93}.

There are multiple observational classifications of radio-loud AGN. Geometrical frameworks have been suggested to unify some of these observational classes into a single type of object \citep[e.g.][]{Urry95}. Invoking both dust obscuration and source orientation, the appearance of a radio-loud AGN can be explained by the line of sight to the observer.

Dust and gas obscuration predominantly affects wavelengths from the near infrared and shorter. Absorption from dust can obscure emission, especially from deeply embedded sources. At longer wavelengths the emission passes through the dust mostly unobscured, allowing direct studies of ``obscured'' regions. Additionally, the shorter wavelength radiation absorbed by the dust is re-radiated in the mid- and far-infrared. This makes the infrared the ideal regime to study bolometric output of an AGN.

The radio galaxy Cygnus A is one of the brightest extragalactic radio sources visible in the sky. Due to its proximity, it can be studied at high spatial resolution, enabling a detailed study of the AGN and its environment. Studies of Cygnus A have indicated there is a powerful AGN at the center, likely comparable to quasars seen in the distant (therefore younger) Universe \citep{Carilli96}. Studying this nearby example can yield significant insight into the processes at work in distant quasars.

Infrared measurements of Cygnus A can test the unification schemes by looking for the signature of an obscured AGN and comparing the observed infrared emission with models. Additionally, the total infrared emission from the AGN can be used to determine the bolometric luminosity, testing the idea that Cygnus A is a nearby example of a powerful quasar.

Cygnus A also shows evidence of significant star formation around the nucleus. Dust enshrouded star formation has its primary signature in the infrared, lending further use to observations in this wavelength regime.

The Spitzer Space Telescope (formerly the Space Infrared Telescope Facility) was launched in 2003 and contains a unique suite of instruments for the mid- and far-infrared. With improved angular resolution and sensitivity over previous infrared satellites, it can provide significant additional detail over previous studies. Spitzer covers a wavelength range of $\sim3-160~\mu$m in a combination of imaging and spectroscopy \citep{Werner04}.

This work presents new mid-infrared spectroscopy of the nuclear regions of Cygnus A. Due to the significant energy output in this wavelength regime, interpreting the infrared emission from Cygnus A is critical to understanding the nature of the AGN and other nuclear activity. Three detailed, physically motivated models were combined to reproduce the observed spectral energy distribution covering 5 dex in frequency.

The new data, plus data compiled from the literature, were modeled to explore the relative contributions from AGN activity and the circumnuclear star formation. The AGN activity is manifested as point-source emission and a relativistic jet. The jet dominates in the radio regime, while the nuclear point source emission is primarily evident in the infrared. The star formation is dust enshrouded so the resulting emission occurs primarily in the thermal infrared.

Chapter \ref{chap:science} explores the motivation for this study by describing the observational zoo of AGN and the possible simplification which arises from a unification scheme. This chapter also describes the current body of knowledge regarding Cygnus A, which in turn provides the justification for the choice of models. The Spitzer Space Telescope and the instruments used to carry out new observations of Cygnus A are discussed in chapter \ref{chap:Observations}. The modeling method and individual models used are described in chapter \ref{chap:Modeling}. Finally, the results of this modeling and interpretation of the results are the subject of chapter \ref{chap:Results}.
 
%

\chapter{Active Galactic Nuclei and Cygnus A}
\label{chap:science}

Active galactic nuclei (AGN) represent some of the most energetic phenomena in the Universe. They can influence the formation and growth of galaxies and galaxy clusters through feedback in the form of winds and jets. Likely powered by the extraction of gravitational energy from objects in the potential of a supermassive black hole ($M_{\bullet}\ge 10^5 M_{\odot}$, $M_{\odot}=1.99\times10^{33}$ g)\footnote{The \emph{CGS} system of units is utilized throughout this work, except where otherwise noted.}, they provide a laboratory to study high energy physical processes otherwise inaccessible \citep{Blandford77}.

\section{Active Galactic Nuclei}

The first (radio--loud) AGN were discovered via radio surveys in the 1950s. The relatively compact radio sources which were detected had no obvious optical counterparts. Follow up observations identified optical counterparts, ranging from optical point sources to giant elliptical galaxies. The optical spectra of these objects implied a large redshift, resulting from a high recession velocity. The large recession velocities indicated these sources were very distant and therefore have high intrinsic luminosities. With efficiencies\footnote{Efficiencies are defined as the fraction of the rest mass energy ($E=mc^2$) released as electromagnetic radiation.} of $\sim0.7\%$, nuclear fusion cannot provide sufficient energy output to power these objects. These objects are now believed to be powered by accretion of matter onto supermassive black holes (SMBH), where efficiencies approach $10\%$. 

There are a variety of observational types of AGN, broadly divided into two categories, ``radio-loud'' and ``radio-quiet'', based on the observed ratio of radio to optical emission. Roughly 10\% of AGN are ``radio-loud'' although this fraction varies with intrinsic power of the AGN.

As the name implies, ``radio-quiet'' AGN are those with minimal radio emission, although they may show weak jets and radio lobes. Typically labeled ``Seyfert Galaxies'' (after Carl Seyfert), their optical spectra often show strong emission lines. In some cases, the emission lines are quite broad ($\sim$ few 1000s of km s$^{-1}$, ``Seyfert I'', ``Type I'' AGN). In sources with only narrow emission lines, they are still broad compared to a quiescent galaxy ($\sim$ few 100s of km s$^{-1}$, ``Seyfert II'', ``Type II'' AGN). 

''Radio-loud'' AGN show strong radio emission. The radio morphology can vary drastically from object to object, from point source through large extended structures up to Mpc scales. Polarization studies of the radio emission, coupled with multiwavelength measurements of the flux, indicate the emission mechanism is synchrotron radiation from a population of relativistic charged particles moving in a magnetic field. ``Radio-loud'' AGN exhibit a similar range of optical spectra as their ``radio-quiet'' counterparts. As noted above, Cygnus A is a radio-loud AGN and its properties will be discussed in $\S$\ref{science:cyga}.

\subsection{Unified AGN Model}
\label{science:agnparadigm}

There have been attempts to unify the zoo of observational AGN types into a consistent framework \citep{Urry95}. Here the focus will be on unification of radio-loud AGN. The two major mechanisms used to explain the variety in observed AGN types are orientation-dependent obscuration and relativistic beaming of jets.

\begin{figure}[h!]
\includegraphics[width=0.4\textwidth]{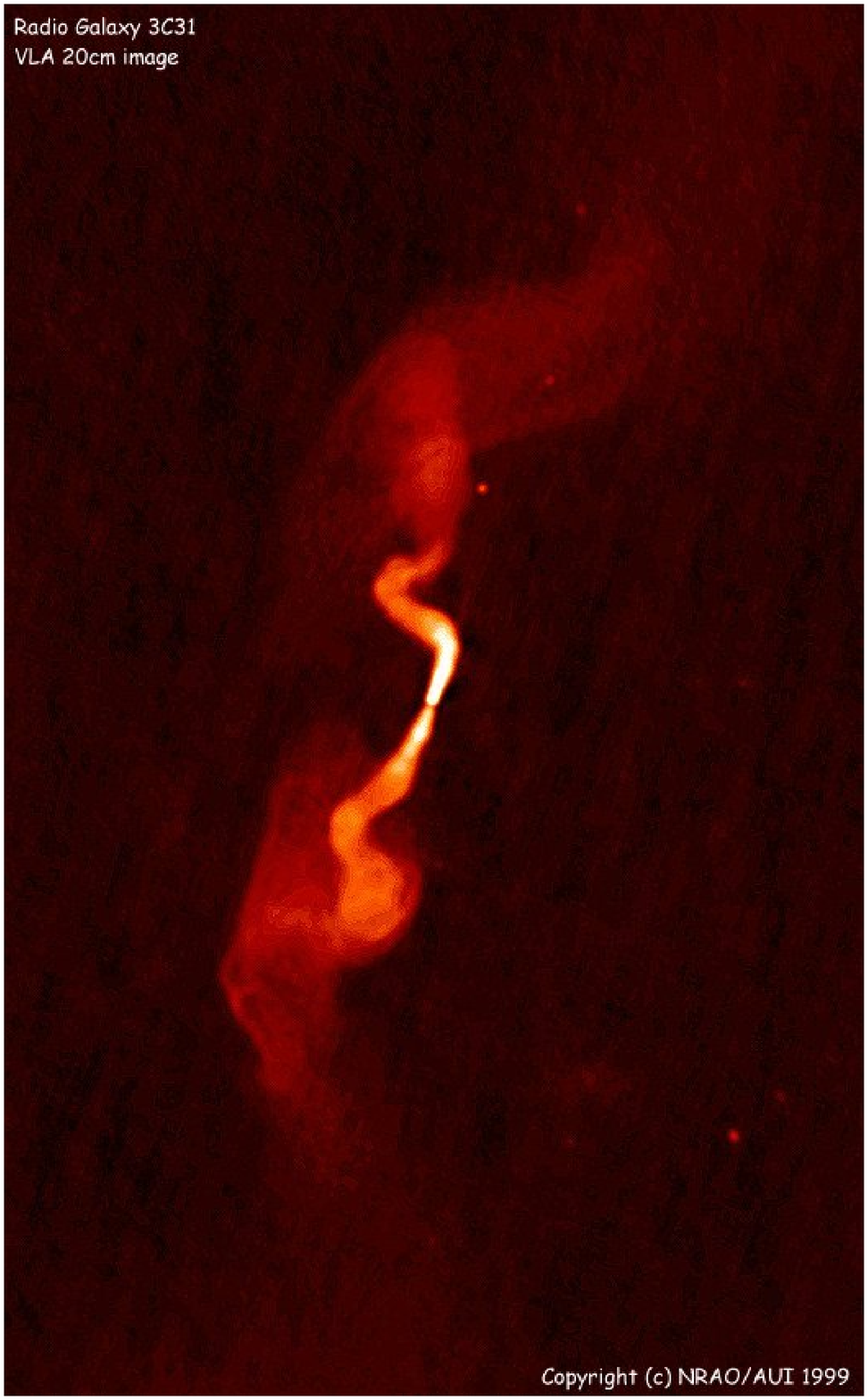}
\includegraphics[width=0.6\textwidth]{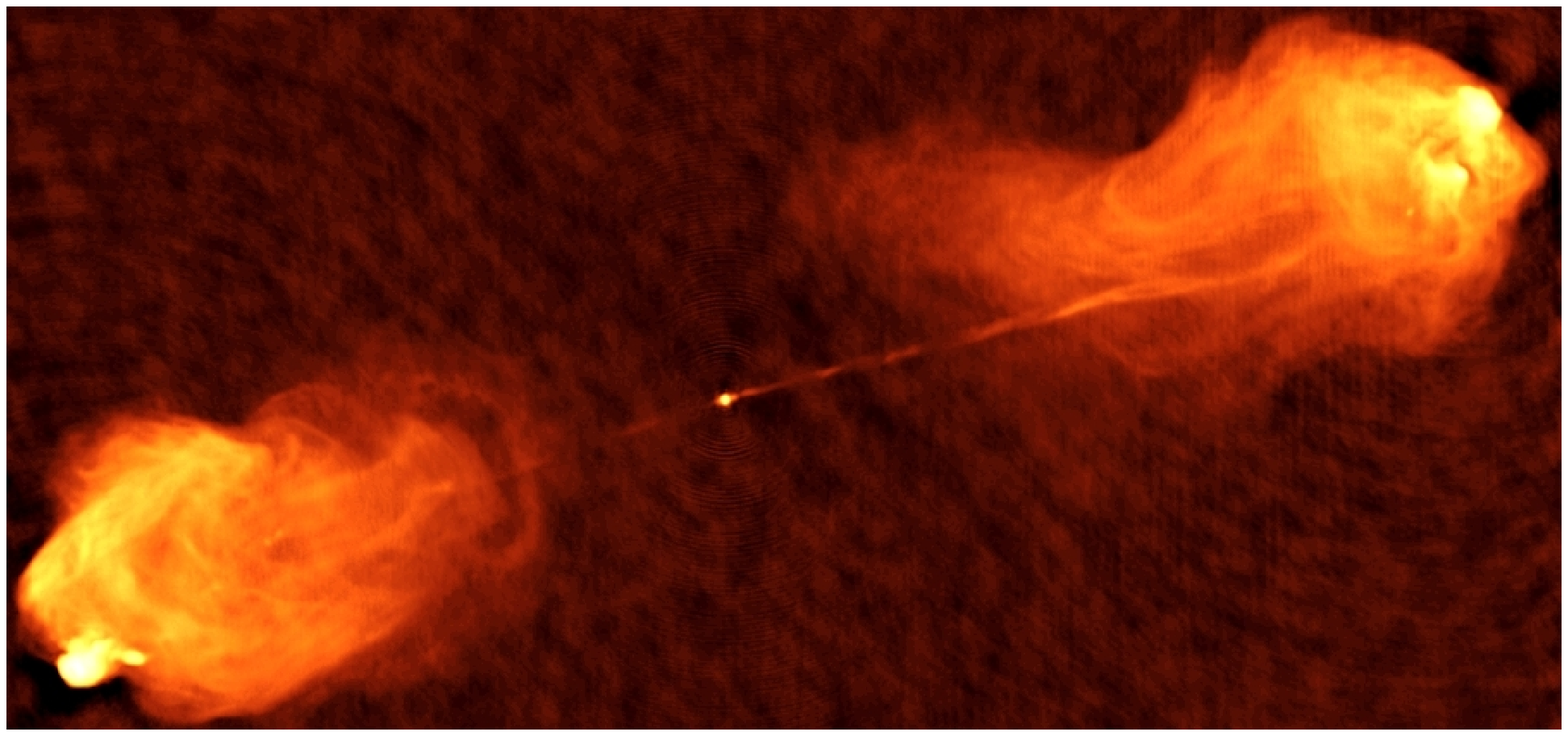}
\caption{Left: 20 cm radio map of FR I radio galaxy 3c31 \citep{Laing08}. Notice the emission tapers off away from the center of the source. Right: FR II source Cygnus A, with the hotspots and lobes clearly visible \citep{Carilli96}.}
\label{science:frtype}
\end{figure}

Radio-loud AGN typically show extended structure at radio wavelengths. The morphology is generally divided into Fanaroff \& Riley (FR) I and II objects \citep{Fanaroff74}. FR I sources are brightest at the center, with emission tapering off away from the radio core. FR II sources are ``edge brightened'', showing distinct lobes with ``hotspots'' of radio emission towards the edges. It is widely believed that FR II radio sources contain relativistic jets, however the jets are often too faint to be seen. Figure \ref{science:frtype} shows radio images of a typical FR I and FR II.

The jet morphology of the FR II sources is evident, in figure \ref{science:frtype}. The overall source morphology is generally attributed to relativistic jets which propagate from the radio core with bulk relativistic motion. These jets likely terminate in the ``hotspots'' of radio emission as the jet runs into the tenuous intergalactic medium. Shocks at the interface likely re-accelerate the electrons and the radio plasma backflows to create the observed lobes of emission.

As noted above there is considerable variation in optical spectra, with some objects showing narrow emission lines, and others with both broad and narrow emission lines. The ``narrow'' emission lines are still broader than lines typically seen in quiescent galaxies. Objects with broad emission lines are classified as ``Type 1'' AGN and those with only narrow lines as ``Type 2''. The variation in optical spectra is attributed to orientation-dependent obscuration of the AGN.

\begin{figure}[h!]
\includegraphics[width=0.5\textwidth]{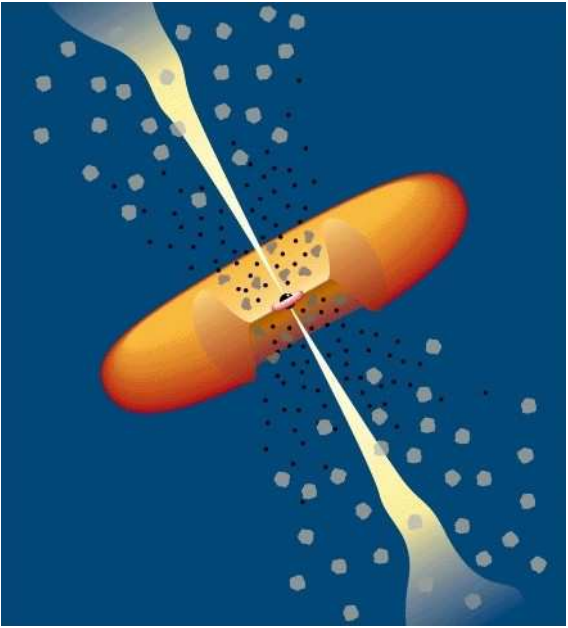}
\caption{Schematic geometric model for the unification of radio loud AGN, from \citet{Urry95}. The SMBH and accretion disk are at the center. The Broad Line Region (BLR) is inside the obscuring torus. The Narrow Line Region (NLR) is exterior to the torus and hence visible at all orientations. The radio jets are also shown, forming in the vicinity of the SMBH and propagating outwards. Figure Copyright Astronomical Society of the Pacific, reprinted by permission of the authors.}
\label{science:urry}
\end{figure}

Figure \ref{science:urry} shows a schematic diagram of the geometric structure invoked to unify radio-loud AGN. The central object (generally believed to be a supermassive black hole) is surrounded by an accretion disk. Viscosity within the disk heats it, resulting in thermal emission. X-rays are either created in the disk, or in a hot corona above the disk. UV and X-ray photons radiate outwards isotropically from the central source. An obscuring torus or disk in the plane of the accretion disk absorbs some of the ionizing radiation, resulting in an anisotropic radiation field.

Broad emission lines arise in the region near the SMBH, within the obscuring torus. Gas clouds orbiting in the deep potential well are moving quickly, and the lines are doppler broadened to $\sim1000$ km s$^{-1}$. This region is called the ``Broad Line Region'' (BLR). Observers looking through the opening in the torus can see the broad line emission from the gas. However, the BLR is obscured for viewers along the plane of the torus. In some cases, broad emission lines are seen only in polarized light. This is interpreted as due to scattering of the broad emission lines off the obscuring torus into the line of sight, implying these objects are at some intermediate orientation. 

Narrow emission lines originate from gas clouds in the ``Narrow Line Region'' (NLR). This region of lower density gas is typically farther from the SMBH and so moves more slowly, resulting in narrower emission lines. The NLR exists primarily outside the obscuring torus, and so is not strongly orientation dependent. Forbidden radiative transitions occur here. Despite the low values of the Einstein $A_{ul}$ coefficients, radiative de-excitations dominate over collisional de-excitations due to the low density.

The specific nature of the obscuring geometry is disputed. Initial suggestions were of a atomic or molecular torus (as illustrated in Figure \ref{science:urry}). However, a uniformly dense obscuring medium was unable to reproduce observations. Modifications to this model resulted in a ``clumpy'' torus formed from discrete clouds \citep[e.g.][]{Nenkova02,Nenkova08}. In modeling the observations of Cygnus A (discussed in Chapter \ref{chap:Modeling}), a model for a clumpy torus was tested. An alternate proposed obscuring geometry arises from a wind of material blown off the accretion disk \citep[e.g.][]{Elvis00}.

The radio structure can also be influenced by the orientation. The jets, moving at relativistic speeds, are doppler boosted, causing them to appear brighter or fainter depending on their relative motion. Evidence for this is seen in VLBI studies of radio cores. When both jets are detected, one jet (the ``jet'') is observed to be brighter than the other (the ``counterjet''). With the assumption that the jet and counterjet are intrinsically the same luminosity luminosity, the inclination of the AGN can be estimated.

The extended radio emission results from the interaction of the jet with the diffuse medium outside the galaxy. The lifetime of relativistic electrons in the hotspots is shorter than the light travel time to the radio core \citep[e.g.][]{Hargrave74}. The electrons might be re-accelerated in the hotspots by 1st order Fermi acceleration before back-flowing to create the diffuse lobes seen \citep[e.g.][]{Scheuer74}. Due to the isotropic nature of these phenomenon, these features are not likely to be relativistically beamed. This can be seen in sources where the hotspot on the side corresponding to the counterjet is brighter than the jet side hotspot. 

In this unification scheme, narrow line radio galaxies (NLRGs) are sources viewed through the plane of the torus. Radio-loud quasars are those sources viewed nearly down the jet axis (ie - through the ``hole'' of the torus). Broad line radio galaxies (BLRGs) are those viewed at intermediate angles, where the broad line region is visible, but the jet is not yet significantly doppler boosted.

\section{Cygnus A}
\label{science:cyga}

Cygnus A is a rare nearby, powerful radio galaxy. At $z=0.056$\footnote{$z$ is the \emph{redshift} of the source, defined as: $1+z=\frac{\lambda}{\lambda_0}$. $\lambda$ is the observed wavelength of an observed (emission) line and $\lambda_0$ is the rest wavelength.} it is the most luminous radio galaxy out to $z\sim1$. Using a cosmology with $H_0=73$ km s$^{-1}$ Mpc$^{-1}$, this puts Cygnus A at a distance of $233.4\pm16.3$ Mpc \citep{Spergel07}. The host galaxy is quite bright at $M_V=-22.6$. It is believed to contain a hidden Type I quasar which is heavily obscured by dust. Inferences from X-ray observations (further discussed below) suggest the AGN's bolometric luminosity is $0.5-2\times10^{46}$ erg s$^{-1}$ \citep{Tadhunter03}.

\subsection{Observational Characteristics}

Due to the significant radio emission, Cygnus A is classified as a ``radio galaxy'' and its intrinsic radio power is comparable to radio-loud AGN visible in the distant Universe ($z\sim1$, $D_L\sim6.6$ Gpc). As a result, it provides an excellent opportunity to study the activity of a powerful AGN and its influence on the host galaxy. Its proximity permits the detailed study of the AGN on spatial scales unachievable for its more distant counterparts.

\begin{figure}[h!]
\includegraphics[angle=270,width=0.8\textwidth]{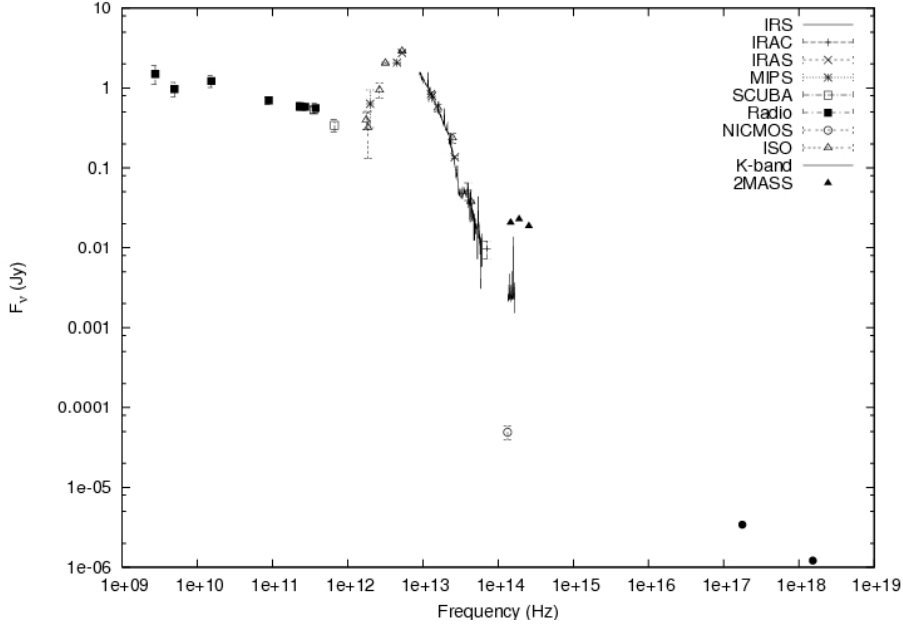}
\caption{Broadband spectral energy distribution for Cygnus A. The data were taken from a variety of sources in the literature. See text for references.}
\label{fig:cyga-sed}
\end{figure}

Cygnus A has been well studied across the electromagnetic spectrum (figure \ref{fig:cyga-sed}). This section summarizes the observational results, focusing on the nuclear regions, where AGN and star formation activity dominate the observed radiation. For a more comprehensive review of Cygnus A, which also describes the host galaxy and the extended radio source see \citet{Carilli96}.

Combining data from the various observations and analyses discussed below, the spectral energy distribution (SED) of the nucleus of Cygnus A can be constructed. Figure \ref{fig:cyga-sed} shows the flux density at the variety of frequencies for which data are available.

\subsubsection{Radio Properties}

Cygnus A has been observed with a variety of radio observatories at many resolutions and frequencies \citep[e.g.][]{Wright84,Alexander84,Meisenheimer01}. The extended radio source is a prototypical, edge-brightened FR II (Fig \ref{science:frtype} right), with both the jet and counter-jet visible. It displays obvious hotspots and radio lobes, similar to those observed in many other triple (core + 2 lobes) radio sources.

Core radio fluxes have been obtained from the literature from a few GHz through the mm \citep[figure \ref{fig:cyga-sed}:][]{Carilli96,Alexander84,Wright84,Eales89,Salter89}. At VLA resolutions the unresolved radio core has a flat spectrum with spectral index $\alpha$=0.18 (f$_{\nu} \propto \nu^{-\alpha}$). However, in high resolution VLBI observations, the core spectrum is inverted between 5 and 43 GHz, possibly due either to synchrotron self-absorption (SSA) or free-free absorption from a molecular torus \citep{Carilli94}. The VLA ''core'' fluxes likely include emission from both the VLBI core and the inner jet. 

Both the jet and counter-jet are detected in VLBI observations, with the counter-jet being fainter. Assuming the jet and counter-jet are intrinsically the same luminosity \citep[e.g.][]{Blandford74}, this indicates the radio source is oriented at some intermediate angle to the line of sight. Application of the Symmetric Twin Relativistic Jet Model to these data constrains the orientation of the radio jets to $50^{\circ} \leq i \leq 85^{\circ}$ \citep{Sorathia96}. Following the unification scheme for radio-loud AGN, this suggests that we are looking in the plane of the obscuring torus. This is consistent with a geometry in which the central SMBH and BLR is hidden from view.

VLBI observations detected broad HI absorption against the nucleus \citep{Conway95}. Corresponding to a column density of N$_{H}=2.54 \pm 0.44 \times 10^{19}$ T$_{spin}$ cm$^{-2}$ (where T$_{spin}$ is the spin temperature of hydrogen), the absorption has been attributed to the circumnuclear torus. The torus is thought to either be atomic (with T$\sim8000$ K) or molecular (T$\sim1000$ K). 

Although molecular absorption was not detected by \citet{Conway95}, the observations do not rule out a molecular torus. Radiative excitation of molecules by the AGN can result in the absence of molecular absorption lines in lower energy transitions.

\subsubsection{Millimeter/Sub-millimeter}

Millimeter and sub-millimeter observations at two epochs using the Submillimetre Common-User Bolometer Array (SCUBA) at the JCMT were presented by \citet{Robson98}. They find a powerlaw core (likely due to synchrotron radiation, $\alpha=0.6$) with a steeper powerlaw than observed at lower frequencies. Their data suggests a break or cutoff in the synchrotron spectrum at the highest frequencies. The cause may either be depletion of energetic electrons, or continuous injection of a powerlaw distribution of electrons, both of which would cause the spectrum to steepen.

An upper limit was established at $350~\mu m$, consistent with thermal emission from dust. However, the $450~\mu m$ flux indicates there is no  significant amount of cold dust present. Based on these fluxes, \citet{Robson98} conclude the emission is dominated by warm dust with $37~K < T < 80~K$.

\subsubsection{Optical Properties}
\label{cyga:optical}

\begin{figure}[h!]
\includegraphics[width=0.4\textwidth]{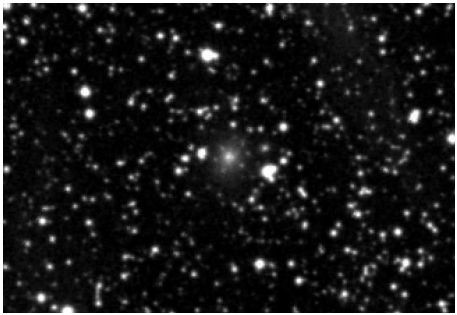}
\caption{Digital Sky Survey (DSS) red image of Cygnus A (center of the frame). }
\label{fig:cyga-dss}
\end{figure}

First identified by \citet{Baade54}, the host galaxy of Cygnus A is a cD elliptical \citep{Matthews64} which is also the brightest member of a cluster (see figure \ref{fig:cyga-dss}\footnote{The Digitized Sky Surveys were produced at the Space Telescope Science Institute under U.S. Government grant NAG W-2166. The images of these surveys are based on photographic data obtained using the Oschin Schmidt Telescope on Palomar Mountain and the UK Schmidt Telescope. The plates were processed into the present compressed digital form with the permission of these institutions.} for an optical image of Cygnus A). Ground based observations show the optical nucleus has a double structure which is interpreted to be a single nucleus with an obscuring dust lane \citep{Osterbrock83,Tadhunter94}.  

\citet{Tadhunter94} obtained high resolution optical spectra of the nuclear region of Cygnus A. Their data were consistent with previous results noting very strong emission lines but a relatively weak continuum. The spectra indicate AGN photoionization is the dominant ionization source, although they conclude shocks may also contribute. However, the coexistence of high and low ionization lines indicates some modifications to AGN photoionization models are necessary. The ionization field is anisotropic (conical morphology) and aligned with the radio axis, consistent with unified models for radio-loud AGN (see $\S$\ref{science:agnparadigm}). Also, in order to explain the observed nitrogen ratios, abundances must be much higher than solar values.

Cygnus A is thought to contain an obscured quasar similar to those seen at high redshift. While \citet{Jackson93} were unable to detect broad H$\alpha$ in polarized light, it was detected in later observations by \citet{Ogle97}. The polarized broad lines were observed within the ionization cone noted above, with a possible scattering origin in the NLR. The detection of broad H$\alpha$ emission lends support for the existence of an obscured BLR.

\citet{Tadhunter03} obtained Hubble Space Telescope (HST) Space Telescope Imaging Spectrograph (STIS) spectra of the nuclear region. Stepping the slit across the nucleus, a velocity gradient indicative of rotation around the radio axis was observed. Modeling the velocity as due to the potential of a SMBH and stellar mass distribution (measured from a $1.6\mu$m NICMOS image) gives a SMBH mass of $2.5\pm0.7\times10^{9}M_{\odot}$. This measurement is consistent with black hole mass -- host galaxy relations \citep{Magorrian98,Ferrarese00,Gebhardt00}.

HST imaging has also revealed star formation in the central region of Cygnus A, which began $< 1$ Gyr ago \citep{Jackson98}. Lying in a 4kpc ring around the nucleus, that is oriented orthogonal to the radio axis, emission from the starburst is detected in the Spitzer observations described below.

\subsubsection{Infrared Properties}

Keck NIR imaging with adaptive optics was presented by \citet{Canalizo03}. They find a secondary point source $0.4\arcsec$ from the nucleus (K$\sim19$). The IR-Optical SED of this off-nucleus point source is consistent with a smaller galaxy which has been tidally stripped of much of its original mass. It is likely a dense core of older stars, which is all that remains from the merging galaxy. The merging scenario can explain the recent ($\sim10^7$ yr) triggering of AGN activity, the disturbed kinematics seen in Cygnus A, and the dust lane across the nucleus.

\citet{Tadhunter99} obtained NICMOS images of Cygnus A from HST. A point source in their image 

Keck K-band spectra of the nuclear region were obtained by \citet{Tadhunter03}. While they did not provide a full analysis, spatially extended Pa$\alpha$ and H$_2$ emission was detected, and fits to these lines were used to constrain the SMBH mass ($M_{\bullet}=(2.5 \pm 0.7) \times 10^{9} M_{\odot}$).

\citet{Bellamy04} presented a more detailed analysis of the K-band spectra, finding complicated emission line properties suggesting an infalling molecular cloud. The location and velocity of the molecular ($H_2$) emission is inconsistent with interpretation as an outflowing cloud. The $H_2$ line ratios are consistent with X-ray heating, indicating the cloud is likely excited directly by the X-rays, via hard X-rays which can penetrate the obscuring torus. The interpretation as an infalling cloud is consistent with the \citet{Canalizo03} picture where Cygnus A is currently undergoing a minor merger. 

Additionally, the $H_2$ line was seen in several components, both red- and blue-shifted relative to the systemic velocity. This was interpreted as emission from a rotating torus. The observed line ratios are consistent with excitation by X-rays (from the AGN), while likely ruling out shocks as a possible excitation method. 

The wide range of flux values observed in the near-IR (figure \ref{fig:cyga-sed}, $\sim10^{14}$ Hz) illustrates the importance of aperture matching when constructing such a broadband SED. The 2MASS aperture encompasses much of the host galaxy, including much stellar emission. The Keck K-band aperture is smaller, while the NICMOS observation is the point source continuum flux associated with the nucleus.

Mid-infrared ($11.7\mu m$) imaging of Cygnus A were obtained by \citet{Whysong04}. Based on the observed nuclear flux, and an assumed spectral shape similar to those for radio-loud Palomar-Green quasars, they estimate the bolometric luminosity of the AGN in Cygnus A to be $\sim1.5\times10^{45}$ erg s$^{-1}$. 

\subsubsection{UV Observations}

HST UV observations of Cygnus A were presented by \citet{Zirbel98}. Undetected at the shortest wavelengths observed (F130M\footnote{Hubble Space Telescope filters are designated by $F$, the central wavelength (in nm), and 'N' for a narrow filter, 'M' for a medium width filter, and 'W' for a wide filter.}, F152M, F170M, \& F220M), the F320M, F342W, and F372M data show no nuclear component in modeling, consistent with the obscuration seen in the optical. Any UV flux from the AGN would be absorbed by the optically thick torus and re-radiated in the infrared.

Using the Faint Object Camera on HST, \citet{Antonucci94} found broad (FWHM $\sim 7500$ km s$^{-1}$) [Mg II] 2799$\buildrel _{\circ} \over {\mathrm{A}}$ emission, supportive of the notion that Cygnus A contains a hidden quasar.

\subsubsection{X-ray Properties}

Cygnus A has been observed with a variety of X-ray instruments, including Uhuru \citep{Gursky72}, ROSAT \citep{Reynolds96}, Chandra \citep{Young02,Evans06}, and INTEGRAL \citep{Beckmann06}. These observations support the existence of the hidden quasar in Cygnus A based on the appearance of an unresolved hard X-ray spectrum coincident with the radio core.

While of relatively poor angular resolution, ROSAT observations by \citep{Reynolds96} were able to resolve the hot ICM around Cygnus A. Spectral fitting gives temperatures of $\sim3-4$ keV over the cluster. Also, evidence for a cooling flow of $\sim250$ M$_{\odot}$ yr$^{-1}$ was seen.

\citet{Young02} conducted Chandra ACIS observations of 32ks and 10ks. The X-ray flux above 2keV was peaked at the nucleus while lower energy flux was more distributed. The hard X-ray nucleus is coincident with the radio nucleus and is less than 0.4" in size. Modeling the emission gives a neutral column density $N_H=2^{+0.1}_{-0.2}\times10^{23}$ cm$^{-2}$ towards the nucleus. 

INTEGRAL observations from 20-100 keV give a luminosity of $log$ $L_x=44.71$ erg s$^{-1}$ \citep{Beckmann06} integrated over the galaxy. The resolution of the imager is $\sim12$ arcmin, and so this value may be subject to contamination from the hot X-ray gas associated with the cluster.

\subsection{Summary: Current Knowledge}
\label{science:SEDoverview}

Based on the abundance of multi-wavelength data, a general picture can be constructed for Cygnus A. It is a large elliptical galaxy in a poor cluster. The galaxy hosts a SMBH with mass typical of SMBHs in other galaxies with similar large-scale properties. It is currently undergoing a minor merger with a smaller galaxy which it has tidally stripped. The existence of a dust lane and infalling gas can be connected to this merger event which is likely linked to the nuclear activity (AGN and starburst). Hard X-ray emission coincident with the nucleus is likely due to emission from the accretion disk surrounding the SMBH. 

While the specifics of physical components derived from modeling will be discussed in Chapter \ref{chap:Results}, it is useful to briefly mention the major features of the SED \ref{fig:cyga-sed}. At the lower frequencies (radio through sub-mm, $10^9-10^{12}$ Hz), the SED clearly follows a powerlaw - indicative of non-thermal synchrotron emission from the radio core. At sufficiently low frequencies, the radiating plasma is expected to become optically thick and synchrotron self-absorption will occur.

The powerlaw is interrupted in the FIR and thermal emission is seen. This thermal bump, which extends almost to the NIR, has an effective temperature of $\sim75$ K \citep{Djorgovski91}. Emission from the circumnuclear starburst and/or torus-reprocessed UV and X-ray radiation from the AGN contributes to this bump. At $\sim10~\mu m$, the extrapolated powerlaw would exceed the emission seen, so the synchrotron spectrum should break somewhere in the IR, likely be due to a depletion of the relativistic electron population at higher energies.

%
%

\chapter{Observations}
\label{chap:Observations}

Infrared observations of AGN are critical in understanding the nature of the emission. In the case of Cygnus A, observations in the mid and far infrared can shed light on the relative importance of the AGN activity and star formation. These phenomena have been studied with previous space-based infrared missions including the Infrared Space Observatory (ISO) and the Infrared Astronomy Satellite (IRAS).

Technological improvements since these missions enable a more detailed study of the nuclear regions of Cygnus A. The following pages discuss the Spitzer Space Telescope and the spectroscopic observations utilized in constructing and modeling the spectral energy distribution of Cygnus A. 

\section{Spitzer Space Telescope}

Launched in 2003, the Spitzer Space telescope improves on previous space-based infrared observatories in both resolution and sensitivity \citep{Werner04}. While its primary mirror is only slightly larger than previous space-borne infrared missions (at 85cm), Spitzer features greatly improved detectors. Additionally, the combination of passive and active cooling should drastically improve the usable lifetime of the telescope. 

There are three instruments on board: the Infrared Array Camera \citep[IRAC,][]{Fazio04}, the Infrared Spectrograph \citep[IRS,][]{Houck04}, and the Multiband Imaging Photometer \citep[MIPS,][]{Rieke04}. These instruments collectively cover a large portion of wavelengths between 3.6 and 160 $\mu m$ in both imaging and spectroscopy modes.

The optical design is a cassegrain telescope, utilizing beryllium optics. The mirror is cooled to $\sim5.5$K. This cooling is achieved through the separation of the satellite into a ``warm assembly'' and a ``cold assembly''. The warm assembly contains the communications equipment and warm electronics for the instruments. The sun shield is also attached here. 

The cold assembly is attached to the rest of the satellite through low thermal conductivity struts and insulated with Mylar thermal blankets and a radiation shield. To conserve cyrogens a ``warm launch'' was utilized in which most of the satellite was launched at ambient temperature. Once in the Earth-trailing orbit, the telescope was allowed to passively cool, after which the cyrogens were used to further cool the instruments to their operating temperature. A satellite orbiting Earth is exposed to both reflected sunlight as well as thermal radiation from the warm Earth. The Earth-trailing orbit allows the satellite to avoid the thermal consequences of close proximity to the Earth, conserving cryogens and extending the mission.

\begin{figure}[h!]
\includegraphics[width=0.75\textwidth]{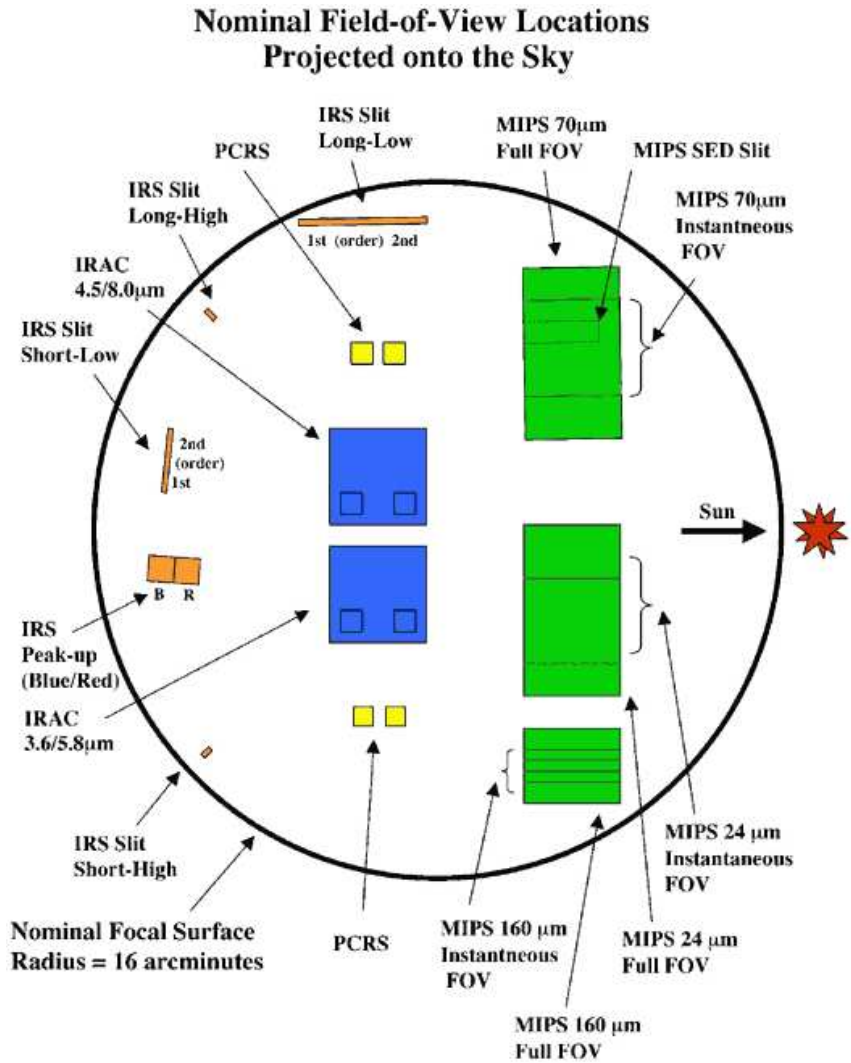}
\caption{A diagram (not to scale) of the focal plane of the Spitzer Space Telescope. From \citet{SOM}.}
\label{ir:spitzerfocalplane}
\end{figure}

The focal plane of Spitzer is shown in Figure \ref{ir:spitzerfocalplane} (not to scale). The slit orientations of both the SL and LL modules are shown, as well as the locations of the various imaging arrays.

\section{The Infrared Spectrograph}

The Infrared Spectrograph (IRS) on Spitzer is primarily designed to obtain spectra from $\sim$5-38$\mu$m. It is capable of medium- (R $\sim$ 600) and low-resolution (R $\sim$ 60-120) spectroscopic observations as well as ``peak-up'' imaging \citep{Houck04}. Here I will only be concerned with the properties of the low-resolution observing mode. Details of the medium-resolution and peak-up imaging modes can be found in the references.

IRS has two modules covering the low-resolution mode. The Short-Low (SL) module observes the short-wavelength ($\sim$5-14 $\mu$m) region while the Long-Low (LL) module observes the long wavelength region ($\sim$14-35 $\mu$m). There is overlap between the SL and LL modules for consistent calibration between the two spectral regions.

The SL slit is 3.7$\arcsec$ wide with a corresponding pixel size of 1.8 $\arcsec$  pixel$^{-1}$. The LL slit is oriented at an angle of 96.5$^{\circ}$ to the SL slit. It is 10.7 $\arcsec$ across with a pixel scale of 5.1 $\arcsec$ pixel$^{-1}$.

Each module covers three spectra orders. Table \ref{irs:orders} lists the wavelength coverage of each order. In both the SL and LL modules, the third order is a ``bonus'' order, included to provide additional data at the overlap between the two primary orders.

\begin{table}[h!]
\caption{Wavelength Coverage of the IRS Orders}
\begin{tabular}{lc}
Order & $\lambda$ ($\mu$m)\\
\hline\hline
SL2     & 5.2 -- 7.7\\
SL1     & 7.4 -- 14.5\\
SL3 (``bonus'')     & 7.3 -- 8.7\\
\hline
LL2     & 14.0 -- 21.3\\
LL1     & 19.5 -- 38.0\\
LL3 (``bonus'')     & 19.4 -- 21.7\\
\end{tabular}
\label{irs:orders}
\end{table}

\subsection{Detectors}

The IRS detector is silicon, doped with arsenide (As) for short wavelengths, and antimony (Sb) for the long wavelengths. The Si:As chips are sensitive from 5-26 $\mu$m and the Si:Sb chips from 14-40 $\mu$m. The detective quantum efficiency is a function of both wavelength and the bias voltage. For a V$_{bias}$=1.5V the DQE for the Si:Sb chips is $\sim$15\% (ranging from 5-25\% over 20-36$\mu$m).

\begin{figure}[h!]
\includegraphics[width=0.5\textwidth]{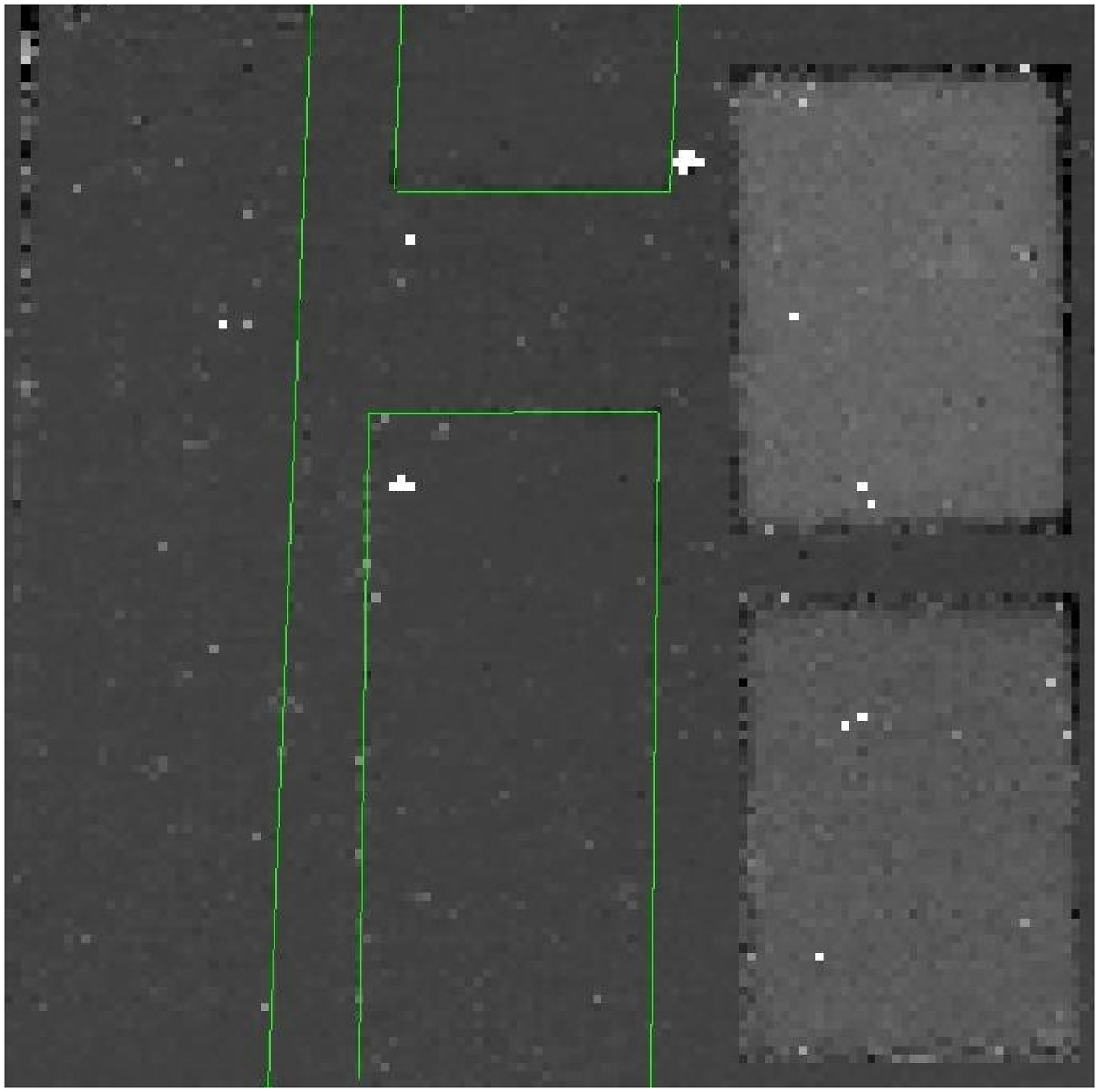}
\includegraphics[width=0.5\textwidth]{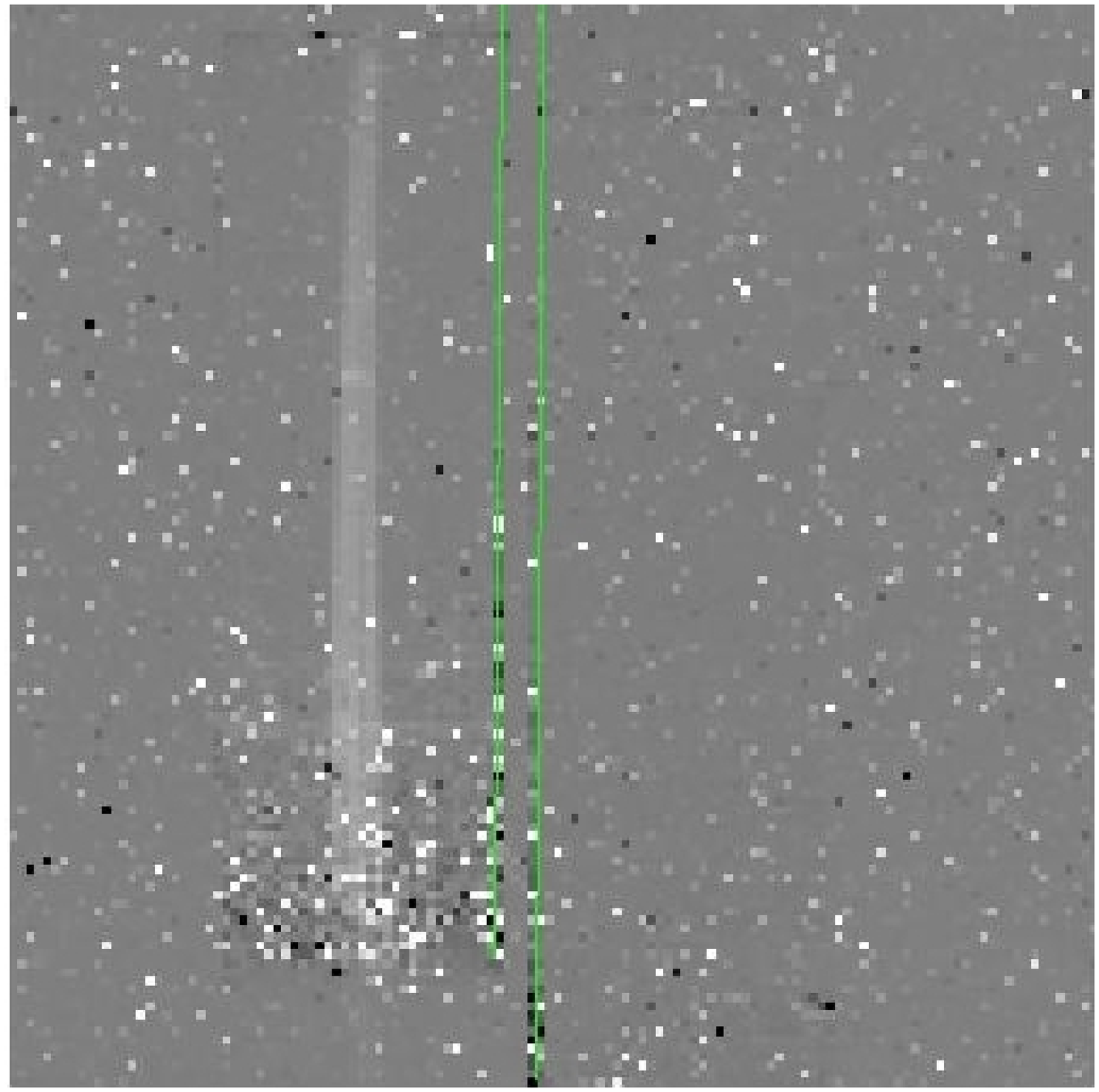}
\caption{Layout of the detector for the low resolution IRS modules. LEFT: SL module. The strip to the left is the SL1 order. The middle section includes the SL2 and SL3 orders. To the right are the two peak-up imaging arrays. Right: LL module.}
\label{irs:chip}
\end{figure}

\begin{figure}[h!]
\includegraphics[width=0.75\textwidth]{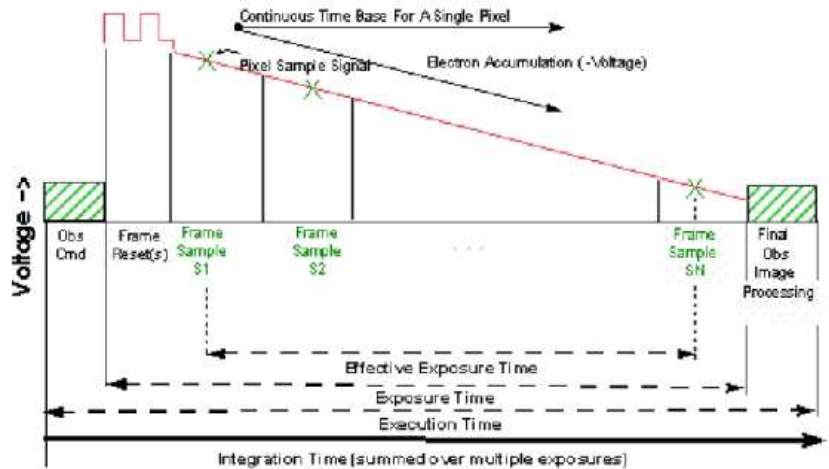}
\caption{Figure demonstrating the data collection method of IRS. See text for details \citep{SOM}.}
\label{irs:readout}
\end{figure}

In order to avoid saturation of the detector and to lessen the impact of cosmic rays (CRs) the detectors are read out in a ramp fashion. The total integration is split up into discrete sample times defined by the user. The pixel values are read out at the end of each sample time and recorded (Figure \ref{irs:readout}). Non-destructive readout of the detector enables the voltages to be recorded multiple times.

If a pixel becomes saturated during an observation (either from a source, or a CR hit), the voltage will remain at a (roughly) constant level for the rest of the observation. Measurements of the ramp slope until saturation enable a determination of the flux. 

Cosmic rays hits that don't saturate the pixel will result in significant drops in voltage, and the slopes on either side of the impact can be used to determine the actual incident flux.

\subsection{IRS Observations}

IRS provides two spectroscopic observing modes: ``staring mode'' and ``mapping mode''. In staring mode, the slit can be positioned directly on the source for a single observation. In ``mapping mode'', the slit is stepped across the source in increments equal to half the slit width. This Nyquist sampling enables reconstruction of the spatial distribution of the emission. The Spitzer Infrared Nearby Galaxy Survey (SINGS) Legacy program has written and provided software to take IRS spectral mapping data and convert into a data cube with both spatial dimensions on the sky (RA \& Dec) as well as a wavelength axis. The ``Cube Builder for IRS Spectra Maps'' (CUBISM) software can use IRS mapping observations to reconstruct the brightness distribution of sources across a $\sim$30 $\mu$m range \citep{Smith07}. In addition, images can be constructed in specific emission lines, spatially localizing the emitting region.

\begin{figure}[h!]
\includegraphics[width=0.5\textwidth]{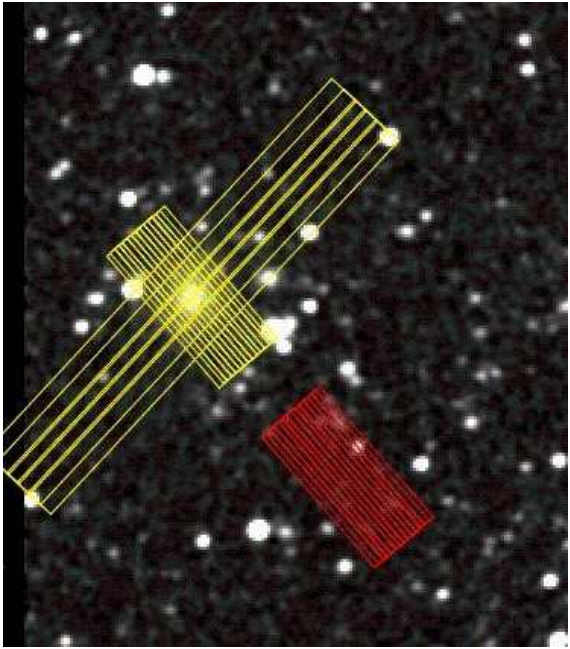}
\caption{2MASS K-band image of Cygnus A with IRS slit positions overlaid. The narrow boxes represent the SL slit, and the wide boxes show the LL slit positions. Note that the slits were stepped across the source in half-slit-width increments. The off-source LL slit positions are not shown, but are to the south-east.}
\label{fig:cygaslit}
\end{figure}

In order to spatially map the MIR emission from Cygnus A, observations were performed in mapping mode. 13 observations were made with the SL module, stepping across the source in $1.8\arcsec$ increments (50\% of the slit width). Similarly, five observations were made with the LL module, again stepping across the source in half-slit width increments ($4.85\arcsec$). In both cases, off source data for background subtraction were obtained roughly $100\arcsec$ perpendicular to the slit stepping direction. These off-source data were obtained automatically as part of the observing program. During an observation utilizing the LL slit on source, off-source observations are simultaneously made with the SL slit, and vice versa. Figure \ref{fig:cygaslit} shows the slit locations across the source.

\section{Data Reduction}

Data from the Spitzer Space Telescope is processed in two stages. First, a software pipeline at the SSC performs standard image processing. These Basic Calibrated Data (BCD) products are then made available to observers for further reduction and analysis through the Spitzer Science Center (SSC). The IRS instrument team and the SSC have provided software tools for final end-user processing.

\subsection{Pipeline Reduction}
\label{reduction:pipeline}

After being downlinked from the satellite individual observations are subject pipeline reduction to remove most instrumental effects and CR hits from the data. I will provide a short overview, however for more detail, see the IRS Data Handbook \citep{IRSmanual}.

 The first step in the pipeline flags pixels affected by CRs. This is done by locating jumps in ramp readouts. Due to the non-destructive readout, the ramp slope can be measured on either side of the jump, and the correct incident flux can be recovered (see Figure \ref{irs:readout} and text above for more details). 

On these detectors the charge on a pixel is affected by the overall flux on the array (called ``droop''). The next step in the pipeline factors in the total flux incident on the array and corrects individual pixel values. Next, dark frames are subtracted from the data, to remove charge due to thermal excitation. As the pixel response to incident flux is generally non-linear, the pixel responses are linearized using a model for each individual pixel.

With cosmic ray hits removed and the pixel response linearized, the incident flux can be determined by measuring the slope of the ramp.

The IRS chips feature 4 readout channels and slightly varying gains in each channel can cause a ``jailbar'' effect in the data. This is corrected next in the pipeline. The final corrections involve removing stray light in the SL regime and applying a flat field to correct for overall pixel sensitivity variations.

The output of the pipeline is a set of files comprising the basic calibrated data (BCD). They are downloaded by the end user and are the starting point for further calibration and analysis. The primary BCD files are:

\begin{enumerate}
\item{\emph{OBS\_IDENTIFIER}\_bcd.fits}
\item{\emph{OBS\_IDENTIFIER}\_func.fits}
\item{\emph{OBS\_IDENTIFIER}\_bmask.fits}
\end{enumerate}

The \emph{OBS\_IDENTIFIER}\_bcd.fits file contains the pipeline processed observations. It is laid out as shown in figure \ref{irs:chip}. \emph{OBS\_IDENTIFIER}\_func.fits contains estimates of the uncertainty in the pixel values. Finally, \emph{OBS\_IDENTIFIER}\_bmask.fits is a mask file of pixels (flagged both by the pipeline and the end user). 

The pipeline also provides post-BCD data products in which spectra have already been pipeline extracted and flux calibrated. However, the post-BCD products have limited flagging of bad pixels and are generally only suitable for a first look at the observations. They can also be used to examine the data for potential issues and plan processing of the BCDs.

In addition the above processed products, the SSC provides the raw data files (pre-pipeline reduction) to enable end-users to process the data in a non-standard fashion.

\subsection{BCD Reduction}
\label{reduction:BCD}

Prior to extracting and calibrating the spectra, additional instrumental effects not addressed by the pipeline processing must be removed. The detector array contains pixels which are defective or have non-linear responses (even after pipeline calibrations). An IDL script \emph{(IRSCLEAN)} is available to assist in eliminating bad pixels from the BCD files. It utilizes both bad pixel masks from the SSC and user flagged pixels to build a mask file used by extraction routines to ignore bad pixels.

Two software products are available for calibration and extraction of IRS spectra. The SSC provides the ``Spitzer IRS Custom Extraction'' (SPICE) tool \citep{SPICEmanual}, and the IRS instrument team at Cornell has developed a set of tools called the ``Spectroscopy Modeling Analysis and Reduction Tool'' (SMART) \citep{Higdon04}. Both tools can be used to extract and flux calibrate IRS spectra.

Once the BCD files have been ``cleaned'' (using \emph{IRSCLEAN}), the background must be subtracted from the images. For the low-resolution modules, this can be accomplished using off-source data. When a source with SL1, the SL2 slit is off source by $\sim100\arcsec$. By subtracting the SL2 off source FITS file from the SL1 on-source FITS file, the local background can be subtracted, leaving just the spectra. 

After the background has been subtracted, the data was loaded into SPICE for extraction and calibration. Three files are required for the extraction: \_bcd.fits, \_func.fits, and \_bmask.fits. The first file contains the observation, the second has uncertainty information, and the final file contains information about bad and ``rogue'' pixels (output from \emph{IRSCLEAN}). ``Rogue'' pixels are those pixels which are not consistently bad, but have anomalous values in a particular exposure. The first step in SPICE is to determine the location of the spectra. In the case of a point source, a profile in the position direction will show a peak at the location of a source. The extraction aperture is centered on this peak. The aperture scales with wavelength to account for the changing size of the point spread function (PSF).

For a point source, the aperture is wavelength-dependent, due to the wavelength dependent size of the PSF. The pixels are summed across this aperture to give counts vs wavelength. An additional wavelength dependent correction is applied to account for light that is scattered out of the slit (slit loss correction). 

This slit loss correction is relatively straight-forward for a point source, where the correction for light not incident on the slit can easily be made. However, in the case of an extended object, flux can also be scattered into the slit. Any correction must include both light scattered out of the slit due to the PSF, as well as light scattered into the slit from the PSF of neighboring source regions. This can only be properly done if the spatial distribution of the source is known. \emph{SPICE} includes an extended source extraction which assumes an extended source of uniform surface brightness. In this case, as much light is scattered into the slit as out, and a ``slit loss correction function'' (SLCF) \emph{uncorrects} the assumed amount of light scattered into the slit in the case of a point source.

Finally, the BCD values are flux calibrated in units of Janskys (1 Jy = $10^{-23}$ erg s$^{-1}$ cm$^{-2}$ Hz$^{-1}$).  The flux calibration is obtained through observations of standard stars. Stellar atmospheres have been modeled for these stars and facilitate the detailed calibration \citep{Decin04}.

This process much be repeated for both orders of each module, resulting in multiple output files which must be stitched together for a final spectrum. The modules overlap in wavelength, so they can be matched in flux (for a point source). The ``bonus'' order in both modules is extracted with the 1$^{st}$ order and can be used to provide an additional constraint.

\begin{figure}[h!]
\includegraphics[angle=270,width=0.75\textwidth]{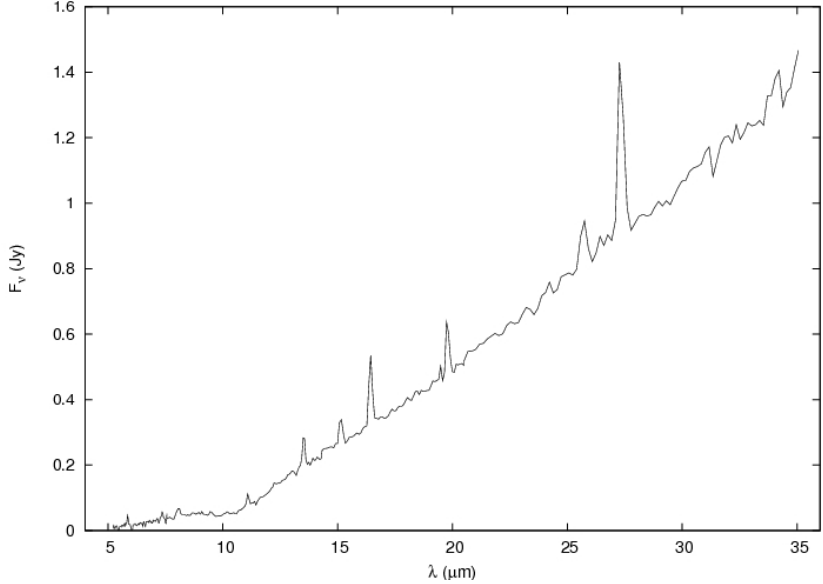}
\caption{Demonstation of the flux discontinuity near 14$\mu$m in SL and LL observations of an extended source (Cygnus A). Note that this jump is present regardless of the extraction method used (point source or extended source).}
\label{IRS:jump}
\end{figure}

As noted earlier, the SL and LL modules have markedly different slit sizes. For point sources, the amount of light scattered out of both slits can be modeled and calibrated. In the case of an extended source, the wider LL slit samples a much larger area on the source. This difference in spatial sampling is evident in a significant jump in flux seen at the intersection of the SL and LL modules (see Fig \ref{IRS:jump}).

Spectral mapping observations offer a solution to this problem. By stepping the slit across the source (at Nyquist limit), a data cube can be constructed (two spatial axes, and one wavelength axis). A spectrum can then be extracted from the cube of any arbitrary aperture at any location. This facilitates the extraction of spectra across the entire wavelength range using the same size aperture.

\section{Cygnus A IRS Data Reduction}

Cygnus A was observed using both the SL and LL modules in mapping mode. The slit was stepped across the source in half-slit width increments to facilitate the construction of a data cube. SL observations were taken at 13 slit locations ($1.8\arcsec$ shifts) and LL observations were taken at 5 slit locations ($4.85\arcsec$ shifts). For all four modules, the effective integration time was 14s at each slit location.

\begin{figure}[h!]
\includegraphics[width=0.75\textwidth]{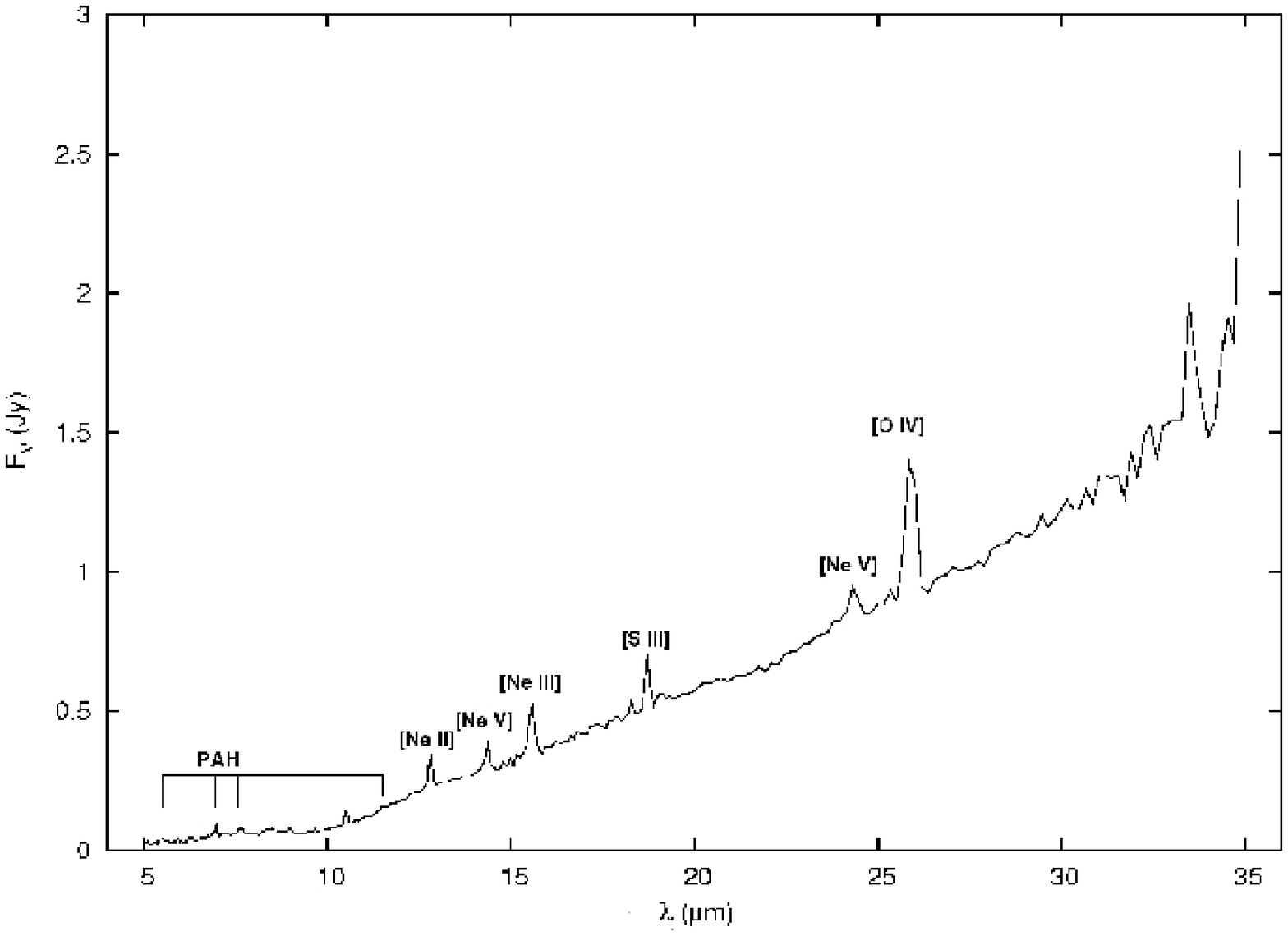}\\
\includegraphics[width=0.75\textwidth]{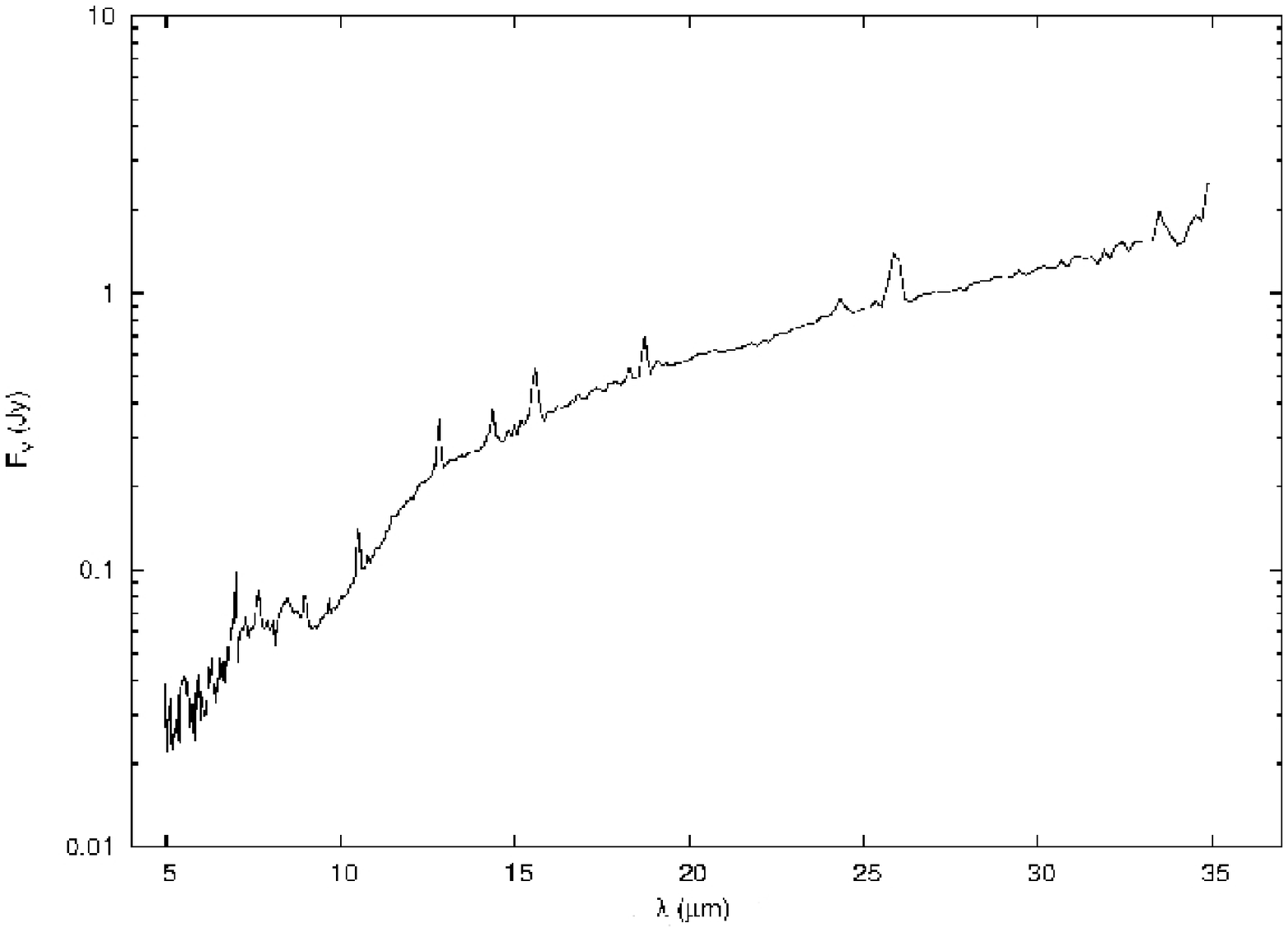}
\caption{Top: IRS spectrum for Cygnus A, as extracted from a $20\arcsec$ aperture centered on the nucleus. Bottom: Same as top, but on a log-linear scale.}
\label{fig:cyga20}
\end{figure}

After pipeline processing at the SSC, the individual BCDs were cleaned of bad and warm pixels using \emph{IRSCLEAN}. The cleaned BCDs were then loaded into \emph{CUBISM} where they were background subtracted and assembled into cubes for the SL and LL modules. A $20\arcsec$ circular aperture was used to extract the IRS spectrum from each cube (shown in figure \ref{fig:cyga20}). A $20\arcsec$ aperture was chosen to cover a well sampled region of sky. The LL slit has a $10.7\arcsec$ width, so the aperture was chosen to be as small as possible while still properly sampling the reconstrctued data cube.

\section{Additional Spitzer Data}

Imaging was also done using Spitzer's Infrared Array Camera (IRAC) at 4.5 and 8.0$\mu$m (PI: Harris). The unpublished data were downloaded from the archive, reduced and calibrated according to the instrument manual and included on SED and in the modeling. The flux densities are given in table \ref{cyga:mips}.

\citet{Shi05} observed Cygnus A using the Multiband Imaging Photometer for SIRTF (MIPS) instrument on Spitzer at 24, 70, and 160$\mu$m. Their 24$\mu$m flux is consistent with our IRS observations. However, their measurements between 70 and 160$\mu$m give a decrease in flux steeper than the Rayleigh-Jeans tail. As the flux in the MIR is probably due to thermal dust emission, such a steep falloff is likely unphysical. The BCDs were downloaded from the SSC and re-reduced.  The new reduction resulted in a higher flux density measurement at 160 $\mu$m, more consistent with thermal emission.  The new flux densities obtained from the archival MIPS data are given in table \ref{cyga:mips}.

\begin{table}
\caption{Additional Infrared Flux Densities from Spitzer}
\begin{tabular}{lc}
$\lambda$ ($\mu m$) & F$_{\nu}$ (Jy) \\
\hline\hline
4.5	& $0.010 \pm 0.003$ \\
8	& $0.054 \pm 0.013$ \\
70	& $2.20 \pm 0.11$\\
160	& $0.668 \pm 0.033$\\
\end{tabular}
\label{cyga:mips}
\end{table}

%

\chapter{Modeling Approach}
\label{chap:Modeling}

As is evident from $\S$\ref{science:SEDoverview} the spectral energy distribution of Cygnus A is quite complicated, featuring both thermal and non-thermal emission from a variety of physical regions on drastically different size scales (ranging from sub-pc to kpc scales). The combination of the various physical processes at work (AGN activity, starburst) creates a possibly complicated infrared spectrum. Here only the continuum emission will be modeled.

Strong emission lines are clearly evident in the IRS spectrum. With the exception of the polycyclic aromatic hydrocarbon (PAH) features, these emission lines generally result from either photoionization from the AGN or shock-ionization resulting from interaction with the radio jet. The emission lines and PAH features were fit using PAHFIT \citep{Smith07}. Using the fits, the emission lines were removed from the spectrum, leaving only the underlying continuum and PAH features.

\begin{figure}[h!]
\includegraphics[width=0.75\textwidth]{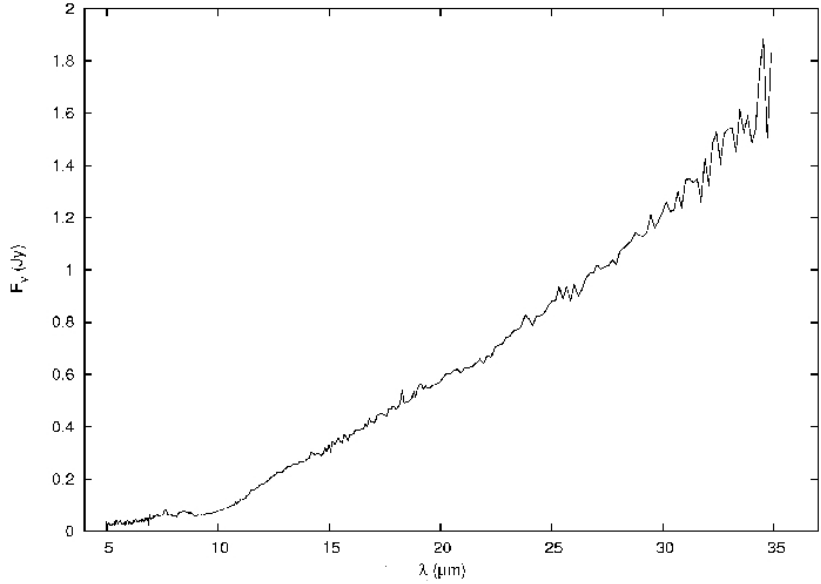}
\caption{Cygnus A IRS spectum with the emission lines removed.}
\label{fig:cyganolines}
\end{figure}

The resulting spectrum of continuum + PAH features was fit using a combination of AGN and Starburst processes as described in $\S$\ref{modeling:models}. This choice of models was motivated by the observed properties previously described (Chapter \ref{chap:science}).

The cleaned IRS spectrum was combined with a variety of data from the literature to produce a spectral energy distribution from the radio through the near IR. All fluxes were measured at the nucleus. Where possible, fluxes were extracted from a $20\arcsec$ aperture to match that of the IRS observations. Data were obtained from: \citet{Eales89}, \citet{Salter89}, \citet{Alexander84}, \citet{Wright84}, and \citet{Robson98}. Table \ref{table:data} lists the values, aperture sizes, and references for the data points.

\begin{table}[h!]
\caption{Cygnus A fluxes adopted from the literature.} 
\begin{tabular}{lccl}
$\nu$ (GHz) & $f_{\nu}$ (Jy) & Ap (\arcsec) & Ref.\\
\hline
2.7	& $1.5\pm0.4$  & 3.7x5.8	& \citet{Alexander84} \\  
5	& $0.95\pm0.2$ & 2.0x3.1	& \citet{Alexander84} \\  
15	& $1.22\pm0.2$ & 0.7x1.0	& \citet{Alexander84} \\  
89	& $0.7\pm0.07$ & 1.5x2.8	& \cite{Wright84} \\ 	  
150	& $1.11\pm0.1$ & 33.5	& \citet{Robson98} \\	  
220	& $0.65\pm0.1$ & 22.9	& \citet{Robson98} \\	  
230	& $0.59\pm0.07$& 11	& \citet{Salter89} \\	  
272	& $0.58\pm0.06$& 19	& \citet{Eales89} \\	  
350	& $0.46\pm0.06$& 14.4	& \citet{Robson98} \\	  
\end{tabular}
\label{table:data}
\end{table}

\section{Model Choices}
\label{modeling:models}

Initially, a simplistic model consisting of blackbody emission with an intervening screen of dust was fit to the data, with very poor results. A two temperature component model was also fit, with similarly poor results. Based on this, I formulated more detailed and specific models.

The choices of models have been dictated by evidence from other observations of the nuclear activity in Cygnus A. The modeling is aimed at determining the relative contributions from each component in the nucleus. Additionally, an estimate of the bolometric luminosity of the AGN can be obtained, testing the hypothesis that Cygnus A harbors an obscured quasar.

Thus for the purposes of modeling the SED of Cygnus A, three components have been included:
\begin{enumerate}
  \item{Clumpy, dusty torus}
  \item{Starburst}
  \item{Synchrotron emission from a jet}
\end{enumerate}
Each component will be discussed in turn, describing the motivation for its selection in the context of Cygnus A and the parameters available to match observations. 

\subsection{Torus Model: CLUMPY}

Cygnus A is a radio-loud active galaxy believed to be powered by a supermassive black hole at its center (Chapter \ref{chap:science}). Observational evidence indicates the inner region of the AGN is obscured, possibly by a geometrical structure similar to that predicted by unification models (see $\S$ \ref{science:agnparadigm}). Models describing the predicted AGN structure can be compared with the observations to test these unification theories.  

In Cygnus A, the radio jets are oriented roughly in the plane of the sky, so doppler (de)boosting effects are roughly equivalent for both the jet and counter-jet. The primary test of unification models in Cygnus A is studying the geometric structure obscuring the central SMBH and accretion disk. The original concept of an obscuring torus was of a thick ``donut'' surrounding the accretion disk, obscuring the accretion disk from certain viewing angles. The stability of such a structure was called into question and simulations indicated that a thick molecular torus would collapse into a disk within a few dynamical timescales.

Improvements in the torus models have resulted in a ``clumpy'' torus, where discrete clouds of gas form a torus like object, surrounding the accretion disk. Radiation pressure can support the torus against gravitational collapse \citep{Pier92}. 

\begin{figure}[h!]
\includegraphics[width=0.5\textwidth]{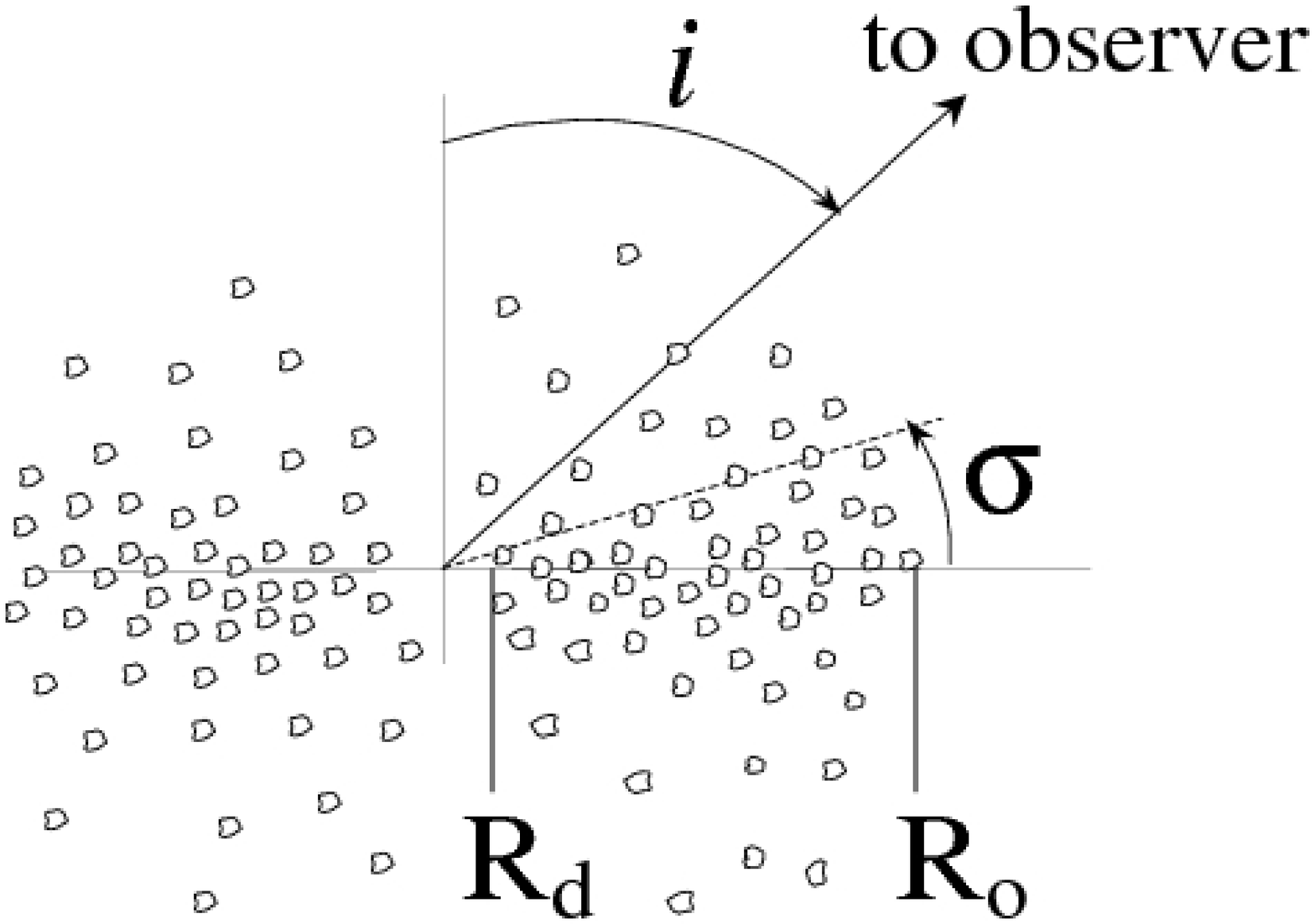}
\caption{Geometry of the CLUMPY torus model as described in \citet{Nenkova02,Nenkova08}. Figure from \citet{Nenkova08}, used with permission.}
\label{fig:clumpy-geometry}
\end{figure}

\citet{Nenkova02} have constructed a set of clumpy torus models using the DUSTY radiative transfer code. Called CLUMPY, their grid of models is available online\footnote{\url{http://newton.pa.uky.edu/~clumpyweb/}}. The geometry of the model is shown in Figure \ref{fig:clumpy-geometry}

The parameters for the model are given here and explained below:
\begin{enumerate}
  \item{$N$: Average number of clumps along the line of sight}
  \item{$Y=\frac{R_o}{R_d}$: Ratio of outer to inne radius of torus}
  \item{$q$: powerlaw index of the radial distribution of clumps}
  \item{$\sigma$ (degrees): characteristic width of the angular distribution of clumps. $\propto e^{\frac{90-i}{\sigma}}$}
  \item{$\tau_V$: Optical depth of a clump at 0.55$\mu$m (all clumps are assumed to have the same optical depth)}
\end{enumerate}

The outer radius of the torus $R_o$ is $Y$ times the inner radius ($R_d$). The inner radius is determined from a dust sublimation temperature of $T=1500$ K ($R_d=0.4*L_{45}^{0.5}$ pc, where $L_{45}$ is the bolometric luminosity of the AGN in units of $10^{45}$ erg s$^{-1}$). All models fit used a torus with a gaussian angular distribution, with $\sigma$ parameterizing the width of the angular distribution from the plane. $q$ is the index of the powerlaw distribution of clumps in the radial direction: $r^{-q}$. Table \ref{table:parameters} gives the range in values for these parameters during the modeling described here.

\begin{figure}[h!]
\includegraphics[width=\textwidth]{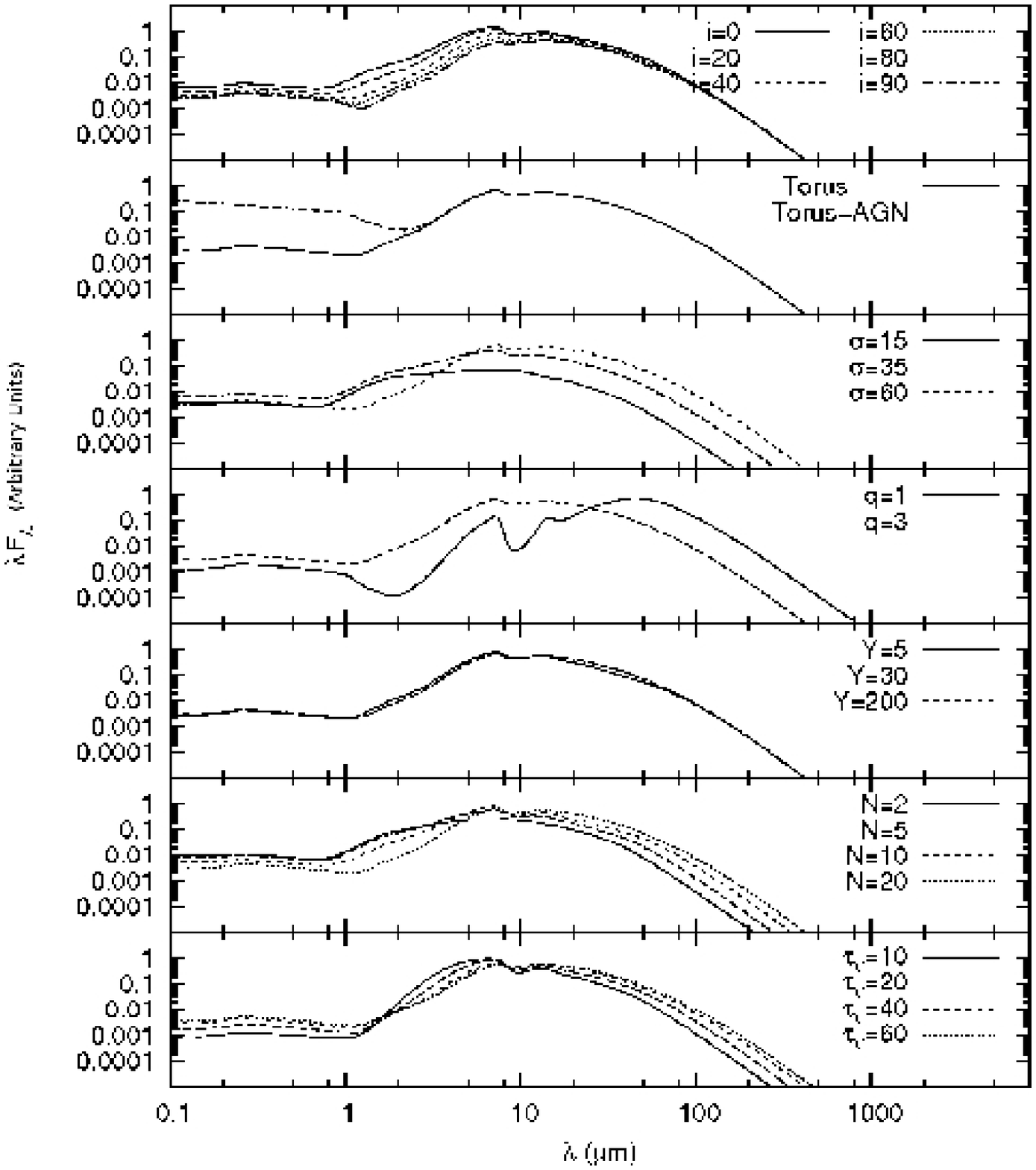}
\caption{$\lambda F_{\lambda}$ vs $\lambda$ SED for the CLUMPY model, showing the effect of changing the various parameters on the output. The base parameters were: $N$=20, $Y$=200, $q$=3, $\sigma$=60, $\tau_{V}$=40, accretion disk powerlaw not visible, and $i$=60. The following parameters are varied in plots, starting at the top: $i$, accretion disk powerlaw visible/not visible, $\sigma$, $q$, $Y$, $N$, $\tau_V$. }
\label{fig:clumpychange}
\end{figure}

In addition to the above parameters the models are provided with and without the AGN emission directly visible. There is a finite probability the AGN emission might be visible through the clumpy torus. Figure \ref{fig:clumpychange} shows the effect of varying each parameter.

The AGN spectrum used in the clumpy models is: 

\begin{equation}
\lambda f_{\lambda} \propto
\begin{cases}
  \lambda^{1.2}	& \lambda \leq \lambda_h \\
  \text{const}	& \lambda_h \leq \lambda \leq \lambda_u \\
  \lambda^{-0.5} & \lambda_u \leq \lambda \leq \lambda_{RJ} \\
  \lambda^{-3}	& \lambda_{RJ} \leq \lambda \\
\end{cases}
\label{eq:AGNspec}
\end{equation}

Where $\lambda_{RJ}$ is the wavelength at which the spectrum becomes Rayleigh-Jeans ($1 \mu m$ here). $\lambda_h$ was taken as $0.01\mu m$ and $\lambda_u=0.1\mu m$. 

The inclination ($i$) determines the degree to which an observer sees the AGN emission absorbed by the torus. As $i$ increases, the observer goes from a (mostly) unobscured view of the central regions to an edge-on view of the torus. Correspondingly, more optical and near-infrared emission is absorbed at higher inclinations. This absorbed radiation is re-emitted isotropically, so the far-infrared emission varies little as the inclination is changed. At wavelengths shorter than 20 $\mu m$ , the clouds begin to become optically thick and the radiation is no longer isotropic.

Two files are provided for each combination of parameters: one with only the torus emission, and one with the torus emission and powerlaw emission from the accretion disk. They are the same at long wavelengths, as the torus dominates. However, in the NIR, the accretion disk emission (if directly visible) begins to dominate.

The characteristic width of the angular distribution ($\sigma$) effectively determines the ``thickness'' of the torus in the vertical direction. Increasing this value results in an increased number of clumps along the line of sight for a given inclination. The additional absorption results in additional radiation being absorbed and re-emitted in the far-infrared.

Varying the powerlaw index of the clump radial distribution ($q$) effectively changes the solid angle subtended by clumps as seen by the AGN. Thus, for a steeply falling distribution (large $q$), the $N$ clumps along a line of sight would tend to be closer to the AGN, and thus subtend a larger solid angle, absorbing more radiation. This is manifested as a shift in the peak flux towards shorter wavelengths as well as a decrease in flux at longer wavelengths with increasing $q$. Also, silicate absorption around $10 \mu$m becomes less pronounced. Clouds closer to the AGN are warmer than clouds out, so a more centrally peaked distribution results in the overall emission peaking towards shorter wavelengths.

Varying the size of the torus ($Y$) has a comparatively smaller effect on the overall spectral shape. Small values of $Y$ have slightly higher flux at shorter wavelengths, with a corresponding decrease in flux at longer wavelengths, when compared to large values of $Y$. Large values of $Y$ provide a larger range in individual cloud temperatures, broadening the emission. The effect is not very pronounced in figure \ref{fig:clumpychange} because of the steep radial distribution ($q=3$) used, resulting in few clouds at large radii.

Varying the number of clumps along a line of sight ($N$) changes the overall spectral shape. Increasing the number of clumps increases the depth of the silicate absorption as well as reprocessing optical and near-infrared radiation into the far-infrared. Increasing the optical depth of a clump ($\tau_V$) has a similar effect. For larger values of $N$ and $\tau_V$, the silicate absorption becomes saturated and begins to fill back in.

For the each set of parameter values, the radiation observed at $i$ between 0$^{\circ}$-90$^{\circ}$ is computed.

\subsection{Starburst Model: Siebenmorgen \& Kr\"{u}gel}

\citet{Siebenmorgen07} have developed a set of starburst models with a variety of physical conditions. Assuming spherical symmetry and an ISM with dust properties characteristic of the Milky Way, they generated a grid of over 7000 models using radiative transfer calculations. The free parameters for the model are:
\begin{enumerate}
  \item{Starburst radius $r$}
  \item{Total Luminosity $L_{tot}$}
  \item{Ratio of bolometric luminosity in O and B stars to total luminosity $f_{OB}$}
  \item{Visual extinction from the center to the edge of the nucleus $A_v$}
  \item{Dust density in hotspots around O and B stars $n$}
\end{enumerate}
Ranges for each parameter are given in table \ref{table:parameters}.

Emission is broken down into two components: an old stellar population uniformly distributed through the volume and hot, luminous O and B stars embedded in dusty hot spots. O and B stars are massive stars ($\sim$4-100 $M_{\odot}$) with high luminosities ($\sim10^2-10^6$ $L_{\odot}$) and effective temperatures ($T_{eff}\sim 2-6\times10^{4}$ K ). As a result they produce a large quantity of ionizing photons.

The density of OB stars is centrally peaked although the model output is the emission integrated over the entire starburst. The model grid is available online\footnote{\url{http://www.eso.org/~rsiebenm/sb\_models/}}. 

\begin{figure}[h!]
\includegraphics[width=\textwidth]{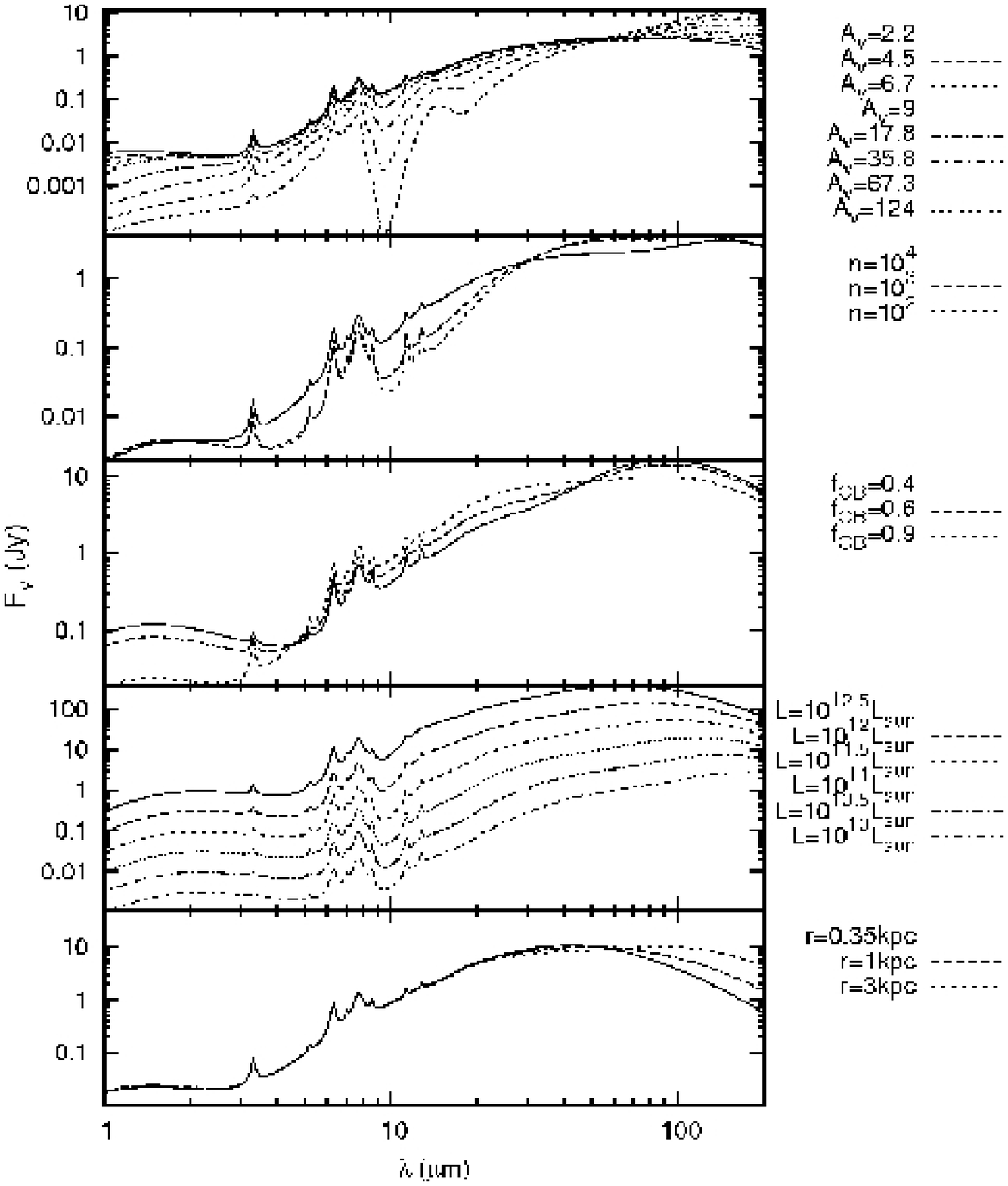}
\caption{Effect of model parameters on the output of the \citet{Siebenmorgen07} models. The following parameters are varied with the constant values in parentheses. From top: $A_v$  ($L_{tot}=10^{10.5}L_{\odot}$, r=3 kpc, $f_{OB}=0.9$, $n=10^4$ cm$^{-3}$); $n$ ($L_{tot}=10^{10.5}L_{\odot}$, r=3 kpc, $A_v=9$, $f_{OB}=0.9$); $f_{OB}$ ($L_{tot}=10^{11.1}L_{\odot}$, r=3 kpc, $A_v=4.5$,$n=10^4$ cm$^{-3}$); $L_{tot}$ (r=3 kpc, $A_v=17$, $f_{OB}=0.6$, $n=10^3$ cm$^{-3}$); $r$ ($L_{tot}=10^{11.1}L_{\odot}$, $A_v=4.5$, $f_{OB}=0.9$, $n=10^4$ cm$^{-3}$). The flux given is the spatially integrated flux at a distance of $D_l$=50 Mpc. }
\label{fig:sbchange}
\end{figure}

Figure \ref{fig:sbchange} shows the model dependence on the various parameters.  The visual extinction ($A_V$) quantifies the radiation absorbed at V-band ($550$ nm). Increasing $A_V$ results in increased absorption at UV to mid-IR wavelengths ($\lambda \leq 70\mu m$) and increased depth of the silicate absorption at $10 \mu m$. The energy absorbed at short wavelengths is re-emitted in the far-infrared, so increasing $A_V$ results in an increase in far-infrared flux.

Increasing the dust density in hot spots ($n$) has a similar effect to increasing $A_V$, however the energy is re-emitted at shorter wavelengths ($\sim 70 \mu m$ compared to $\sim > 100 \mu m$). 

The fraction of OB stars has a smaller effect on the spectrum, with an increased $f_{OB}$ resulting in increased mid-infrared flux and a decrease in optical and far-infrared flux.

The integrated luminosity of the starburst ($L_{tot}$) determines primarily the vertical position of each model. There is some effect on the releative strength of the emission features as a function of $L_{tot}$, with shorter wavelength features becoming stronger at higher luminosities. The radius of the starburst ($r$) has no effect on the spectrum at wavelengths shorter than $\sim 20 \mu m$. At longer wavelengths, increasing $r$ increases the far-infrared flux. This can be attributed to a larger volume of warm dust.

\subsection{Synchrotron Emission}
\label{sec:sync-model}

Extrapolating the powerlaw emission in the radio portion of the SED for Cygnus A to the mid-IR indicates the powerlaw must break somewhere in mid-IR. The break frequency is apparently hidden by the thermal bump, and the contribution of the synchrotron radiation to the sub-mm and IR flux is uncertain. This component is modeled as a self-absorbed power-law spectrum which breaks at high frequencies due to a cutoff in the distribution of particle energies. At lower frequencies the emitting plasma becomes optically thick and results in synchrotron self-absorption (SSA). This turnover at lower frequencies is not evident so this portion was not modeled. The accepted functional form is:

\begin{equation}
F_{\nu} \propto \left( \frac{\nu_{o}}{\nu} \right)^{-\alpha_1} \left( 1 - e^{-\left(\frac{\nu_{o}}{\nu} \right)^{\alpha_1 - \alpha_2}} \right) e^{-\frac{\nu}{\nu_c}} e^{-\tau_r(\nu)} 
\label{eq:synchrotron}
\end{equation}

where $\nu_{o}$ is the frequency at which the optical depth of the emitting plasma is equal to unity, $\nu_c$ is the frequency corresponding to the cutoff in the energy distribution of the particles, $\alpha_1$ is the spectral index in the optically thick regime, $\alpha_2$ is the spectral index in the optically thin portion, and $\tau_r$ is the dust screen between the synchrotron emitting region and the observer. 

The turnover due to self-absorption of synchrotron radiation occurs at frequencies lower than those observed, so $\alpha_2$ and $\nu_o$ were not free parameters for fitting. In Cygnus A, the spectral index $\alpha_2$ can be measured from core radio fluxes. Additionally, the amplitude of the powerlaw is fixed from the same observations. The only unconstrained parameter is the break frequency $\nu_c$ which will depend on the contributions from the previous two model components to the mid-IR spectrum.

An alternate model for the synchrotron emission at higher frequencies is a broken powerlaw. Aging of the population of relativistic electrons results in a broken powerlaw whose spectral index steepens at frequencies higher than the break frequencies \citep{Kardashev62}. The functional form adoped is.

\begin{equation}
F_{\nu}  \propto
\begin{cases} 
  \nu^{-\alpha_1}  & \nu < \nu_{break},\\ 
  \nu^{-\alpha_2}  & \nu \geq \nu_{break}
 \end{cases}
\label{eq:brokenpw}
\end{equation}

In the case of Cygnus A, the spectral index post-break would be $\alpha_2=1.24$.

Fits were run using either the exponential cutoff or the spectral break. Each implies fundamentally different populations of relativistic electrons.

To simplify references to these two fits, they will be referred to as:

\begin{itemize}
  \item{Case I: Synchrotron spectrum with an exponential cutoff at high energy}
  \item{Case II: Synchrotron spectrum with a broken powerlaw - aging population}
\end{itemize}

\section{Modeling Implementation}

The radio through mid-IR data were modeled using a with a linear combination of the three models listed above (CLUMPY torus, starburst, and synchrotron emission). The modeled SED covers $\sim10^9 - 10^{14}$ Hz ($3 \mu m - 30~cm$) in the rest frame of Cygnus A. The best fit combination (using the reduced $\chi^{2}$) was determined using Levenberg–-Marquardt minimization. Some parameters are variables within the code and can be continuously varied while others must be changed by loading a new text file corresponding to that parameter set and model.

An output from the CLUMPY code was loaded with a specific set of parameter values. The same was done for the Siebenmorgen starburst models. An initial guess was made at the bolometric AGN flux ($F_{AGN}$), the synchrotron cutoff frequency ($\nu_c$), and the optical depth of a dust slab attenuating the radio ($\tau_r$). The flux was calculated for these combination of parameters, and Levenberg-Marquardt least squares minimization was used to converge on the minimum $\chi^{2}$ value. The ``best'' values of $F_{AGN}$, $\nu_c$, and $\tau_r$ were recorded with the parameters from the loaded CLUMPY and starburst models, along with the reduced $\chi^{2}$. Then, one of the parameters was varied in either the CLUMPY or starburst model (and the appropriate files loaded), and the process was repeated. The full list of parameters and their type (variable vs fixed) is given in table \ref{table:parameters}. If a parameter is limited to discrete values, the range of values is given in the table. Otherwise ``yes'' is listed, indicating the variable is continuous.

As there were a wide variety of parameters among the various models included, it was advantageous to constrain the parameters as much as possible prior to fitting. Some had constraints from previous studies or our data. The inclination of Cygnus A is known to be between $50^{\circ}\leq i \leq 85^{\circ}$, constraining the choices of orientation for the CLUMPY model. 

The core radio fluxes specify the power-law index before the break as well as the zero point. A powerlaw was fit to the radio core fluxes and used to determine the spectral index ($\alpha=0.18$) and the flux scaling. 

The result of the combination of models was a grid of roughly $4\times10^6$ models. Due to computational constraints, it was not possible to fit each individual model combination to the data. Accordingly the model grid was sampled in all parameters to yield a more manageable selection. The sampled grid consisted of approximately 842,000 model combinations.

\begin{table}[h!]
\caption{Parameter List}
\begin{tabular}{l|l|ll}
Model   & Parameter     & Type  & Continuous?/Range \\
\hline
\multirow{8}{*}{CLUMPY} & $F_{AGN}$     & variable & yes \\
        & $N$           & variable & 1-25 \\
        & $Y$           & variable & 5, 10, 30, 100, 200 \\
        & $q$           & variable & 0-5 \\
        & $\sigma$      & variable & 10-85 (degrees, in 5$^{\circ}$ increments) \\
        & $\tau_{V}$    & variable & 10, 20, 30, 40, 60, 80, 100, 150, 200, 300, 500 \\
        & $i$           & variable & 0-90 (degrees, in 10$^{\circ}$ increments) \\
        & AGN PL visible? & variable & on/off \\
\hline
\multirow{5}{*}{Starburst}      & $r$           & variable & 0.35, 1, 3, 9, 15 (kpc) \\
        & $L_{tot}$     & variable & 10$^{10}$ - 10$^{14}$ (L$_{\odot}$) \\
        & $f_{OB}$      & variable & 0.4, 0.6, 0.9 \\
        & $A_V$         & variable & 2.2, 4.5, 7, 9, 18, 35, 70, 120 \\
        & $n$           & variable & 10$^{2}$ - 10$^{4}$ (cm$^{-3}$) \\
\hline
\multirow{6}{*}{Synchrotron} & $f_o$ & fixed    & NA \\
        & $\nu_o$       & fixed & NA \\
        & $\nu_c$ / $\nu_{break}$       & variable & yes \\
        & $\alpha_1$    & fixed & NA \\
        & $\alpha_2$    & fixed & NA \\
        & $\tau_r$      & variable & yes \\
\end{tabular}
\label{table:parameters}
\end{table}

%

\chapter{Results}
\label{chap:Results}

The synthesis of physical models (each with multiple parameters) into a coherent picture of the physical nature of an object is unlikely to be as obvious as desired. A degree of degeneracy can be expected (and indeed is seen) when attempting to fit a $\sim$15 parameter model. However, even with the ambiguity, the fits still shed light on the physical processes and conditions in the nucleus of Cygnus A. 

\begin{figure}[h!]
\includegraphics[width=0.9\textwidth]{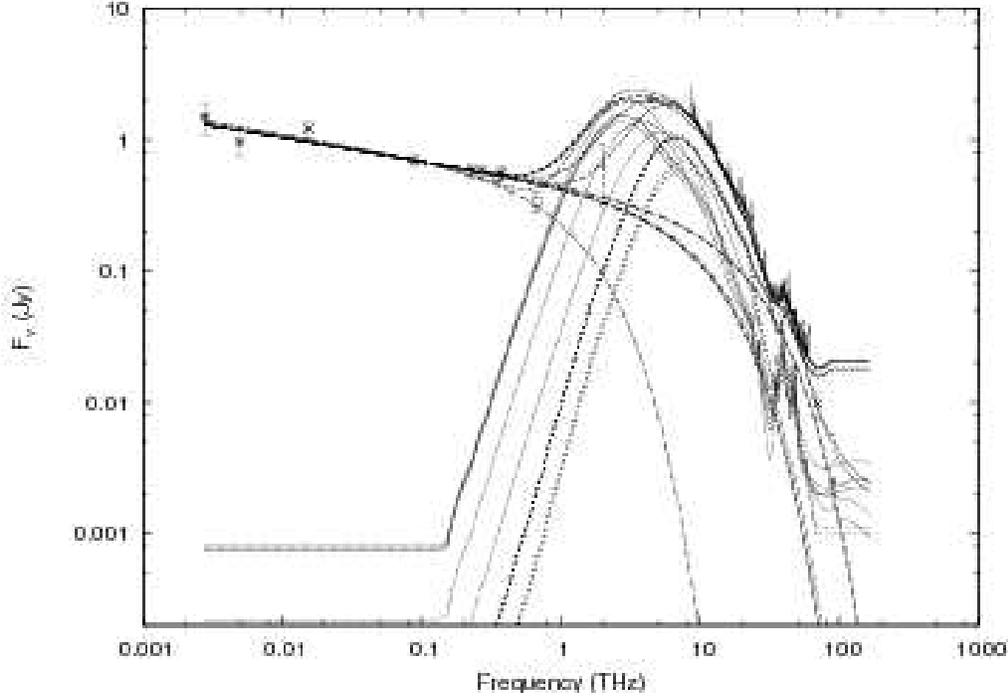}
\caption{Cygnus A SED from radio through 4.5$\mu$m with the 10 best fit models for $i=80^{\circ}$, using a synchrotron spectrum with an exponential cutoff (Case I). Data from this work and references given in Ch \ref{chap:Modeling}.}
\label{fig:results80}
\end{figure}

\begin{figure}[h!]
\includegraphics[width=0.9\textwidth]{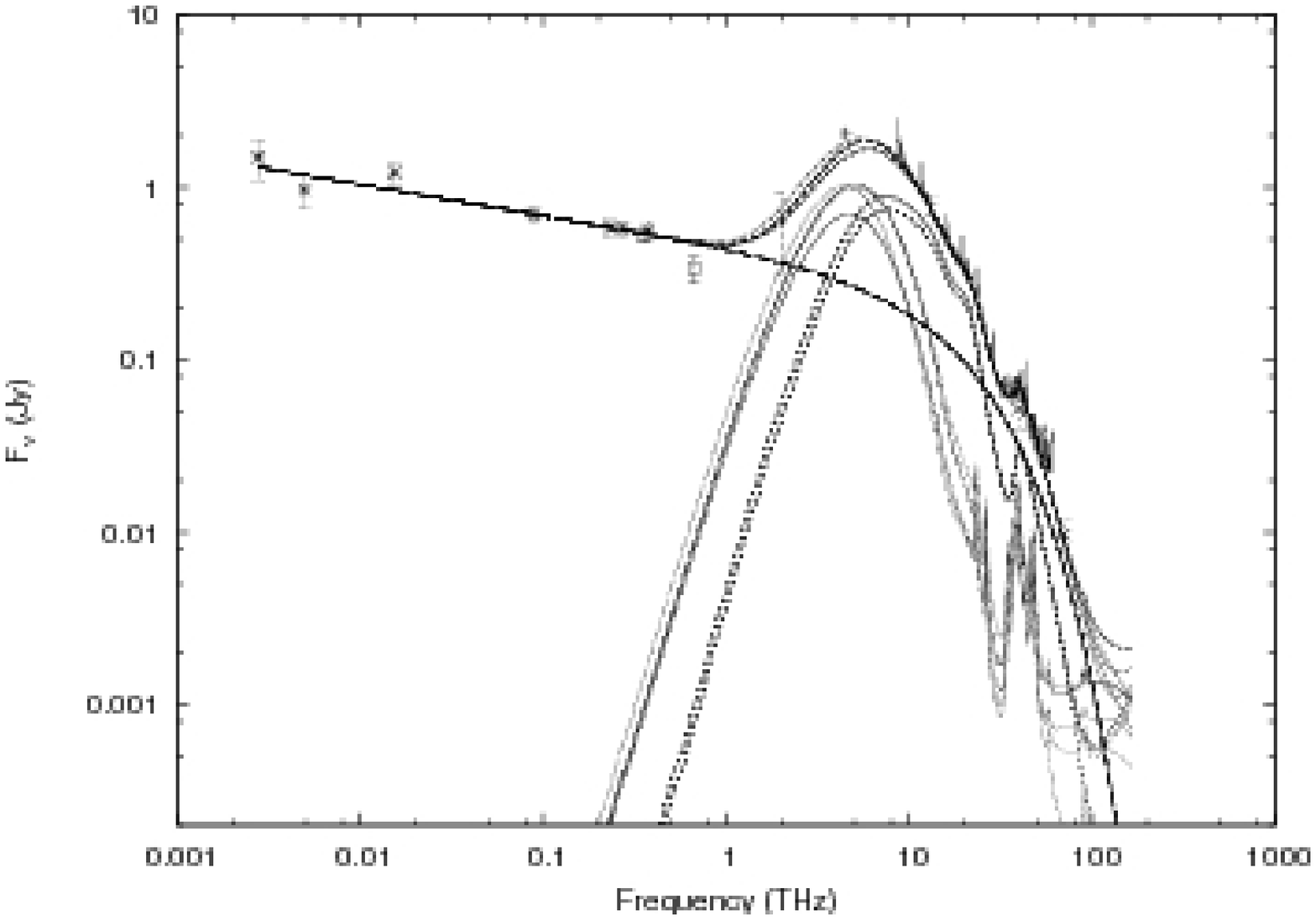}
\caption{Same as figure \ref{fig:results80} but for $i=70^{\circ}$.}
\label{fig:results70}
\end{figure}

\begin{figure}[h!]
\includegraphics[width=0.9\textwidth]{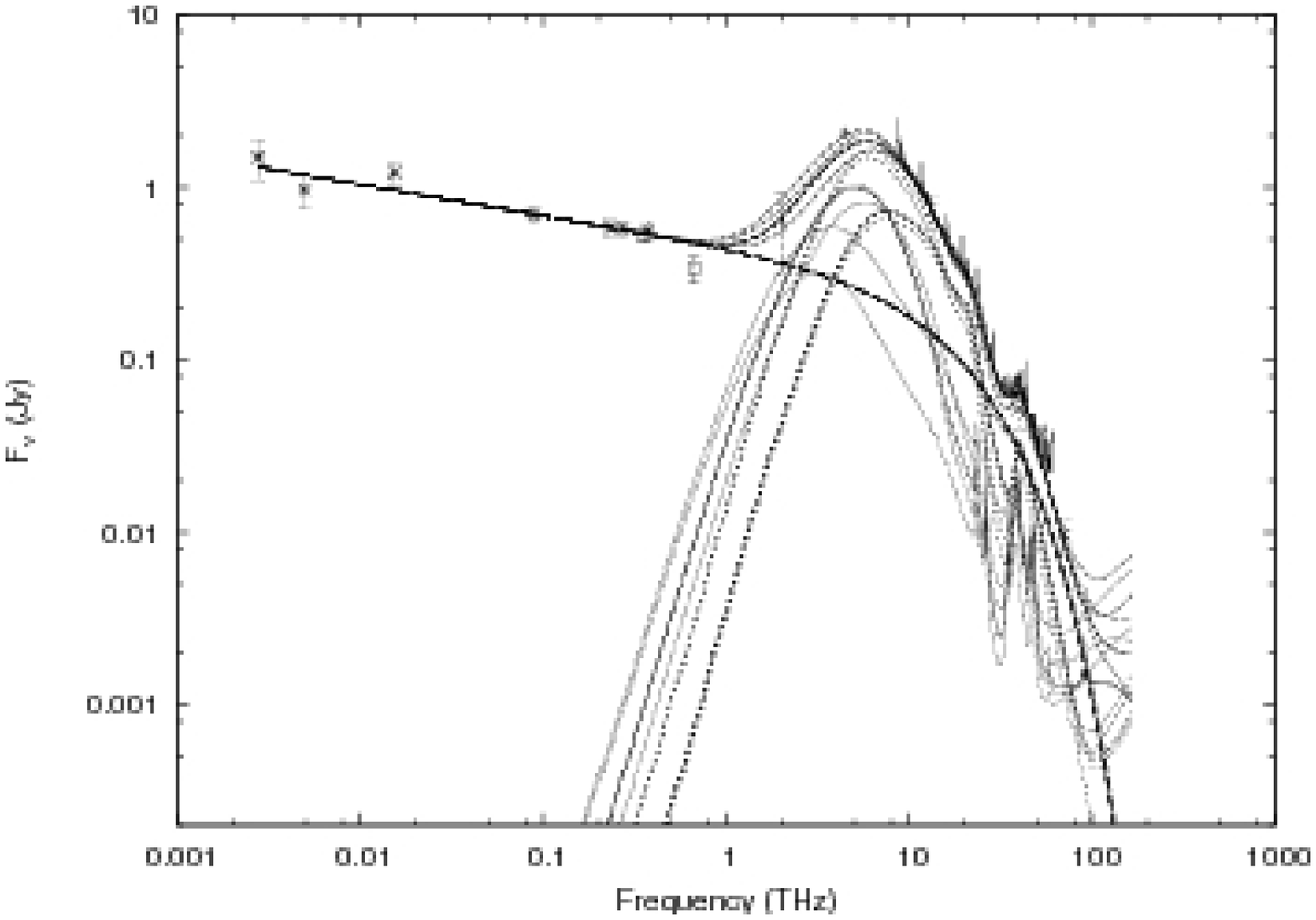}
\caption{Same as figure \ref{fig:results80} but for $i=50^{\circ}$.}
\label{fig:results50}
\end{figure}

\begin{figure}[h!]
\includegraphics[width=0.9\textwidth]{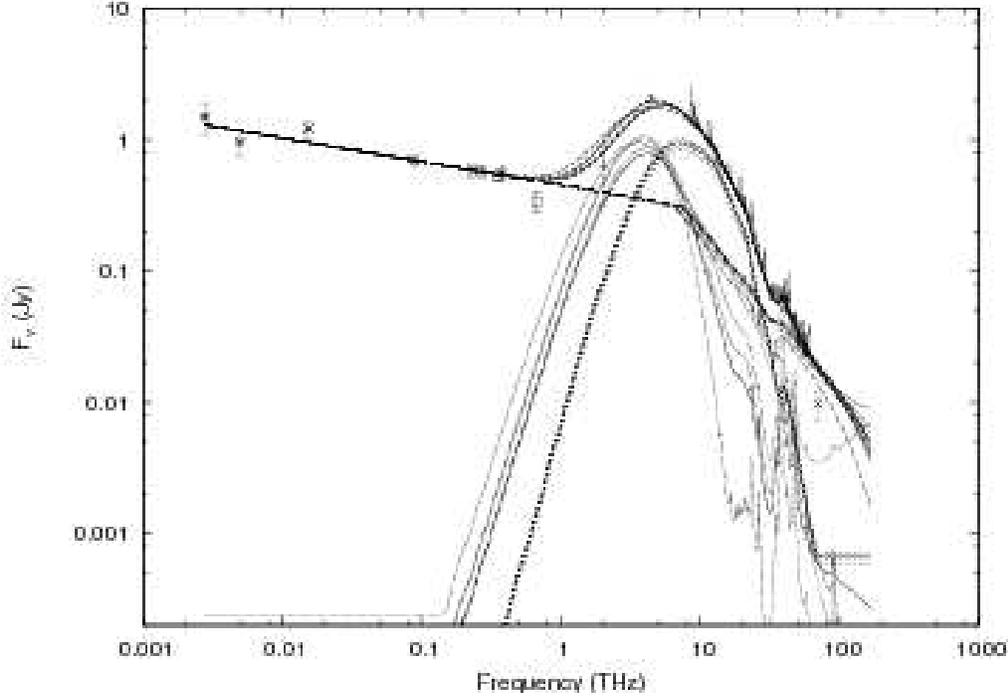}
\caption{Cygnus A SED from radio through 4.5$\mu$m with the 10 best fit models for $i=80^{\circ}$, using a synchrotron spectrum with a broken powerlaw resulting from aging of the relativistic electron population without particle injection (Case II).  Data from this work and references given in Ch \ref{chap:Modeling}.}
\label{fig:results80-b}\end{figure}

\begin{figure}[h!]
\includegraphics[width=0.9\textwidth]{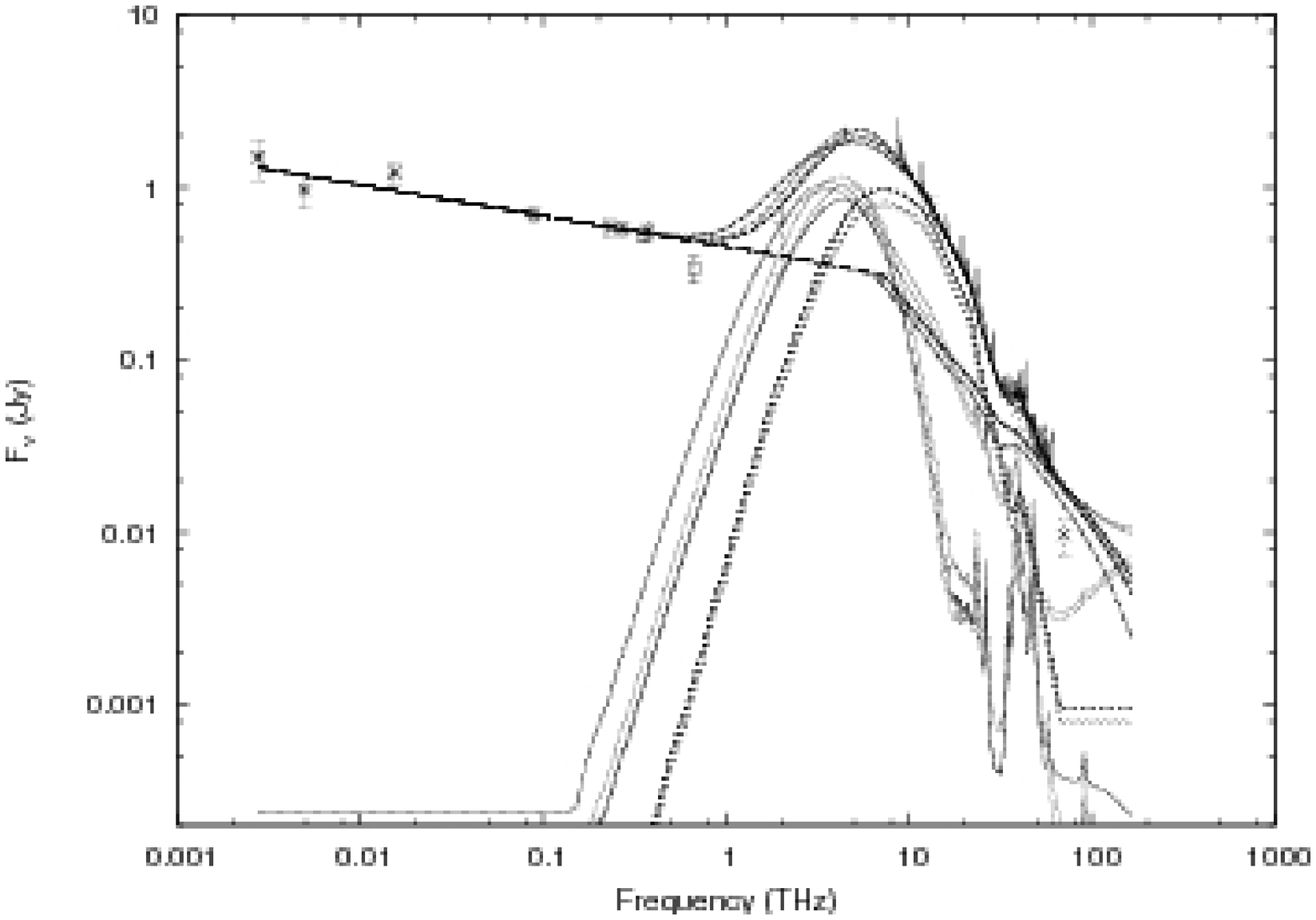}
\caption{Same as figure \ref{fig:results80-b} but for $i=70^{\circ}$.}
\label{fig:results70-b}
\end{figure}

\begin{figure}[h!]
\includegraphics[width=0.9\textwidth]{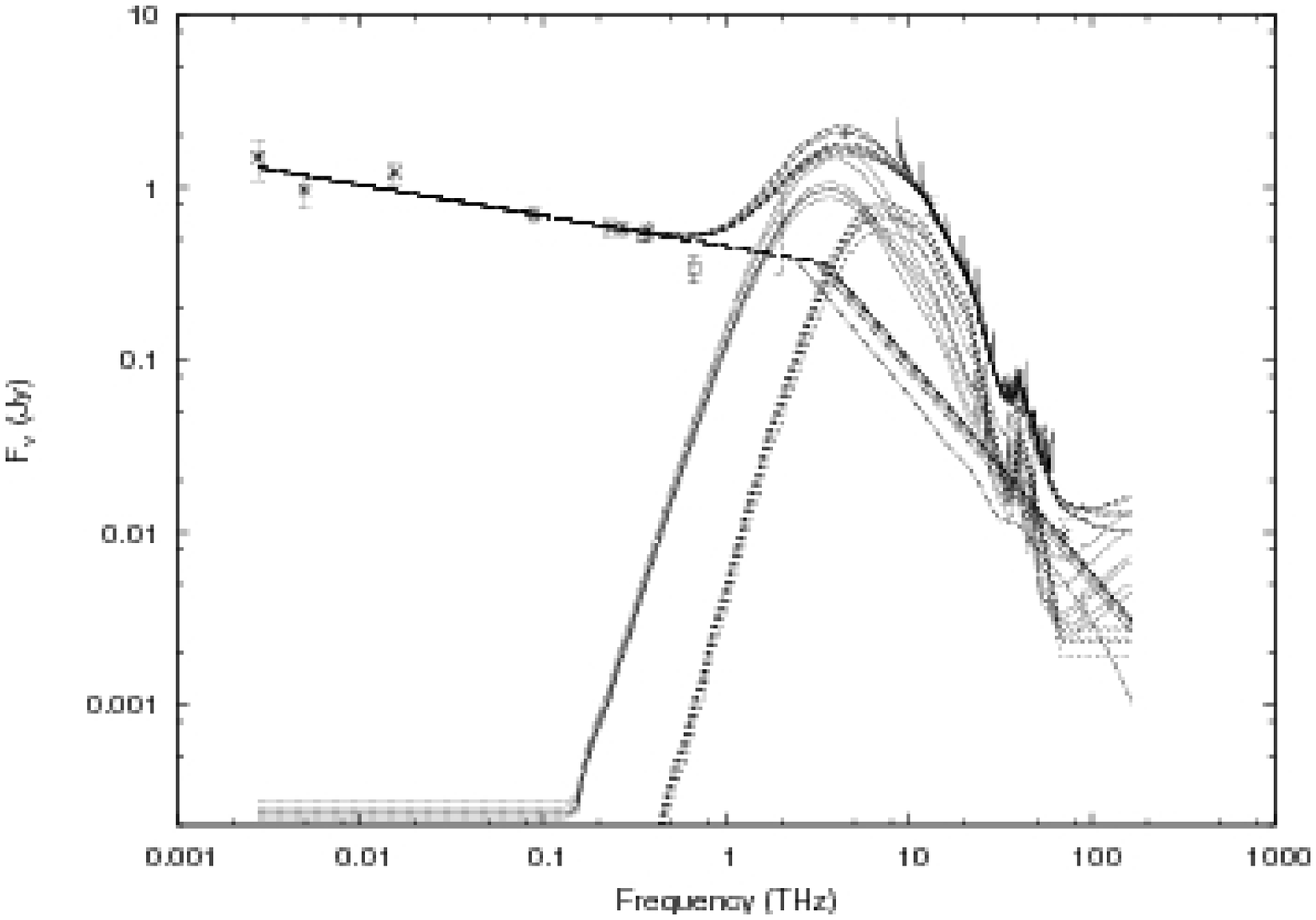}
\caption{Same as figure \ref{fig:results80-b} but for $i=50^{\circ}$.}
\label{fig:results50-b}
\end{figure}

Figures \ref{fig:results80} -- \ref{fig:results50-b} show the best 10 fits to the data for $i=80,70, 50^{\circ}$ with either an exponential cutoff (Case I) for the synchrotron spectrum or a break in the powerlaw (Case II). All fits shown have a $\chi^2/DOF < 2.3$. The models accurately reproduce the radio, sub-mm, and IR emission seen in the nuclear region. 

Each collection of model fits managed to fit the data somewhat accurately. Table \ref{table:chi} contains the average reduced $\chi^2$ value for the top 10 fits in each class.

\begin{table}[h!]
\caption{Average reduced $\chi^2$ values for the top 10 fits in each set.}
\begin{tabular}{l|cc}
$i$	& Case I	& Case II	\\
\hline
50 & 1.969 & 2.062 \\
70 & 1.828 & 1.317 \\
80 & 2.266 & 1.307 \\
\end{tabular}
\label{table:chi}
\end{table}

\section{Radio}
\label{sec:radio}

The source of the power-law emission in Cygnus A has been understood for some time as non-thermal synchrotron emission from a population of relativistic charged particles. This dominates from radio frequencies through several hundred GHz. At higher frequencies thermal emission begins to dominate the observed SED.

\section{Infrared}
\label{sec:infrared}

The non-thermal synchrotron radiation obvious below a few hundred GHz contributes to the total FIR flux, but drops off quickly after $\sim$10 THz, when the break in the particle energy distribution is reached. However, even before the  break, thermal emission from the torus and starburst dominates the synchrotron. 

The model components selected successfully reproduce the observed infrared emission (Figures \ref{fig:results80} -- \ref{fig:results50-b}). The details of each component will be discussed below, focusing on the overall results and implications of the model components. $\S$ \ref{subsec:sync-exp} discusses Case I fits (those with an exponential cutoff to the synchrotron population). $\S$ \ref{subsec:sync-break} shows results from Case II fits (those with a break in the powerlaw due to aging of the relativistic electron population).

\subsection{Synchrotron Properties and Contribution -- Case I}
\label{subsec:sync-exp}

The synchrotron emission amplitude was fixed using the non-thermal emission from the radio core, assuming it to be a point source at all frequencies observed. As a result, there were only two parameters fit: spectrum cutoff and extinction due to dust. None of the best fits had significant attenuation of the synchrotron due to dust. 

The AGN is behind a dust lane with significant extinction \citep[$A_V=50\pm30, A_K=5.6\pm3.3$][]{Djorgovski91}. Though some attenuation of the synchrotron flux is expected at higher frequencies, this alone is not sufficient to explain the cutoff. With extinction alone, the flux densities at 30 and 10$\mu m$ would be similar for the spectral index given. This would exceed the measured infrared flux at 10$\mu m$, so there must be a break in the electron population.

In general, the cutoff frequency varies with the inclination angle of the torus, with some scatter for a given inclination. Edge on inclinations generally show a cutoff around 10 THz (30$\mu$m). At this cutoff frequency, non-thermal emission contributes $\sim$10\% of the observed flux in the FIR and MIR. 

Less extreme inclinations ($i\sim$50-70) show slightly higher cutoff frequencies, on the order of 18-20 THz (15-16 $\mu$m). At these inclinations, the synchrotron again contributes $\sim$10\% of the observed flux in the FIR and MIR. However, at frequencies higher than the cutoff frequency, the synchrotron can contribute over 50\% of the observed flux (for example, around 30 THz, 10$\mu$m). With such a significant synchrotron component, variability around $10\mu~m$ might be present.

Assuming each electron emits only at its critical frequency, the gamma factor for an electron is given by eq \ref{eq:e-gamma} (where $B$ is the magnetic field strength, $\nu$ is the frequency).

\begin{equation}
\gamma = \sqrt{\frac{4\pi m_e c \nu}{3eB}}
\label{eq:e-gamma}
\end{equation}

Plugging in numerical values, this becomes eq \ref{eq:e-gamma2} (for $\nu$ in THz and B in $\mu$ Gauss. The energy of the electron can then be computed using $E=\gamma m c^2$.

\begin{equation}
\gamma = 4.884 \times 10^{5} \left ( \frac{\nu}{THz} \right )^{\frac{1}{2}} \left ( \frac{B}{\mu G} \right )^{-\frac{1}{2}}
\label{eq:e-gamma2}
\end{equation}

\begin{table}[h!]
\caption{Synchrotron Cutoff Frequencies}
\begin{tabular}{l|ccccc}
$i$ & Median $\nu_c$ & Median $\lambda_c$ & $\sigma_{\nu_{cutoff}}$ & $\gamma_{cutoff}$ &  $L_{sync}$\\
deg & THz & $\mu m$ & & B=$100 \mu G$ & $\nu \sim 10^9-10^{14} Hz$; erg s$^{-1}$\\
\hline
80 & 10.62	& 28.05 & 5.46 & 1.59 $\times$ 10$^{5}$ & $2\times10^{44}$\\
70 & 19.84	& 15.02 & 0.14 & 2.17 $\times$ 10$^{5}$ & $4\times10^{44}$\\
50 & 19.19	& 15.52 & 0.42 & 2.14 $\times$ 10$^{5}$ & $4\times10^{44}$\\
\end{tabular}
\label{table:synchrotron}
\end{table}

\begin{figure}[h!]
\includegraphics[width=0.5\textwidth]{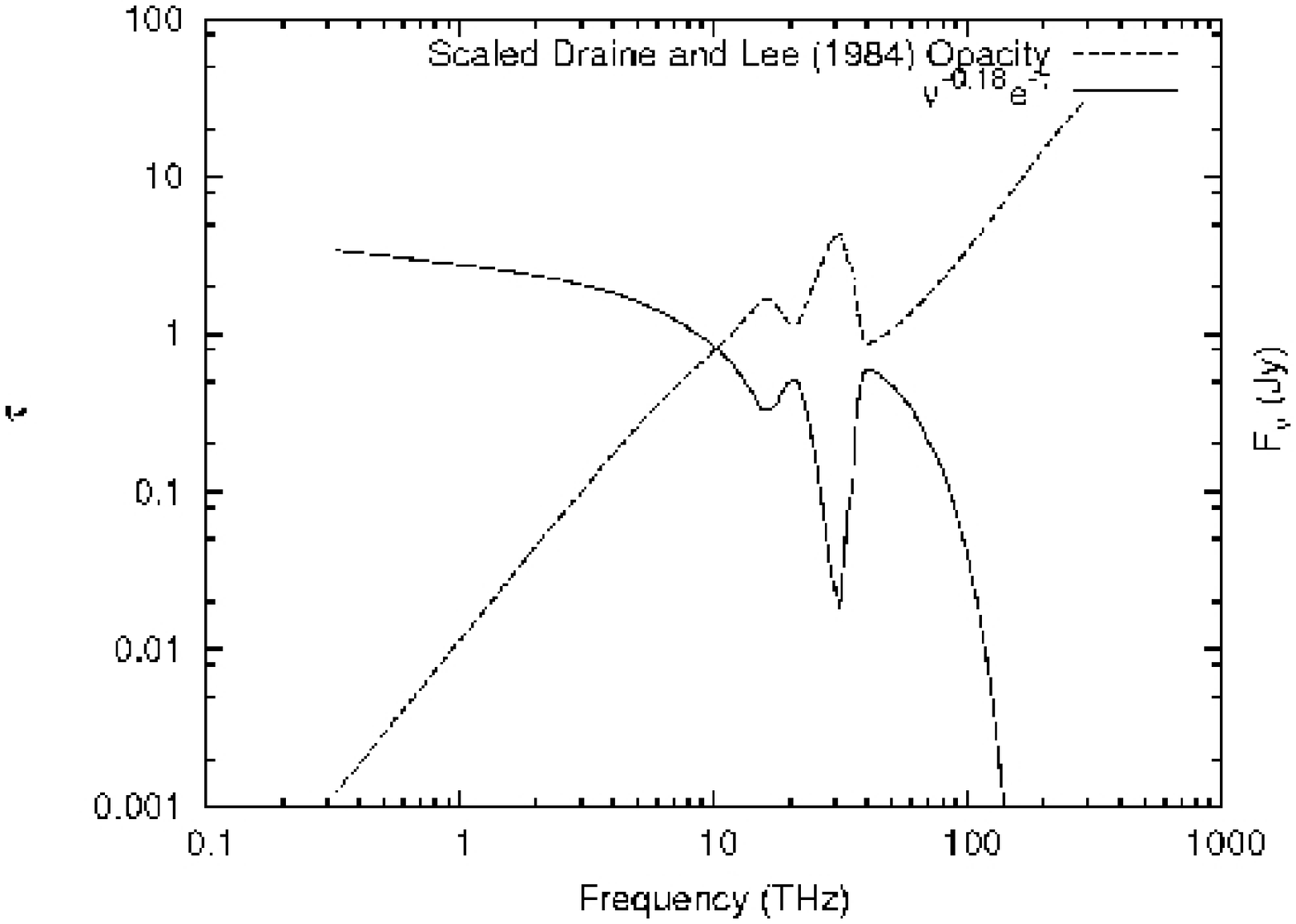}
\includegraphics[width=0.5\textwidth]{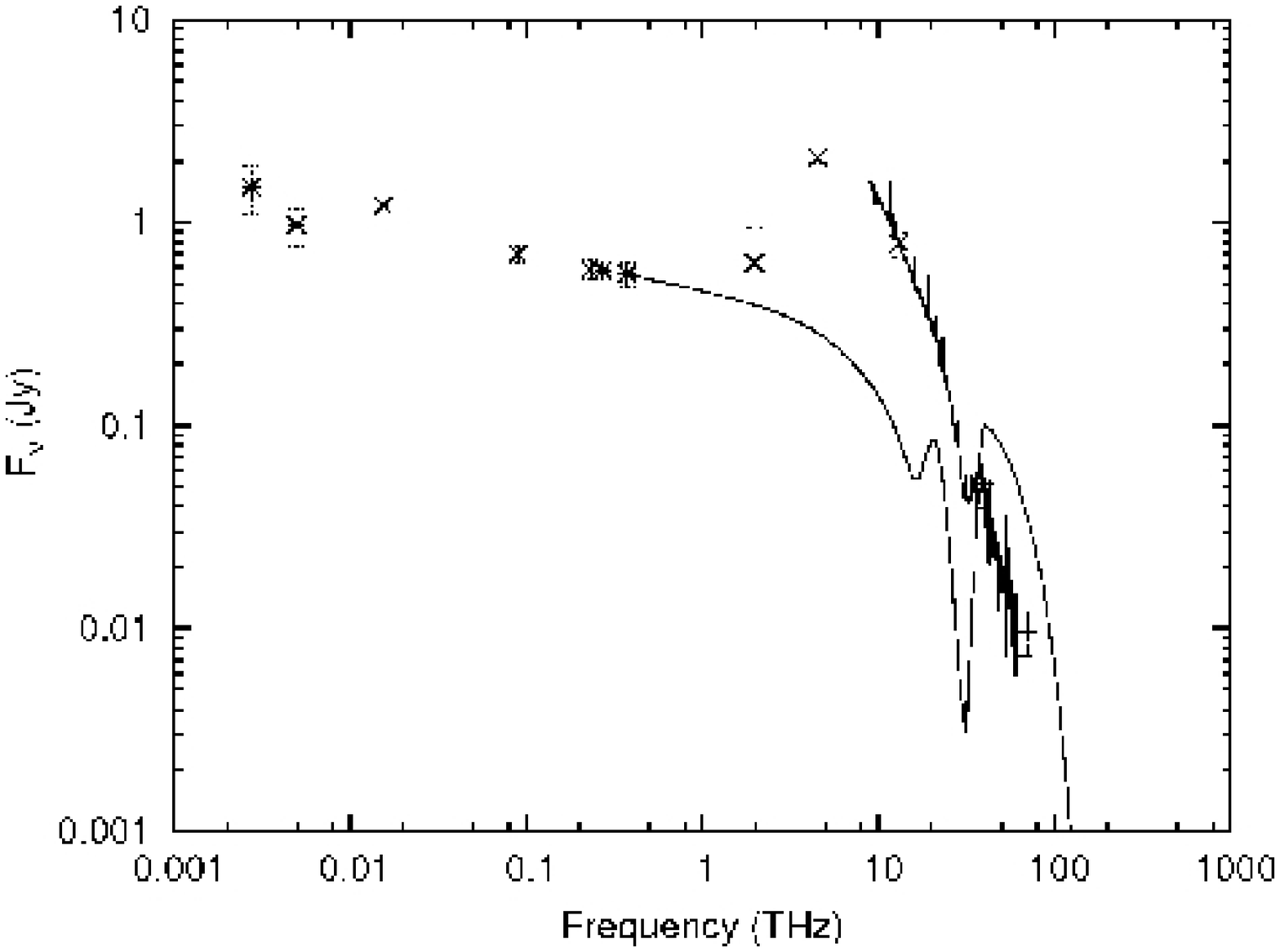}
\caption{Left: The \citet{Draine84} opacity curve scaled to $A_K$ given by \citet{Djorgovski91} (dashed line). Also plotted is a $F_{\nu}\propto \nu^{-0.18}e^{-\tau}$ (solid line), representing the measured synchrotron spectrum of the core attenuated by a dust slab. Right: The same powerlaw, overplotted with the SED of Cygnus A. An absorbing slab consistent with that is seen in Cygnus A is not sufficient to explain the observed properties of the synchrotron emission.}
\label{fig:cyga-ext}
\end{figure}

There is clearly significant extinction in the optical due to the dust lane across the nucleus. \citet{Djorgovski91} derive $A_K=5.6\pm3.3$ for the nucleus. One possible explanation for the break in the synchrotron flux is simply absorption due to the dust lane. Figure \ref{fig:cyga-ext} (LEFT) shows the opacity scaled to $A_K=5.6$. Figure \ref{fig:cyga-ext} (RIGHT) plots a synchrotron spectrum with $\alpha=0.18$ behind a dust lane with $A_K=5.6$. Comparison of the attenuated powerlaw with the emission around 50 THz indicates absorption alone cannot account for the observed spectrum of Cygnus A. The powerlaw must break somewhere in the mid-infrared. 

The best fit synchrotron parameters for a powerlaw with an exponential cutoff are listed in table \ref{table:synchrotron}. $\gamma$ is calculated for a $100~\mu$G field. Measured values of magnetic field are $\sim130~\mu$G in knots in the jet \citep{Carilli96}. These calculations assume the ``minimum energy'' condition where the energy in relaivistic particles is roughly equivalent to the energy in magnetic fields. $100~\mu$G was adopted as a round number on the order of the magnetic field strength in the kpc-scale jet. Note that magnetic field strengths measured in the VLBI radio core are higher, $0.16$ G. The lower value measured using VLA observations was chosen as the VLA ``core'' includes emission from both the VLBI core and the jet.

Energy loss in synchrotron is proportional to both the square of the magnetic field strength and the square of the particle energy. For a given particle energy, a synchrotron cooling time can be defined as in eq \ref{eq:sync-cooling}.

\begin{equation}
\frac{\tau_{sync}}{seconds} = 635 \left ( \frac{B}{Gauss} \right ) ^{-2} \left ( \frac{E}{erg} \right )^{-1}
\label{eq:sync-cooling}
\end{equation}

For the magnetic field assumed above ($100 \mu G$) and a particle energy at the cutoff (given by $\gamma$ above), typical synchrotron cooling times are roughly 12,300 years.

\subsubsection{Inverse Compton Scattering}

In addition to producing synchrotron radiation, a population of relativistic electrons will scatter photons to higher energies through inverse compton scattering with power given in eq \ref{eq:ic}. $\sigma_T$ is the Thomson cross section, $\beta=v/c$ and $U_{rad}$ is the energy density in the radiation field.

\begin{equation}
P_{IC}=\frac{4}{3} \sigma_T c \beta^2 \gamma^2 U_{rad}
\label{eq:ic}
\end{equation}

For comparison, synchrotron losses for relativistic electrons follow a similar relation, with the radiation energy density replaced by the magnetic energy density (eq \ref{eq:syncpower}, where $U_B$ is the energy density in the magnetic fields).

\begin{equation}
P_{syn}=\frac{4}{3} \sigma_T c \beta^2 \gamma^2 U_{B}
\label{eq:syncpower}
\end{equation}

Thus, the ratio of the energy lost through synchrotron to the energy lost to IC is simply given by the ratio of the energy densities. The Cosmic Microwave Background (CMB) sets a minimum value for the radiation energy density. Using the magnetic field strength assumed above ($100 \mu$ G), and $T_{CMB}=2.73$K for the CMB, synchrotron losses are dominant by a factor of $\sim1000$. A larger magnetic field would further enhance this, so IC losses from scattering of CMB photons have a negligible effect on the population of relativistic particles.

The resulting frequency of IC scattered CMB photons is proportional to $\gamma^2$. A $T_{CMB}=2.73$K blackbody has an emission peak at $\nu_{CMB}\sim160$ GHz. CMB photons at the peak which scatter off relativistic electrons emitting at the synchrotron cutoff frequency would be scattered to frequencies of $\nu_{IC}\approx8.5\times10^{21}$ Hz (from eq \ref{eq:icfreq}).  This corresponds to photon energies of $\sim35$ MeV. IC scattering off a powerlaw distribution of relativistic electrons (as in Cygnus A) results in IC emission with the same spectral index ($\alpha=0.18$).

\begin{equation}
\nu_{IC}=\frac{4}{3}\gamma^2\nu_{rad}
\label{eq:icfreq}
\end{equation}

In addition to scattering of CMB photons, the population of relativistic electrons can interact with the synchrotron radiation via IC in what is known as ``Synchrotron Self-Compton'' (SSC). Synchrotron radiation at the cutoff frequency would be scattered to frequencies of $\nu\sim6\times10^{23}$~Hz ($E\sim2.4$~GeV). 

\subsection{Synchrotron Properties and Contribution -- Case II}
\label{subsec:sync-break}

An alternate mechanism for limiting the influence of synchrotron emission at shorter wavelengths is for the spectrum to break at some frequency (see $\S$ \ref{sec:sync-model}). An identical process to that above was used to fit a broken powerlaw synchrotron spectrum (plus torus and starburst models) to the Cygnus A data. 

Fits used a pre-break spectral index of $\alpha_1=0.18$ and a post-break spectral index of $\alpha_2=1.24$, consistent with aging of the relativistic population \citep[without injection of additional particles;][]{Kardashev62}. As the flux density decreases in a slower fashion when compared to an exponential cutoff, the powerlaw must break at lower frequencies to ensure the observed midIR flux is not exceeded. Table \ref{table:sync-break} is analogous to Table \ref{table:synchrotron}, but for the broken powerlaw.

\begin{table}[h!]
\caption{Synchrotron Break Frequencies}
\begin{tabular}{l|ccccc}
$i$ & Median $\nu_{break}$ & $\lambda_{break}$ & $\sigma_{\nu_{break}}$ & $\gamma_{break}$ & $L_{sync}$\\
deg & THz & $\mu m$ & & B=$100 \mu G$ & $\nu \sim 10^9-10^{14} Hz$; erg s$^{-1}$\\
\hline
80 &  7.44      & 40.04 & 0.42 & 1.33 $\times$ 10$^{5}$  & $5\times10^{44}$\\
70 &  6.91      & 43.11 & 0.33 & 1.28 $\times$ 10$^{5}$  & $5\times10^{44}$\\
50 &  3.16      & 94.28 & 0.55 & 0.868 $\times$ 10$^{5}$ & $3\times10^{44}$\\
\end{tabular}
\label{table:sync-break}
\end{table}

\subsubsection{Inverse Compton Scattering}

Similarly to Case I, IC from a relativistic electron emitting at $\sim5$ THz will be scattered to $\nu_{IC}\sim8.0\times10^{22}$ Hz ($E\sim330$ MeV). Potentially higher energy photons could result from scattering off electrons in the post-break distribution. 

\subsection{AGN/Torus Properties and Contribution}

An estimate for the inclination of the AGN was given based on previous work in the radio regime. This constraint was used in fitting the SED. However, acceptable fits were obtained for all inclinations within the range, so no further constraint can be placed on $i$ using this analysis. For the each of the inclinations, the properties of the best fit torus model will be discussed below. Summary tables are provided for both Case I and Case II fits following the discussion of the individual models (Tables \ref{table:torus-result-I} \& \ref{table:torus-result-II}).

\subsubsection{i=80$^{\circ}$}

\begin{figure}[h!]
\includegraphics[width=\textwidth]{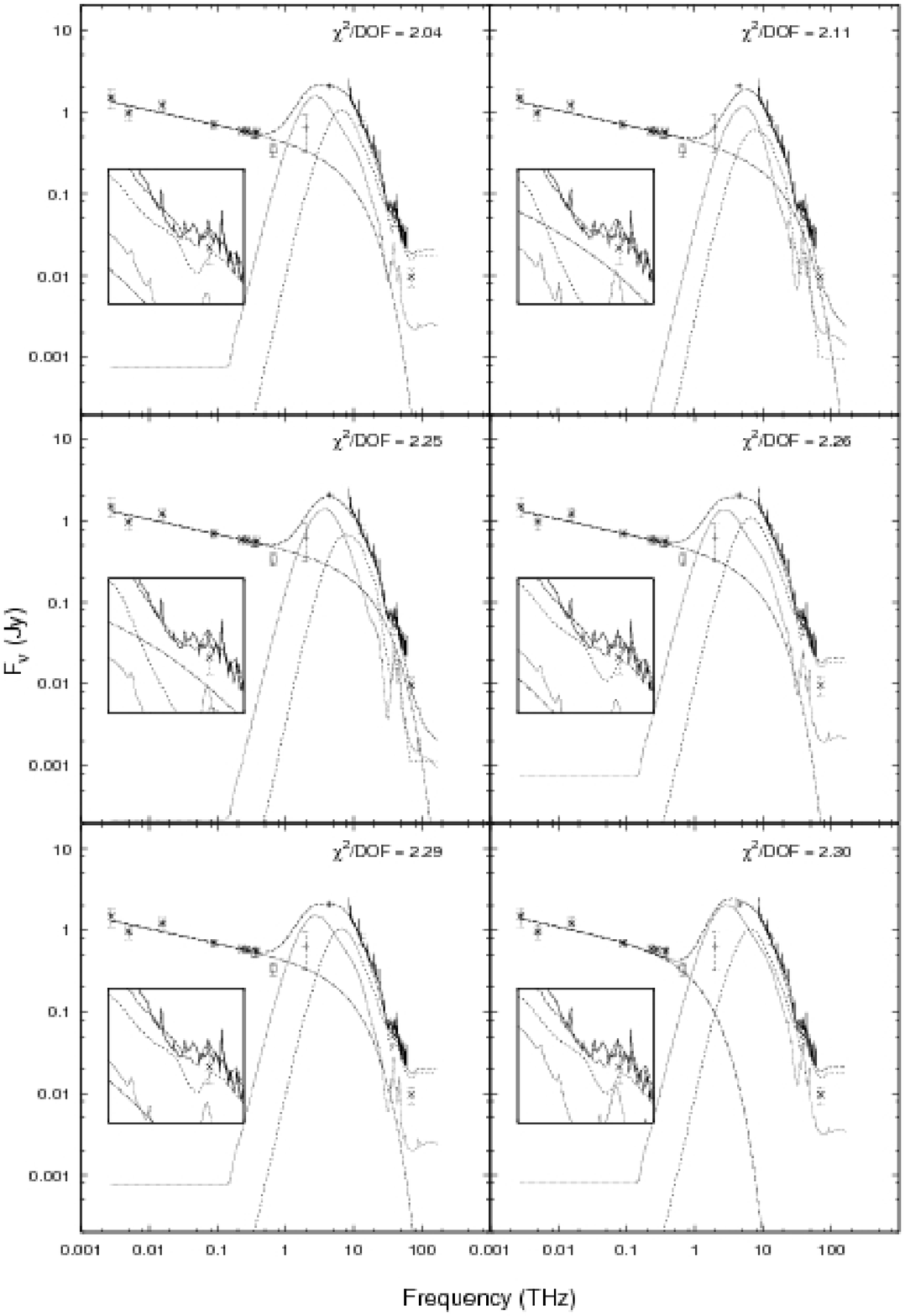}
\caption{Individual plots of the best fit models in figure \ref{fig:results80}. Legend is the same as figure \ref{fig:results80}.}
\label{fig:results80a}
\end{figure}

\begin{figure}[h!]
\includegraphics[width=\textwidth]{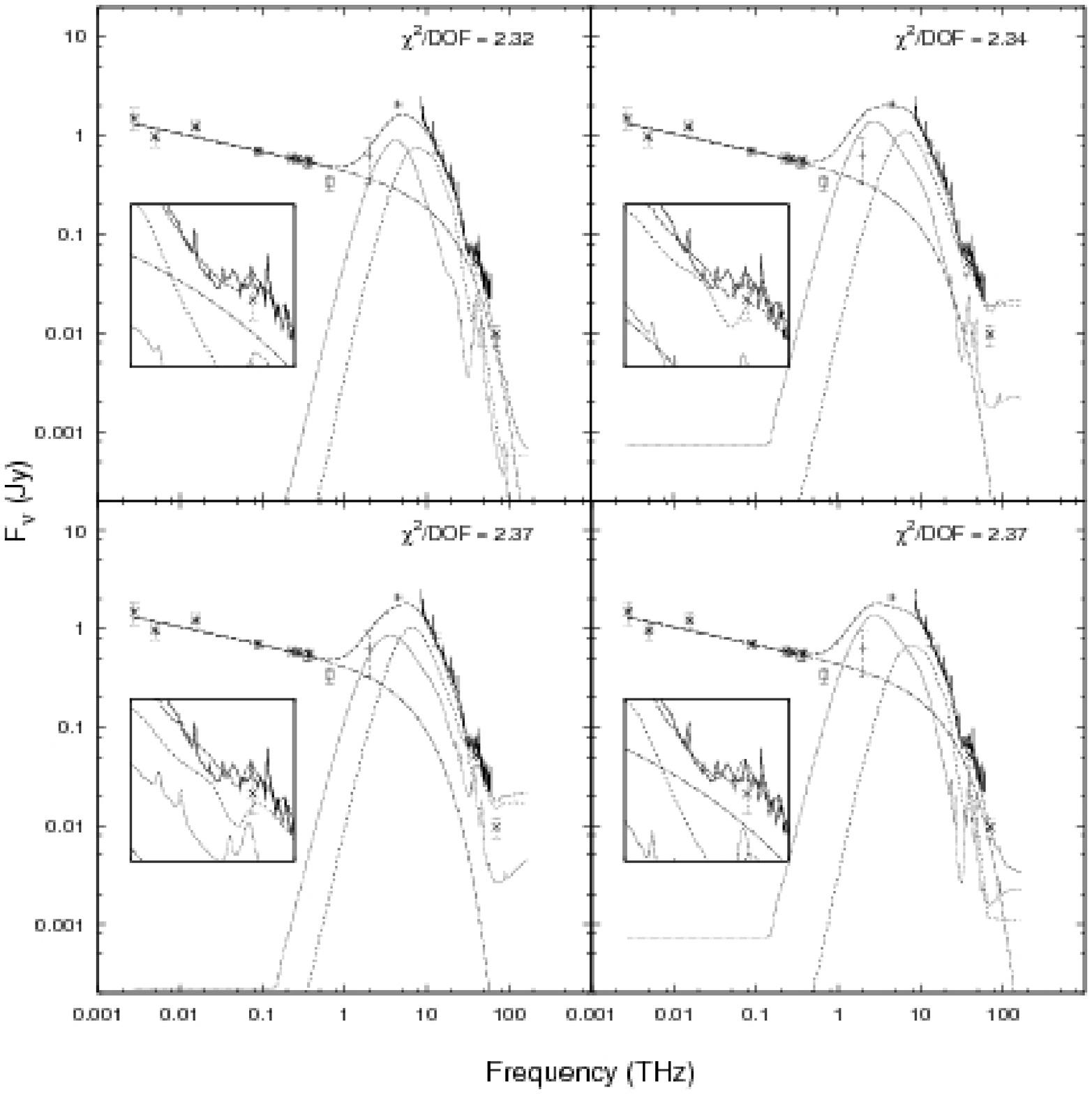}
\end{figure}

Figure \ref{fig:results80} shows the Cygnus A SED with the 10 best fit Case I models at $i=80$. These best fits are individually plotted in \ref{fig:results80a}. All models had $1.802 \leq \chi^2/DOF \leq 1.865$. In general, the best fit torus model did not have contributions from the AGN powerlaw. 

For model combinations with the AGN directly visible, the bolometric AGN flux is $\sim680\times10^{-10}$ erg s$^{-1}$ cm$^{-2}$. At a luminosity distance of 237 Mpc ($7.31\times10^{26}$ cm) the luminosity of the AGN is $L_{agn}=4.6\times10^{47}$ erg s$^{-1}$. Without the AGN directly visible, the AGN flux is $\sim14\times10^{-10}$ erg s$^{-1}$ cm$^{-2}$, giving $L_{agn}=9.4\times10^{45}$ erg s$^{-1}$. The inner radius of the torus is then $8.6$ pc for the fits with the AGN powerlaw visible and $1.22$ pc for fits where the powerlaw isn't visible.

The CLUMPY torus model which best fits the data at $i=80$ includes a visible AGN powerlaw. Due to the clumpy nature of the torus, there is a finite probability of directly viewing the AGN. These fits involve a relatively thin torus ($\sigma=15$) with a large number of clumps along the line of sight ($N=20$), each with moderately high optical depth ($\tau_v=40$). The torus is also quite broad in the radial direction ($Y=200$) and the density of clumps does not drop very rapidly. 

These models with the AGN visible require a much more luminous AGN (by at least one order of magnitude, sometimes two). The bulk of the energy output ($>90\%$) in these fits is in the optical and UV, which is not constrained here. Given such optical AGN continuum is not observed, these fits do not seem physically plausable for Cygnus A.

Mid-infrared observations of NGC 1068 (also a type II AGN) indicate the torus is contained within a 15 pc region \citep{Mason06}. Fits at $i=80$ with the AGN powerlaw visible require an outer torus radius for Cygnus A of $\sim1.7$ kpc! This is unlikely to be a realistic result, as a $1.7$ kpc torus would have easily been observed (at the distance of Cygnus A, 1" corresponds to $\sim1$ kpc, so a structure of this size would be well within range of HST and some ground-based observatories). Additionally, these fits require a bolometric AGN luminosity at least a factor of 10 larger than implied by x-ray observations \citep{Tadhunter03}, and a factor of over 100 larger than the bolometric luminosity of the AGN implied by \citet{Whysong04}. Finally, the torus model component for these fits significantly overshoots the nuclear K-band spectra presented by \citet{Tadhunter03}.

For the model in which the AGN is not visible, the torus model was markedly different in physical configuration, being much smaller ($Y=30$; $\sigma=60$), but having otherwise similar properties. 

The NIR provides an opportunity to break the degeneracy between models with the AGN powerlaw visible and those without the AGN powerlaw. The AGN powerlaw results in an upturn in the SED around 100 THz (3$\mu$m), while other models have a flux density which continues to fall. Preliminary fitting using 2MASS points indicates the AGN powerlaw exceeds the observed flux in the NIR. 

\begin{figure}[h!]
\includegraphics[width=\textwidth]{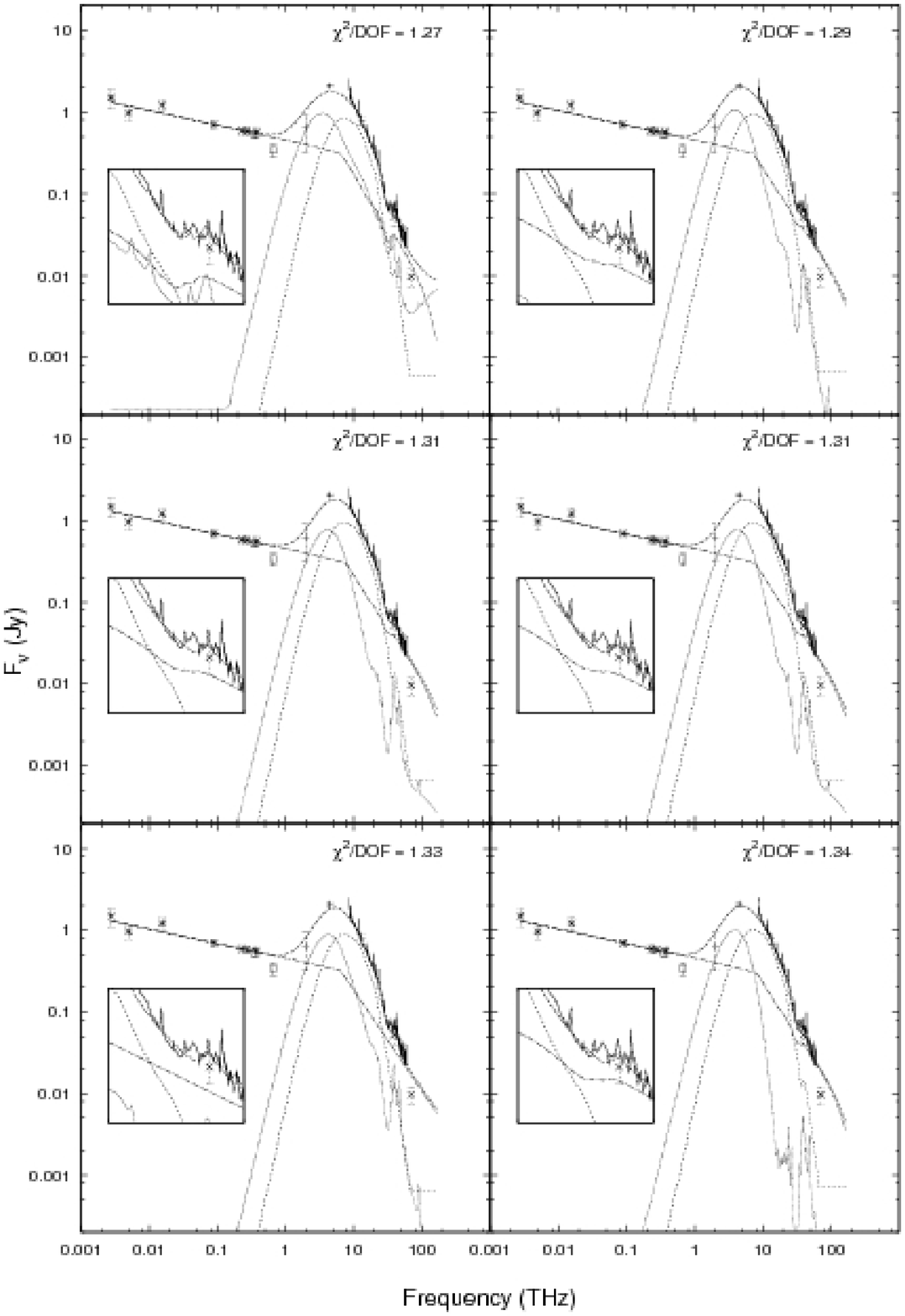}
\caption{Individual plots of the best fit models in figure \ref{fig:results80-b}. Legend is the same as figure \ref{fig:results80-b}.}
\label{fig:results80-ba}
\end{figure}

Figure \ref{fig:results80-b} shows the top 10 Case II fits at $i=80$. These models had $1.273 \leq$ $\chi^2/DOF \leq 1.336$. The torus model fits are very similar to the Case I fits where the AGN powerlaw is not visible, with an AGN that is $\sim15\%$ more luminous. Individual plots of the fits are in figure \ref{fig:results80-ba}

\begin{figure}[h!]
\includegraphics[width=\textwidth]{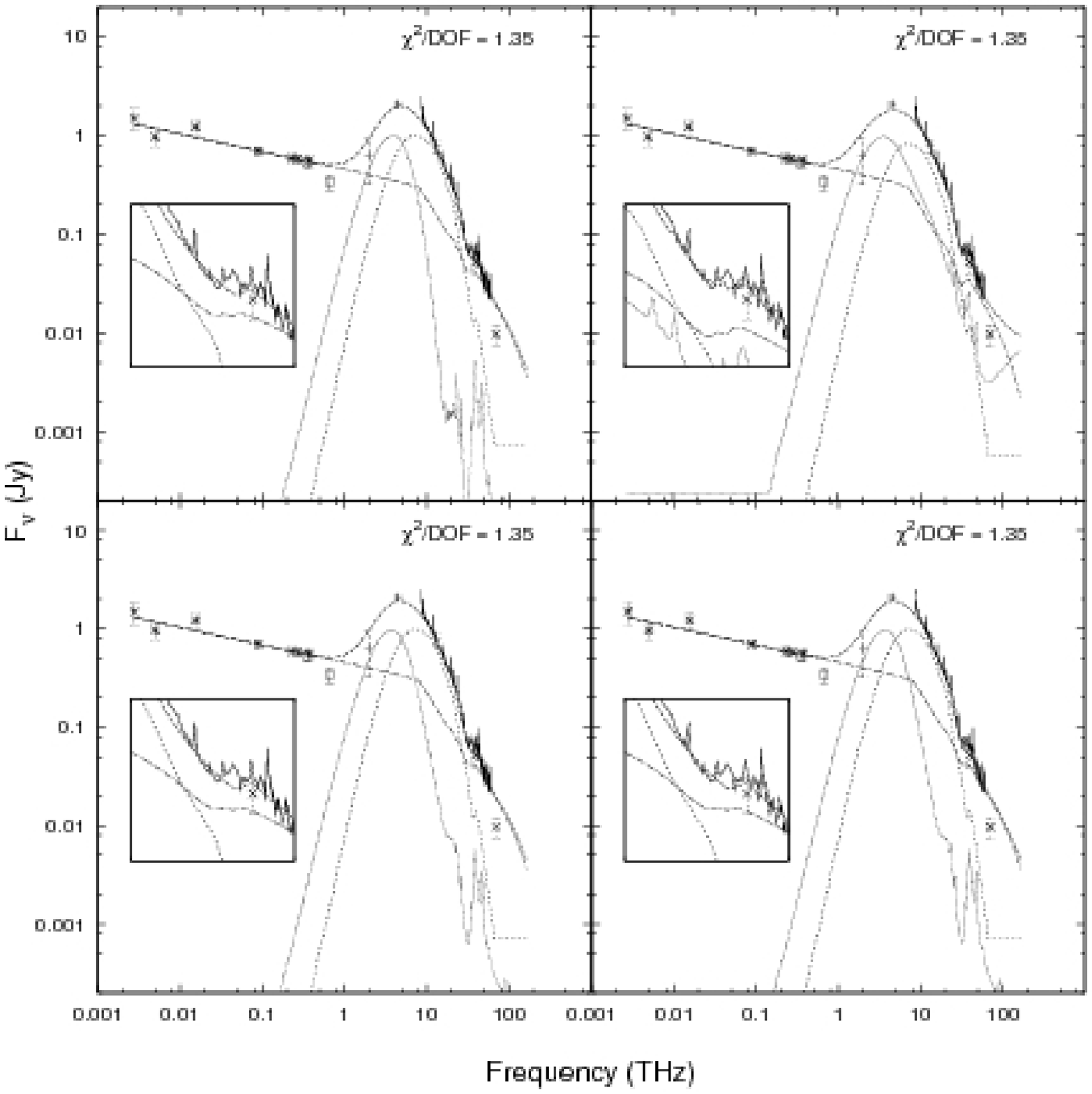}
\end{figure}

\subsubsection{i=70$^{\circ}$}

\begin{figure}
\includegraphics[width=\textwidth]{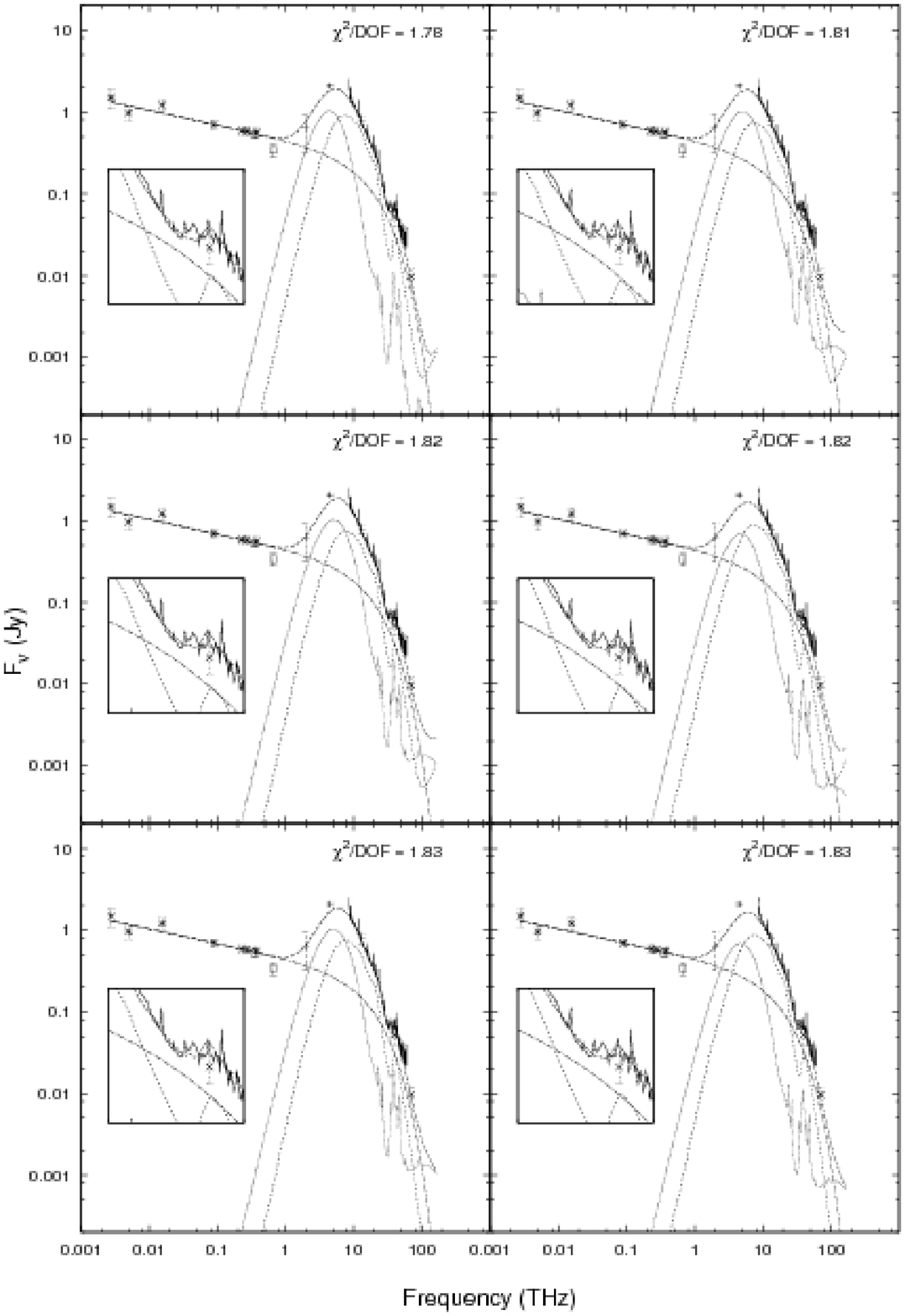}
\caption{Individual plots of the best fit models in figure \ref{fig:results70}. Legend is the same as figure \ref{fig:results70}.}
\label{fig:results70a}
\end{figure}

\begin{figure}
\includegraphics[width=\textwidth]{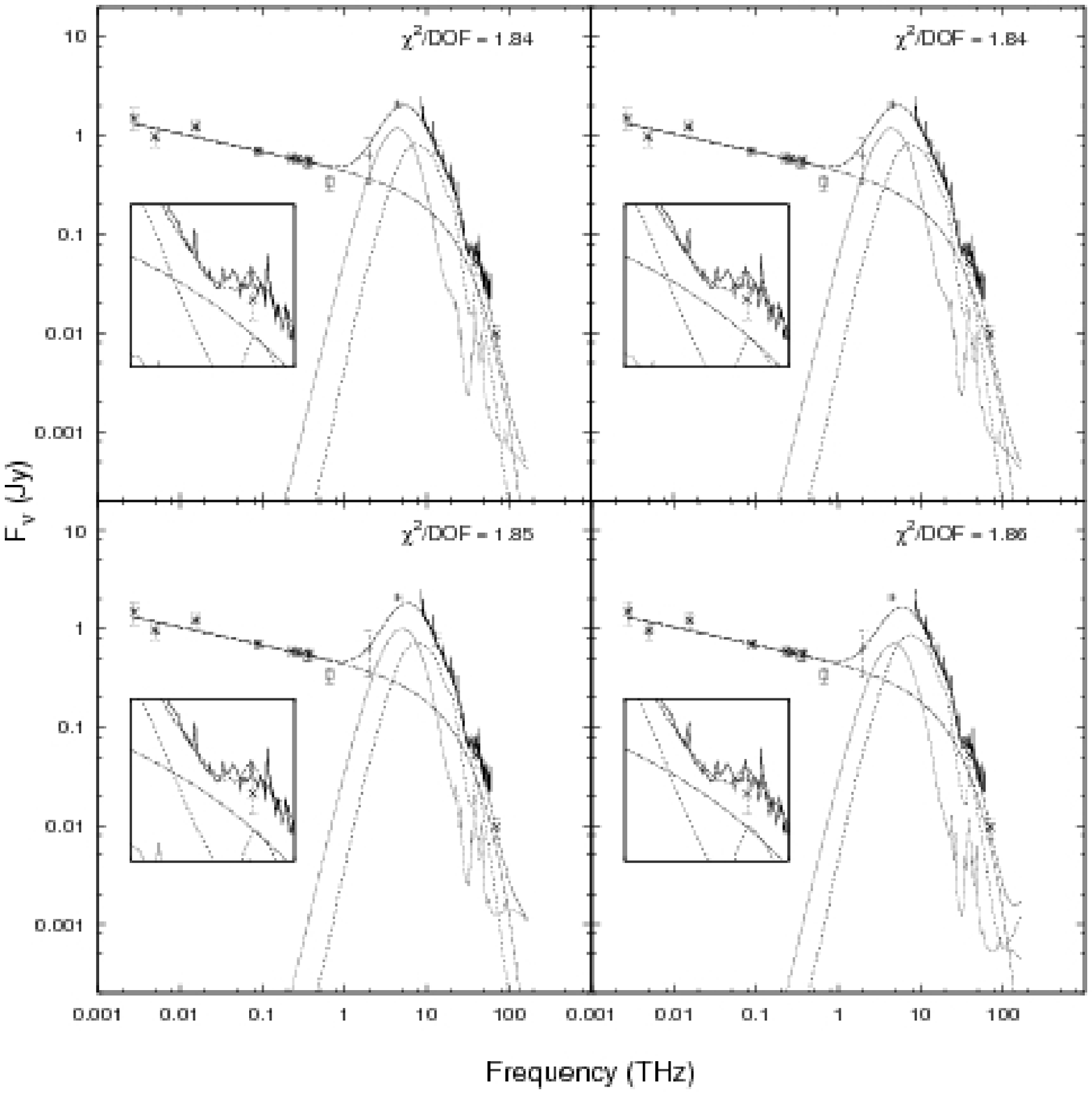}

\caption{Individual plots of the best fit models in figure \ref{fig:results70}. Legend is the same as figure \ref{fig:results70}.}
\label{fig:results70b}
\end{figure}

At the slightly shallower inclination of $i=70$ a much less luminous AGN is required to fit the data (Case I fits shown in figures \ref{fig:results70} and \ref{fig:results70a}). Typical bolometric AGN fluxes were $\sim1.15\times10^{-10}$ erg s$^{-1}$ cm$^{-2}$, corresponding to a luminosity of $\sim7.7\times10^{44}$ erg s$^{-1}$. $\chi^{2}$ values were $\sim1.8$. The corresponding inner radius of the torus is $\sim0.35$ pc. Reduced $\chi^2$ values are between 1.780 and 1.858.

As with $i=80$, torus models with and without the AGN powerlaw provide acceptable fits to the available data. However, in contrast, they both predict similar bolometric fluxes for $i=70$. 
The best fit torus models have a relatively small opening angle, with $\sigma \sim 60-65^{\circ}$. The other fit parameters are $Y=30$, $N=22$, $q=2$ and $\tau_v=60$. 

\begin{figure}[h!]
\includegraphics[width=\textwidth]{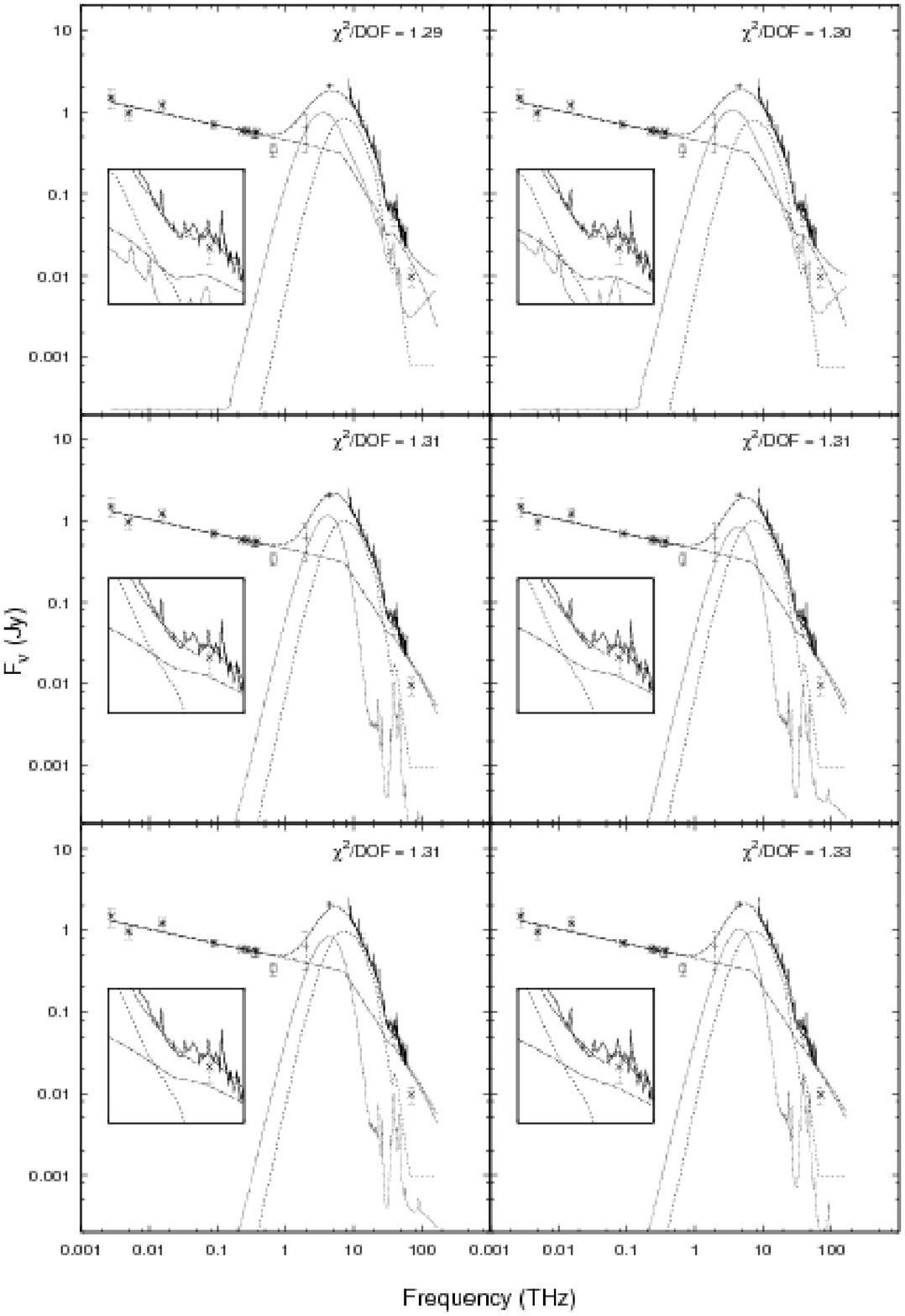}
\caption{Individual plots of the best fit models in figure \ref{fig:results70-b}. Legend is the same as figure \ref{fig:results70-b}.}
\label{fig:results70-ba}
\end{figure}

Case II fits (figures \ref{fig:results70-b} \& \ref{fig:results70-ba}) select torus models which are effectively identical to those in Case I fits in the physical characteristics. The only difference occurs in the luminosities, which tend to be $\sim50\%$ higher in the Case II fits. Reduced $\chi^2$ values are between 1.290 and 1.345.

\begin{figure}[h!]
\includegraphics[width=\textwidth]{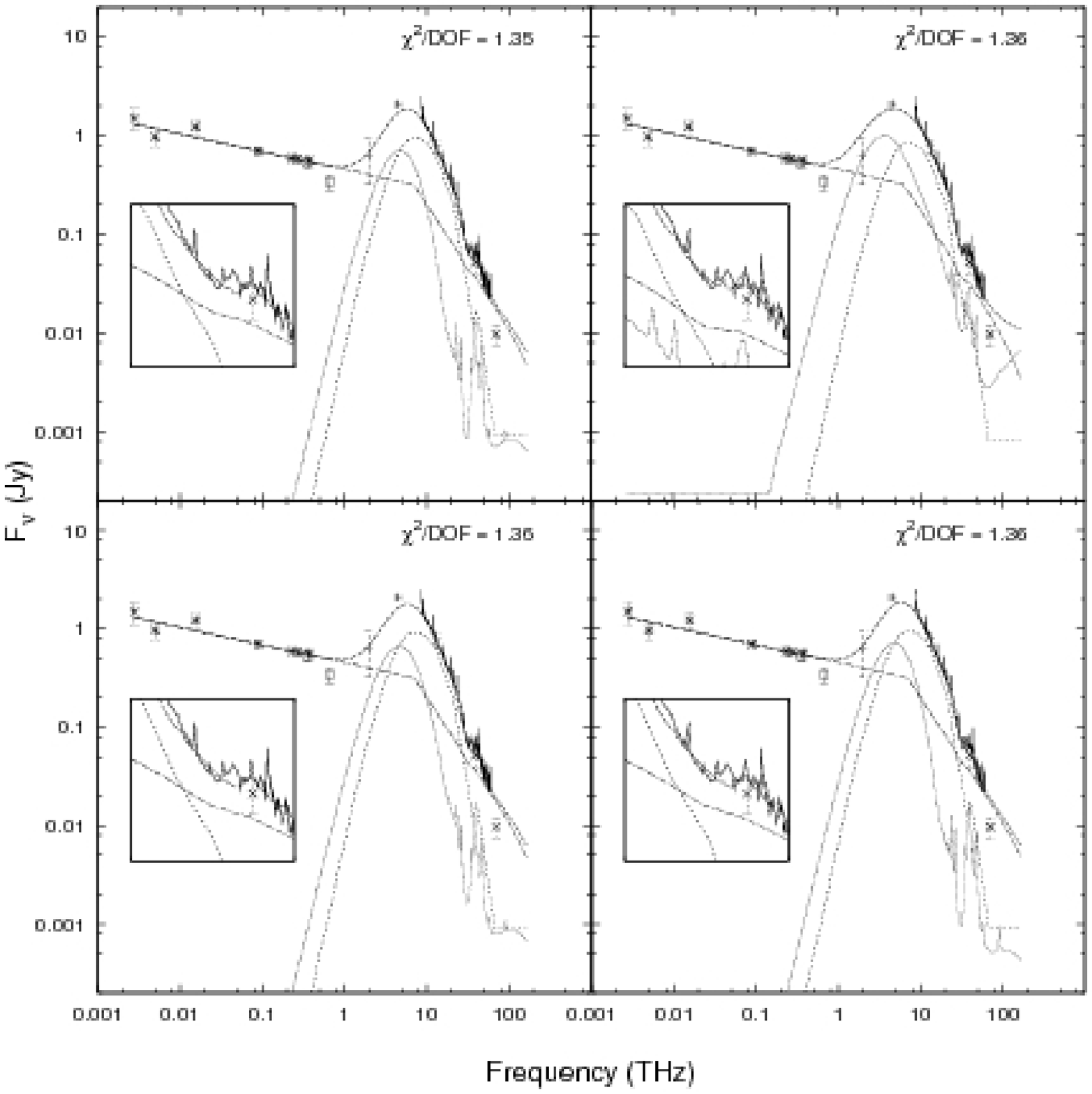}
\end{figure}

\subsubsection{i=50$^{\circ}$}

\begin{figure}
\includegraphics[width=\textwidth]{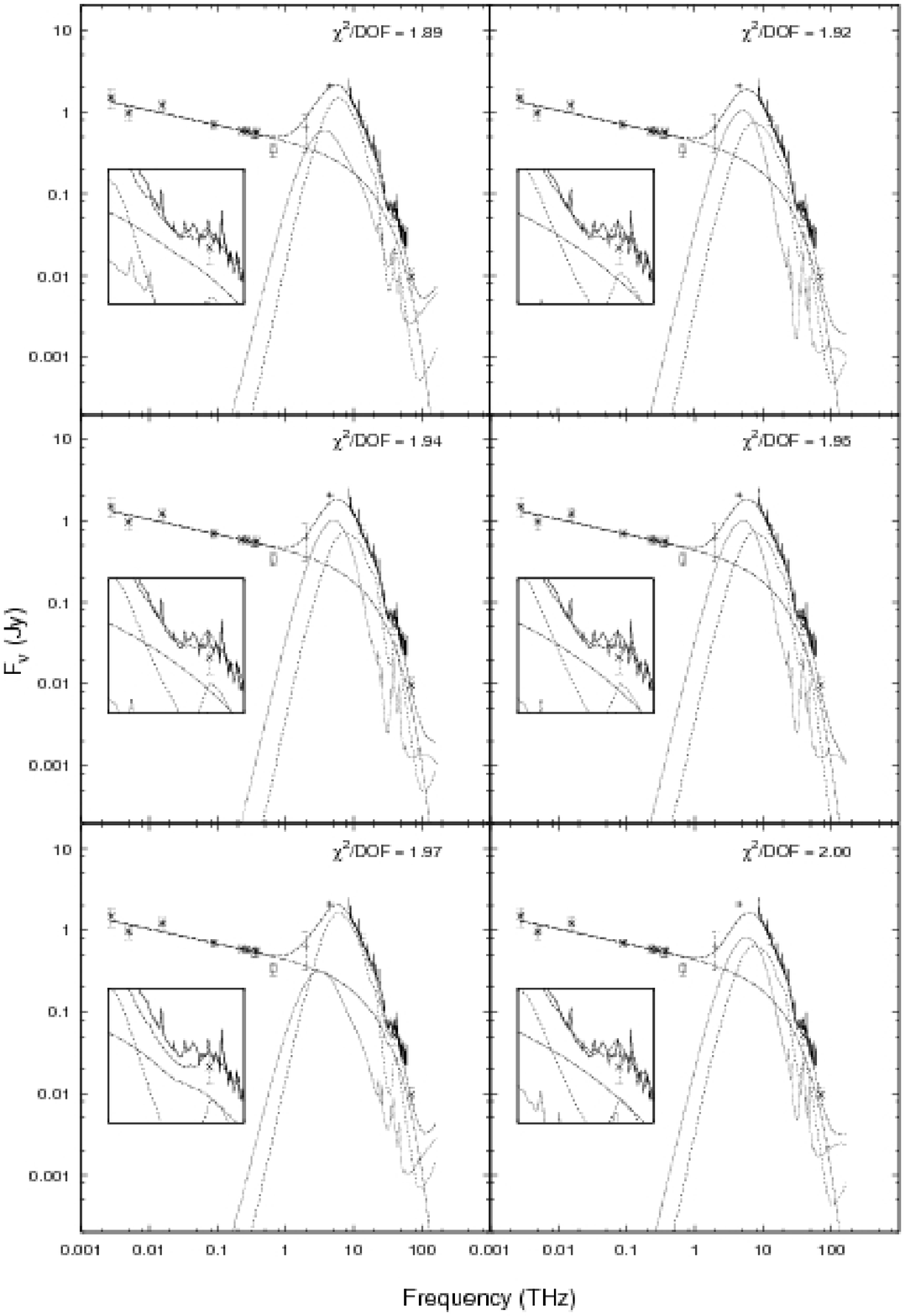}
\caption{Individual plots of the best fit models in figure \ref{fig:results50}. Legend is the same as figure \ref{fig:results50}.}
\label{fig:results50a}
\end{figure}

\begin{figure}
\includegraphics[width=\textwidth]{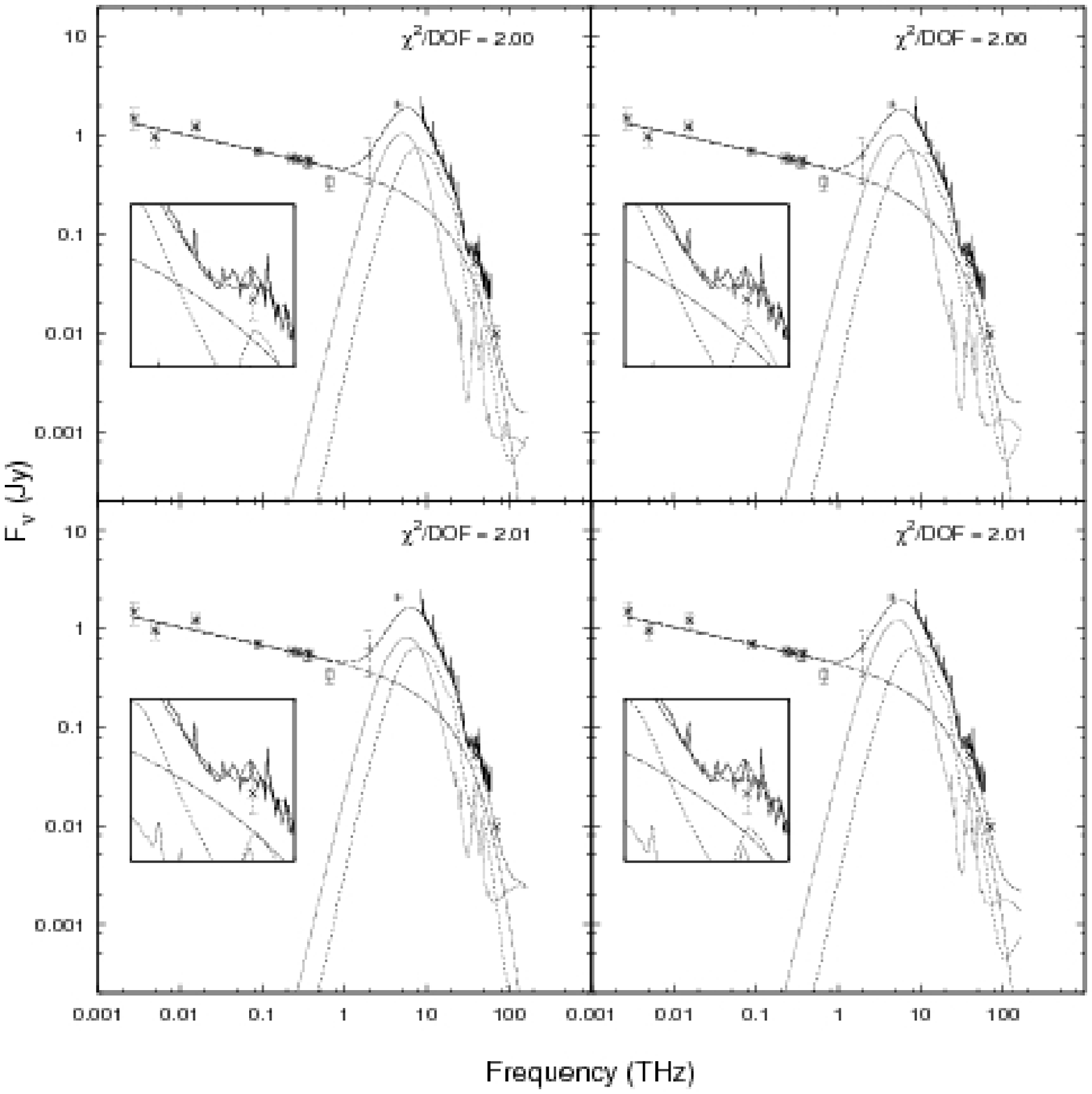}
\end{figure}

Case I fits using an inclination angle of $i=50$ imply a bolometric AGN flux similar to that of $i=70$: $\sim(0.85-1.6)\times 10^{-10}$ erg s$^{-1}$ cm$^{-2}$, corresponding to a luminosity of $(5.8-11)\times10^{44}$ erg s$^{-1}$. The favored torus model is fairly consistent among the best fits. At this luminosity the inner radius of the torus is $\sim0.3-0.4$ pc. The reduced $\chi^2$ values are between 1.888 and 2.006.

\begin{figure}[h!]
\includegraphics[width=\textwidth]{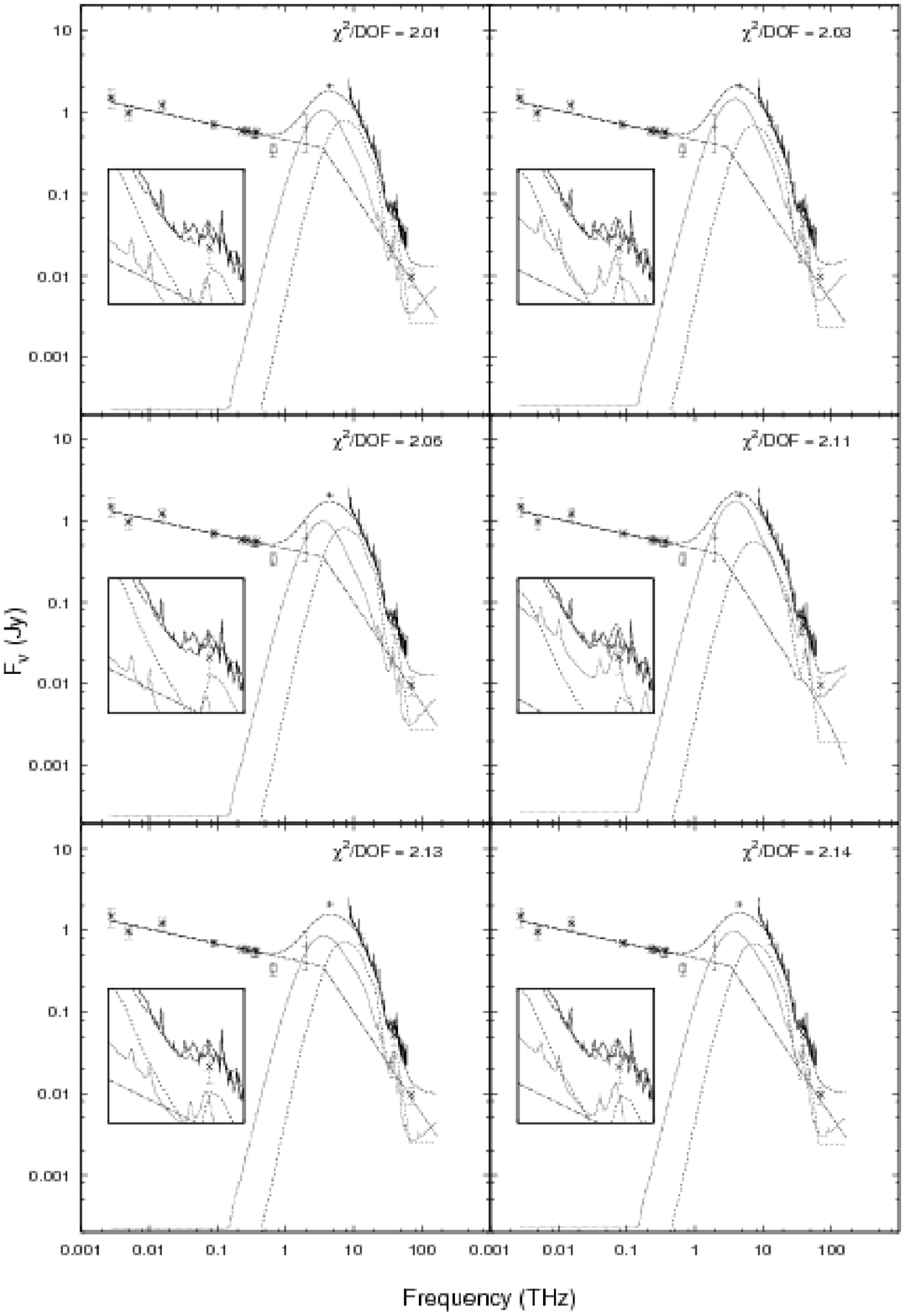}
\caption{Individual plots of the best fit models in figure \ref{fig:results50-b}. Legend is the same as figure \ref{fig:results50-b}.}
\label{fig:results50-ba}
\end{figure}

The best fit torus parameters for $i=50$ are similar to those for $i=70$, with slight fluctuation in the powerlaw distribution of clumps ($q$ between 1 and 2), and a slightly wider range in the angular distribution of clumps ($\sigma$ between 55 and 65$^{\circ}$). 
As with the other inclinations, Case II fits result in similar torus properties as Case I fits for $i=50$ (Figures \ref{fig:results50-b} \& \ref{fig:results50-ba}). In this case the best fit luminosities are similar. Reduced $\chi^2$ values are between 1.997 and 2.122 for the top ten fits.

\begin{figure}[h!]
\includegraphics[width=\textwidth]{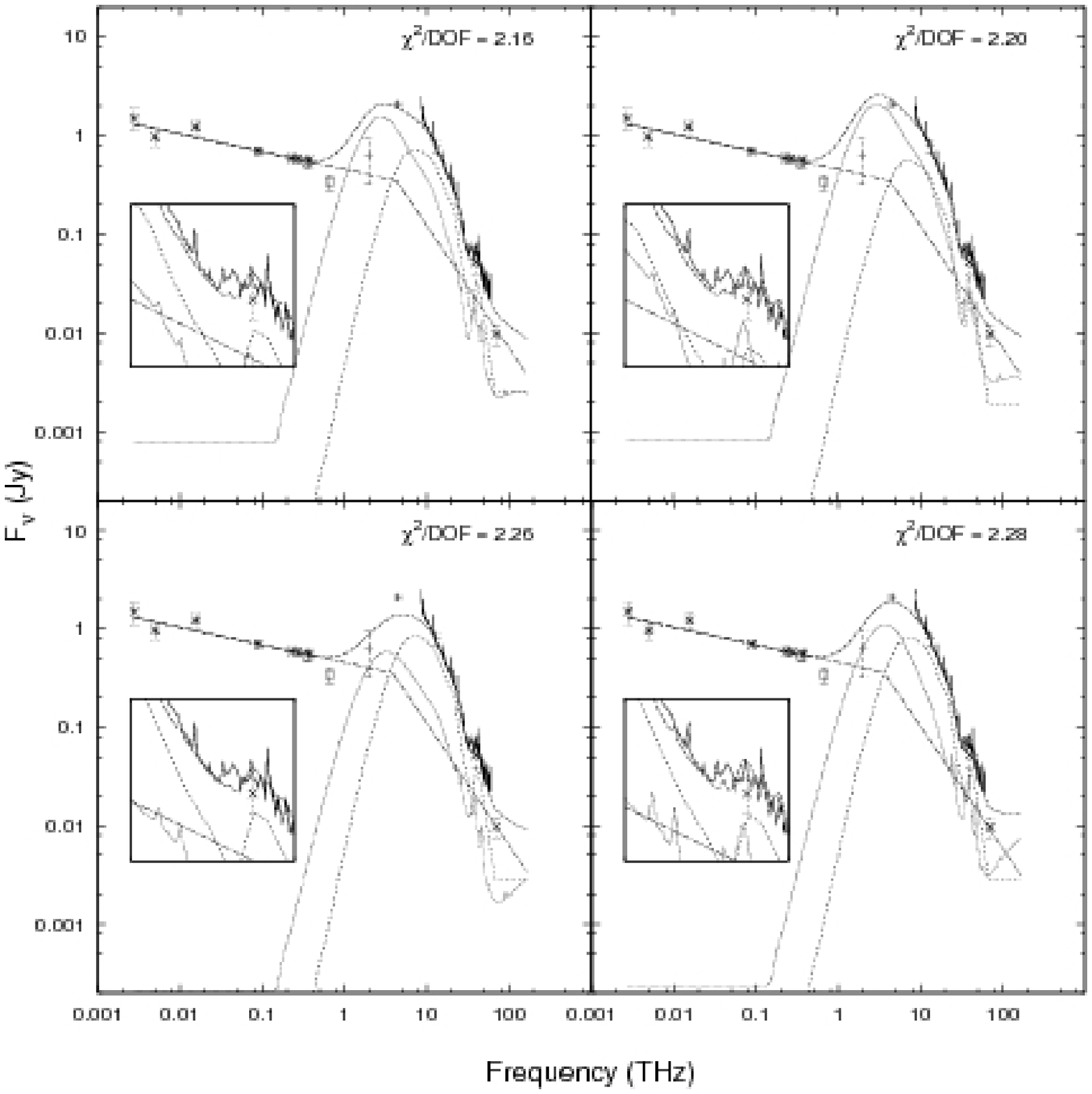}
\end{figure}

The best fit torus parameters for the inclinations are summarized for Case I in Table \ref{table:torus-result-I} and Case II in Table \ref{table:torus-result-II}.

\begin{table}[h!]
\caption{Summary of Torus Model Results: Case I -- Exponential Cutoff}
\begin{tabular}{l|l|ccccccc}
$i$		& AGN?	& $Y$	& $\sigma$	& $N$	& $q$	& $\tau_v$ & $L_{agn}$ (erg s$^{-1}$)	& $\eta=L_{AGN}/L_{Edd}$ \\
\hline 
$80^{\circ}$ &	yes	& 200	& 15$^{\circ}$	& 20	& 1	& 40	& $4.6\times10^{47}$  & 0.56\\
$80^{\circ}$ &  no	& 30	& 60$^{\circ}$	& 20	& 1	& 40	& $9.4\times10^{45}$ & $1.1\times10^{-2}$ \\
$70^{\circ}$ &  yes \& no	& 30	& 60-65$^{\circ}$	& $\sim$20	& 2	& 60	& $7.7\times10^{44}$ & $9.3\times10^{-4}$ \\
$50^{\circ}$ &  yes \& no       & 30    & 55-65$^{\circ}$  & $\sim$20    & 2     & 60 & (5.8-11)$\times10^{44}$ &  $7.0-13.\times10^{-4}$\\
\end{tabular}
\label{table:torus-result-I}
\end{table}

\begin{table}[h!]
\caption{Summary of Torus Model Results: Case II -- Broken Powerlaw}
\begin{tabular}{l|l|ccccccc}
$i$             & AGN?  & $Y$   & $\sigma$      & $N$   & $q$   & $\tau_v$ & $L_{agn}$ (erg s$^{-1}$)  & $\eta=L_{AGN}/L_{Edd}$ \\
\hline 
$80^{\circ}$ &  yes \& no      & 30    & 60$^{\circ}$  & 20    & 1     & 40    & $1.1\times10^{45}$ & $1.3\times10^{-3}$ \\
$70^{\circ}$ &  yes \& no       & 30    & 60$^{\circ}$       & 20      & 1     & 60    & $1.1\times10^{45}$ & $1.3\times10^{-3}$\\
$50^{\circ}$ &  yes \& no       & 30    & 60$^{\circ}$  & 20    & 1     & 60 & $7.3\times10^{44}$ & $8.8\times10^{-4}$\\
\end{tabular}
\label{table:torus-result-II}
\end{table}

\subsubsection{Physical Interpretation}

Clearly the best fit models have settled on a torus with a small opening angle and a shallow falloff in clump density with radius. The optical depth of each clump is also fairly high, but not excessively so. Additionally, with the exception of $i=80$ + visible AGN, the luminosities are similar to the bolometric AGN luminosity inferred from X-ray and mid-IR observations (giving $L_{bol}=0.1-2\times10^{46}$ ergs s$^{-1}$) \citep{Tadhunter03,Whysong04}.

\citet{Young02} found $N_H=2.0\times10^{23}$ cm$^{-3}$ towards the nucleus from X-ray observations. For a standard gas-to-dust ratio, this column density corresponds to $A_V\sim100$ mag. Torus model fits find a somewhat lower $A_V\sim65$. 

There is little difference in the Case I and Case II fits with respect to torus models. Torus emission is dominant in the midIR, where the synchrotron arguably has the smallest effect in either case. 

\subsection{Starburst Properties and Contribution}

The best fit starburst component varies significantly between the inclinations, however remains relatively consistent for the various fits at a given inclination. A star formation rate is calculated using $L_{FIR}$ following \citet{Kennicutt98b} (eq \ref{eq:sfr}). For a review of SFR estimates see \citet{Kennicutt98a}.

\begin{equation}
SFR~(M_{\odot}~yr^{-1}) = 4.5\times10^{-44}~\frac{L_{FIR}}{1~erg~s^{-1}}=1.72\times10^{-10}~\frac{L_{FIR}}{L_{\odot}}
\label{eq:sfr}
\end{equation}

At nearly edge on inclinations $i=80$, a starburst of $L_{sb}\sim10^{11.2}$ $L_{\odot}$ with $A_v\sim18$ is the best fit (with scatter in $A_v$ from 4.5 through 72). The size is typically 3kpc (ranging from 0.35-3 kpc), similar to the size of the circumnuclear ring known to exist in Cygnus A (4kpc). The dust density in hot spots is typically $n\sim5\times10^3$ cm$^{3}$, with OB stars contributing roughly 60\% of the luminosity. Using eq \ref{eq:sfr}, the SFR rate is $\sim$27 M$_{\odot}$ yr$^{-1}$.

With $i=70$, the best fit starburst model changes slightly, favoring a smaller burst ($r=0.35$ kpc), and a slightly lower $f_{OB}=40\%$. The total luminosity decreases somewhat when compared to fits for an torus orientation of $i=80$: $L_{sb}\sim10^{11}$ $L_{\odot}$, while the extinction increases to $A_v=35$ (scatter from 18-72). The dust density in the hotspots ranges from $1-10.\times10^{3}$ cm$^{-3}$, and the SFR is 17 M$_{\odot}$ yr$^{-1}$.

The fits with a torus inclination of $i=50$ are very similar to those of $i=70$ with a smaller range in the density in the hotspots of the OB stars ($1-2.5\times10^{3}$ cm$^{-3}$. The luminosity is also slightly higher: $L_{sb}\sim10^{11.1}$ $L_{\odot}$, corresponding to a SFR of 22 M$_{\odot}$ yr$^{-1}$.

In contrast with the torus results, the best fit starburst properties do change somewhat between Case I and Case II fits.

Table \ref{table:starburst-result-I} summarizes the results of the starburst fits for Case I. Table \ref{table:starburst-result-II} corresponds to Case II fits.

\begin{table}[h!]
\caption{Summary of Starburst Model Results for Case I}
\begin{tabular}{l|cccccc}
$i$	& $r$ (kpc) & $f_{OB}$ & $L_{tot}$ ($L_{\odot})$ & $A_v$	& $n$ (cm$^{-3}$) & SFR ($M_{\odot} yr^{-1}$) \\
\hline
80	& 3 	& 60\%	& $1.5\times10^{11}$	& 18	& $5\times10^{3}$ & 27\\
70	& 0.35	& 40\%	& $0.8\times10^{11}$	& 35	& $2.5\times10^{3}$ & 17\\
50	& 0.35	& 40\%	& $0.8\times10^{11}$      & 35    & $1\times10^{3}$ & 22 \\
\end{tabular}
\label{table:starburst-result-I}
\end{table}

\begin{table}[h!]
\caption{Summary of Starburst Model Results for Case II}
\begin{tabular}{l|cccccc}
$i$     & $r$ (kpc) & $f_{OB}$ & $L_{tot}$ ($L_{\odot})$ & $A_v$        & $n$ (cm$^{-3}$) & SFR ($M_{\odot} yr^{-1}$) \\
\hline
80      & 0.35     & 40-90\%  & $1\times10^{11}$    & 72    & $1\times10^{2}$ & 17\\
70      & 0.35  & 40-90\%  & $(0.8-1.7)\times10^{11}$      & 72    & $\sim5\times10^{3}$ & 14-30\\
50      & 3  & 40\%  & $2\times10^{11}$      & 4.5    & $5\times10^{3}$ & 34 \\
\end{tabular}
\label{table:starburst-result-II}
\end{table}

\subsubsection{Physical Interpretation}

The best-fit starburst models at all inclinations are very similar in most areas. Roughly half the luminosity is from OB stars enshrouded in hot spots with dust densities of  a few$\times10^{3}$ cm$^{-3}$ and having a total luminosity of $\sim10^{11}$ L$_{\odot}$ ($3\times10^{44}$ erg s$^{-1}$). Star formation rates are roughly the same in all fits, ranging between 20 and 30 $M_{\odot}~yr^{-1}$ for Case I and Case II fits. 

The integrated far infrared luminosities of the starburst component gives significant star formation rates, indicating Cygnus A would still be bright in the infrared, even in the absence of an AGN. The luminosities are comparable to those found in luminous infrared galaxies (LIRGs). 

\subsection{Far IR Analysis}

Around 1 THz ($\sim300\mu$m) the three model components are of similar flux density, with the synchrotron dying out and the thermal contributions from star formation and a clumpy torus strengthening. Unfortunately, data at this location is quite limited, and of relatively poor spatial resolution, leaving little hope of spatially disentangling the emission from the starburst and emission from the AGN at this frequency until the availability of the Atacama Large Millimeter Array \citep[ALMA; eg][]{Carilli05}. 

The thermal emission seen in the far-IR dominates at frequencies higher than 1 THz ($\sim300\mu$m). The reprocessed emission from the AGN exceeds the starburst emission just shy of 10 THz ($\sim30\mu$m). At this point, the flux density of the observed SED is decreasing, and the peak flux density of the torus contribution is lower than the peak flux density of the starburst.

\subsection{Mid IR}

The MIR portion of the spectrum contains many (dust) emission features which are expressed in the models. Additionally, there are numerous forbidden lines from highly ionized species as well as $H_2$ molecular lines in the IRS spectrum. These are likely from the inner NLR, not the AGN/torus region (BLR), and discussion of them is beyond the scope of this study.

An obvious feature of the MIR spectra is the broad absorption around $10\mu m$. This silicate absorption is well matched by the combination of torus and starburst models (see inserts in figs \ref{fig:results80a} - \ref{fig:results50-ba}).

As noted earlier, the continuum flux in this regime requires the synchrotron powerlaw to break at an unknown frequency. Based on the above modeling results, the break likely occurs in the mid-infrared ($\sim15 \mu m$). In contrast to Cygnus A, some AGN exhibit optical synchrotron jets \citep[eg. 3C133, 3C273, M87; ][]{Floyd06,Jester07,Perlman07}, indicating the distribution of the relativistic electrons might extend to higher energies in these sources. 

Approaching the NIR, all model components continue to rapidly decrease in flux density. In fits where the AGN powerlaw is visible, this component begins to dominate into the NIR. However, the stellar contribution to the NIR have not been taken into account.  

\section{Overall Contributions to the Luminosity in Cygnus A}

The three physical processes included in modeling the radio through mid-IR SED of Cygnus A all contribute to the luminosity. Of the three, torus reprocessed AGN radiation dominates the energy output of the nuclear region. Table \ref{table:ratios} lists the ratios of each model component.

\begin{table}[h!]
\caption{Ratio of Luminosity for Model Components}
\begin{tabular}{l|lccc}
 	& $i$	& $L_{AGN}$	& $L_{SB}/L_{AGN}$	& $L_{sync}/L_{AGN}$ \\
\hline
\multirow{4}{*}{Case I}&	80	& $4.6\times10^{47}$ & $1.2\times10^{-3}$   & $8.7\times10^{-4}$ \\
&        80	& $9.4\times10^{45}$ & $0.061$ & $0.043$ \\
&	70	& $7.7\times10^{44}$ & $0.39$  & $0.52$ \\
&	50	& $\sim8\times10^{44}$ & $0.37$ & $0.25$ \\
\hline
\multirow{3}{*}{Case II}&        80      & $1.1\times10^{45}$ & $0.38$ & $0.27$ \\
&       70      & $1.1\times10^{45}$	& $0.43$ & $0.45$ \\
&       50      & $7.3\times10^{44}$	& $1.0$ & $0.68$ \\
\end{tabular}
\label{table:ratios}
\end{table}

In most cases, the re-radiated emission from the torus dominates the luminosity by a factor of a few. In addition, the energy emitted in the core through synchrotron is roughly the order of that in the starburst, and factors of a few less than the emission from the torus.

%

\chapter{Conclusions}
\label{chap:conclusion}

Using a combination of new mid-infrared spectra from the Spitzer Space Telescope and radio data from the literature, the nuclear emission of Cygnus A has been modeled as a combination of powerlaw emission from a synchrotron jet, reprocessed AGN emission from a dusty torus, and reprocessed emission from a circumnuclear starburst. This is one of the first detailed fits to a broad SED, covering $\sim5$ dex in frequency. The addition of the mid-IR data is critical in constraining the relative contributions of each component to the bolometric luminosity of Cygnus A. It places limits on the amount of warm dust emission present in the inner regions, constraining the contribution from the torus component. 

The data is well fit by a combination of these three models. The fits suggest the nuclear emission at these frequencies is primarily due to AGN activity (synchrotron radiation \& torus reprocessed AGN emission) and star formation. 

The modeled jet shows evidence for a cutoff in the mid-infrared. This is in contrast with some radio galaxies and quasars which showing jets emitting from the radio through X-rays. Statistically acceptable fits were found for both an exponential cutoff in the population of relativistic electrons (Case I) as well as synchrotron losses (Case II). The cutoff/break frequencies tended to be within the $3-30$ THz range. It is possible the synchrotron emission is a significant component of the mid-IR flux, which could be tested using polarimetry.

The nuclear AGN emission was modeled as a clumpy torus, reprocessing the emission from and around the accretion disk. The fits suggest a bolometric AGN luminosity of $\sim10^{45}$ erg s$^{-1}$. The torus is predicted to have an outer radius of $\sim7$ pc ($7$ mas at the distance of Cygnus A). 

The star formation rate determined for Cygnus A is on the order of $10-30$ M$_{\odot}$ yr$^{-1}$, well above the rate for quiescent galaxies. The IR luminosity of this component is large enough that this galaxy would be classified as a LIRG, even in the absence of AGN activity.

The reprocessed torus radiation dominates the bolometric luminosity of Cygnus A, typically a factor of $\sim3$ greater than the luminosity associated with star formation. While the dominant emission at radio wavelengths, the synchrotron emission is a small fraction of the emission in the far-IR. 

It was noted that Cygnus A is in the midst of a merger. The existence of the merging system, star formation, and AGN activity is consistent with suggestions that mergers funnel gas towards the nucleus where it forms stars and feeds a SMBH at the center.

\section{Further Work}

There remains some ambiguity in the fits primarily related to the visibility of the AGN powerlaw. Inclusion of data in the near infrared should resolve this issue. Preliminary use of 2MASS data indicates the AGN powerlaw is not visible. However, this remains to be properly verified by taking into account stellar contribution in the near-infrared.

The contribution of synchrotron self-compton (SSC) to the observed X-ray flux remains unquantified. Synchrotron photons scattering off the relativistic particles that emitted them can potentially be a significant contribution to the X-ray luminosity in Cygnus A.

The Herschel Space Observatory, set to launch in May 2009, will provide spectral and imaging coverage in the far infrared and sub-mm. Low- and medium- resolution spectroscopy ($R\sim40-5000$, depending on instrument and wavelength) will cover a wavelength range of $55-672\mu m$, probing a relatively unexplored wavelength regime \citep{Poglitsch06,Griffin06}. This will likely put additional constraints on the physical processes contributing in the far infrared. Also, atomic and molecular lines at the longer wavelengths will provide additional information on the physical conditions present in the nuclear regions.

The sub-mm is a wavelength regime in which there is a comparable dearth of information. Historically plagued with poor angular resolution, future instruments, particularly ALMA will enable deeper, higher-resolution studies of the nuclear regions. It will be able to resolve and image the nuclear region, potentially spatially disentangling the circumnuclear starburst and AGN, removing some ambiguity in the modeling. However, Cygnus A is at a high declination and might prove difficult to observe with ALMA.

The launch of the James Webb Space Telescope (JWST) will facilitate deeper and more sensitive observations in the near- and mid-infrared. The Mid-Infrared Instrument (MIRI) will cover a wavelength range similar to that of IRS while offering increased resolution (both spatial and spectral) and sensitivity \citep{Wright04}. This would enable a study of the nuclear region which resolves physical separation between the starburst ring and the AGN, testing the conclusions from the model fitting shown here.

\bibliographystyle{astron}
\bibliography{thesis}

\begin{thebibliography}{}

\bibitem[\protect\astroncite{IRS}{2006}]{IRSmanual}
IRS \ 2006,
\newblock {\em {Infrared Spectrograph Data Handbook}},
\newblock {Spitzer Science Center}

\bibitem[\protect\astroncite{SPI}{2006}]{SPICEmanual}
SPI \ 2006,
\newblock {\em {SPICE\ Spitzer IRS Custom Extraction}},
\newblock {Spitzer Science Center}

\bibitem[\protect\astroncite{SOM}{2006}]{SOM}
SOM \ 2006,
\newblock {\em {Spitzer Observer's Manual}}, \newblock {Spitzer Science Center}

\bibitem[\protect\astroncite{{Alexander} et~al.}{1984}]{Alexander84}
{Alexander}, P., {Brown}, M.~T., \& {Scott}, P.~F.\ 1984,
\newblock {\em \mnras} {\bf 209}, 851

\bibitem[\protect\astroncite{{Antonucci} et~al.}{1994}]{Antonucci94}
{Antonucci}, R., {Hurt}, T., \& {Kinney}, A.\ 1994,
\newblock {\em \nat} {\bf 371}, 313

\bibitem[\protect\astroncite{{Antonuccio-Delogu} \&
  {Silk}}{2008}]{Antonuccio-Delogu08}
{Antonuccio-Delogu}, V. \& {Silk}, J.\ 2008,
\newblock {\em \mnras} {\bf 389}, 1750

\bibitem[\protect\astroncite{{Baade} \& {Minkowski}}{1954}]{Baade54}
{Baade}, W. \& {Minkowski}, R.\ 1954,
\newblock {\em \apj} {\bf 119}, 206

\bibitem[\protect\astroncite{{Beckmann} et~al.}{2006}]{Beckmann06}
{Beckmann}, V., et~al.\ 2006,
\newblock {\em \apj} {\bf 638}, 642

\bibitem[\protect\astroncite{{Bellamy} \& {Tadhunter}}{2004}]{Bellamy04}
{Bellamy}, M.~J. \& {Tadhunter}, C.~N.\ 2004,
\newblock {\em \mnras} {\bf 353}, 105

\bibitem[\protect\astroncite{{Blandford} \& {Rees}}{1974}]{Blandford74}
{Blandford}, R.~D. \& {Rees}, M.~J.\ 1974,
\newblock {\em \mnras} {\bf 169}, 395

\bibitem[\protect\astroncite{{Blandford} \& {Znajek}}{1977}]{Blandford77}
{Blandford}, R.~D. \& {Znajek}, R.~L.\ 1977,
\newblock {\em \mnras} {\bf 179}, 433

\bibitem[\protect\astroncite{{Canalizo} et~al.}{2003}]{Canalizo03}
{Canalizo}, G., et~al.\ 2003,
\newblock {\em \apj} {\bf 597}, 823

\bibitem[\protect\astroncite{{Carilli}}{2005}]{Carilli05}
{Carilli}, C.~L.\ 2005,
\newblock in A. {Wilson} (ed.), {\em ESA Special Publication}, Vol. 577 of {\em
  ESA Special Publication}, pp 47--54

\bibitem[\protect\astroncite{{Carilli} et~al.}{1994}]{Carilli94}
{Carilli}, C.~L., {Bartel}, N., \& {Diamond}, P.\ 1994,
\newblock {\em \aj} {\bf 108}, 64

\bibitem[\protect\astroncite{{Carilli} \& {Barthel}}{1996}]{Carilli96}
{Carilli}, C.~L. \& {Barthel}, P.~D.\ 1996,
\newblock {\em \aapr} {\bf 7}, 1

\bibitem[\protect\astroncite{{Conway} \& {Blanco}}{1995}]{Conway95}
{Conway}, J.~E. \& {Blanco}, P.~R.\ 1995,
\newblock {\em \apjl} {\bf 449}, L131+

\bibitem[\protect\astroncite{{Decin} et~al.}{2004}]{Decin04}
{Decin}, L., et~al.\ 2004,
\newblock {\em \apjs} {\bf 154}, 408

\bibitem[\protect\astroncite{{Djorgovski} et~al.}{1991}]{Djorgovski91}
{Djorgovski}, S., et~al.\ 1991,
\newblock {\em \apjl} {\bf 372}, L67

\bibitem[\protect\astroncite{{Draine} \& {Lee}}{1984}]{Draine84}
{Draine}, B.~T. \& {Lee}, H.~M.\ 1984,
\newblock {\em \apj} {\bf 285}, 89

\bibitem[\protect\astroncite{{Eales} et~al.}{1989}]{Eales89}
{Eales}, S.~A., {Alexander}, P., \& {Duncan}, W.~D.\ 1989,
\newblock {\em \mnras} {\bf 240}, 817

\bibitem[\protect\astroncite{{Elvis}}{2000}]{Elvis00}
{Elvis}, M.\ 2000,
\newblock {\em \apj} {\bf 545}, 63

\bibitem[\protect\astroncite{{Evans} et~al.}{2006}]{Evans06}
{Evans}, D.~A., {Worrall}, D.~M., {Hardcastle}, M.~J., {Kraft}, R.~P., \&
  {Birkinshaw}, M.\ 2006,
\newblock {\em \apj} {\bf 642}, 96

\bibitem[\protect\astroncite{{Fanaroff} \& {Riley}}{1974}]{Fanaroff74}
{Fanaroff}, B.~L. \& {Riley}, J.~M.\ 1974,
\newblock {\em \mnras} {\bf 167}, 31P

\bibitem[\protect\astroncite{{Fazio} et~al.}{2004}]{Fazio04}
{Fazio}, G.~G., et~al.\ 2004,
\newblock {\em \apjs} {\bf 154}, 10

\bibitem[\protect\astroncite{{Ferrarese} \& {Merritt}}{2000}]{Ferrarese00}
{Ferrarese}, L. \& {Merritt}, D.\ 2000,
\newblock {\em \apjl} {\bf 539}, L9

\bibitem[\protect\astroncite{{Floyd} et~al.}{2006}]{Floyd06}
{Floyd}, D.~J.~E., et~al.\ 2006,
\newblock {\em \apj} {\bf 643}, 660

\bibitem[\protect\astroncite{{Gebhardt} et~al.}{2000}]{Gebhardt00}
{Gebhardt}, K., et~al.\ 2000,
\newblock {\em \apjl} {\bf 539}, L13

\bibitem[\protect\astroncite{{Griffin} et~al.}{2006}]{Griffin06}
{Griffin}, M., et~al.\ 2006,
\newblock in {\em Society of Photo-Optical Instrumentation Engineers (SPIE)
  Conference Series}, Vol. 6265 of {\em Presented at the Society of
  Photo-Optical Instrumentation Engineers (SPIE) Conference}

\bibitem[\protect\astroncite{{Gursky} et~al.}{1972}]{Gursky72}
{Gursky}, H., et~al.\ 1972,
\newblock {\em \apjl} {\bf 173}, L99+

\bibitem[\protect\astroncite{{Hargrave} \& {Ryle}}{1974}]{Hargrave74}
{Hargrave}, P.~J. \& {Ryle}, M.\ 1974,
\newblock {\em \mnras} {\bf 166}, 305

\bibitem[\protect\astroncite{{Higdon} et~al.}{2004}]{Higdon04}
{Higdon}, S.~J.~U., et~al.\ 2004,
\newblock {\em \pasp} {\bf 116}, 975

\bibitem[\protect\astroncite{{Houck} et~al.}{2004}]{Houck04}
{Houck}, J.~R., et~al.\ 2004,
\newblock {\em \apjs} {\bf 154}, 18

\bibitem[\protect\astroncite{{Jackson} et~al.}{1998}]{Jackson98}
{Jackson}, N., {Tadhunter}, C., \& {Sparks}, W.~B.\ 1998,
\newblock {\em \mnras} {\bf 301}, 131

\bibitem[\protect\astroncite{{Jackson} \& {Tadhunter}}{1993}]{Jackson93}
{Jackson}, N. \& {Tadhunter}, C.~N.\ 1993,
\newblock {\em \aap} {\bf 272}, 105

\bibitem[\protect\astroncite{{Jester} et~al.}{2007}]{Jester07}
{Jester}, S., et~al.\ 2007,
\newblock {\em \mnras} {\bf 380}, 828

\bibitem[\protect\astroncite{{Kardashev}}{1962}]{Kardashev62}
{Kardashev}, N.~S.\ 1962,
\newblock {\em Soviet Astronomy} {\bf 6}, 317

\bibitem[\protect\astroncite{{Kennicutt}}{1998a}]{Kennicutt98a}
{Kennicutt}, Jr., R.~C.\ 1998a,
\newblock {\em \araa}

\bibitem[\protect\astroncite{{Kennicutt}}{1998b}]{Kennicutt98b}
{Kennicutt}, Jr., R.~C.\ 1998b,
\newblock {\em \apj}

\bibitem[\protect\astroncite{{Laing} et~al.}{2008}]{Laing08}
{Laing}, R.~A., et~al.\ 2008,
\newblock {\em \mnras} {\bf 386}, 657

\bibitem[\protect\astroncite{{Magorrian} et~al.}{1998}]{Magorrian98}
{Magorrian}, et~al.\ 1998,
\newblock {\em \aj} {\bf 115}, 2285

\bibitem[\protect\astroncite{{Mason} et~al.}{2006}]{Mason06}
{Mason}, R.~E., et~al\ 2006,
\newblock {\em \apj} {\bf 640}, 612

\bibitem[\protect\astroncite{{Matthews} et~al.}{1964}]{Matthews64}
{Matthews}, T.~A., {Morgan}, W.~W., \& {Schmidt}, M.\ 1964,
\newblock {\em \apj} {\bf 140}, 35

\bibitem[McCarthy(1993)]{McCarthy93} McCarthy, P.~J.\ 1993, \araa, 31, 639 

\bibitem[\protect\astroncite{{Meisenheimer} et~al.}{2001}]{Meisenheimer01}
{Meisenheimer}, et~al.\ 2001,
\newblock {\em \aap} {\bf 372}, 719

\bibitem[\protect\astroncite{{Nenkova} et~al.}{2002}]{Nenkova02}
{Nenkova}, M., {Ivezi{\'c}}, {\v Z}., \& {Elitzur}, M.\ 2002,
\newblock {\em \apjl} {\bf 570}, L9

\bibitem[\protect\astroncite{{Nenkova} et~al.}{2008}]{Nenkova08}
{Nenkova}, M., et~al.\ 2008,
\newblock {\em \apj} {\bf 685}, 160

\bibitem[\protect\astroncite{{Nesvadba} et~al.}{2008}]{Nesvadba08}
{Nesvadba}, N.~P.~H., et~al.\ 2008,
\newblock {\em \aap} {\bf 491}, 407

\bibitem[\protect\astroncite{{Ogle} et~al.}{1997}]{Ogle97}
{Ogle}, P.~M., et~al.\ 1997,
\newblock {\em \apjl} {\bf 482}, L37+

\bibitem[\protect\astroncite{{Osterbrock}}{1983}]{Osterbrock83}
{Osterbrock}, D.~E.\ 1983,
\newblock {\em \pasp} {\bf 95}, 12

\bibitem[\protect\astroncite{{Perlman} et~al.}{2007}]{Perlman07}
{Perlman}, E.~S., et~al.\ 2007,
\newblock {\em \apj} {\bf 663}, 808

\bibitem[\protect\astroncite{{Pier} \& {Krolik}}{1992}]{Pier92}
{Pier}, E.~A. \& {Krolik}, J.~H.\ 1992,
\newblock {\em \apjl} {\bf 399}, L23

\bibitem[\protect\astroncite{{Poglitsch} et~al.}{2006}]{Poglitsch06}
{Poglitsch}, A., et~al.\ 2006,
\newblock in {\em Society of Photo-Optical Instrumentation Engineers (SPIE)
  Conference Series}, Vol. 6265 of {\em Presented at the Society of
  Photo-Optical Instrumentation Engineers (SPIE) Conference}

\bibitem[\protect\astroncite{{Reynolds} \& {Fabian}}{1996}]{Reynolds96}
{Reynolds}, C.~S. \& {Fabian}, A.~C.\ 1996,
\newblock {\em \mnras} {\bf 278}, 479

\bibitem[\protect\astroncite{{Rieke} et~al.}{2004}]{Rieke04}
{Rieke}, G.~H., et~al.\ 2004,
\newblock {\em \apjs} {\bf 154}, 25

\bibitem[\protect\astroncite{{Robson} et~al.}{1998}]{Robson98}
{Robson}, E.~I., et~al.\ 1998,
\newblock {\em \mnras} {\bf 301}, 935

\bibitem[\protect\astroncite{{Salter} et~al.}{1989}]{Salter89}
{Salter}, C.~J., et~al.\ 1989,
\newblock {\em \aap} {\bf 220}, 42

\bibitem[\protect\astroncite{{Sanders} et~al.}{1988}]{Sanders88}
{Sanders}, D.~B., et~al.\ 1988,
\newblock {\em \apj} {\bf 325}, 74

\bibitem[\protect\astroncite{{Scheuer}}{1974}]{Scheuer74}
{Scheuer}, P.~A.~G.\ 1974,
\newblock {\em \mnras} {\bf 166}, 513

\bibitem[\protect\astroncite{{Shi} et~al.}{2005}]{Shi05}
{Shi}, Y., et~al.\ 2005,
\newblock {\em \apj} {\bf 629}, 88

\bibitem[\protect\astroncite{{Siebenmorgen} \&
  {Kr{\"u}gel}}{2007}]{Siebenmorgen07}
{Siebenmorgen}, R. \& {Kr{\"u}gel}, E.\ 2007,
\newblock {\em \aap} {\bf 461}, 445

\bibitem[\protect\astroncite{{Smith} et~al.}{2007}]{Smith07}
{Smith}, J.~D.~T., et~al.\ 2007,
\newblock {\em \pasp} {\bf 119}, 1133

\bibitem[\protect\astroncite{{Sorathia} et~al.}{1996}]{Sorathia96}
{Sorathia}, B., et~al.\ 1996,
\newblock {\em {The parsec-scale jet \& counterjet in Cygnus A}}, pp 86--+,
\newblock Cygnus A -- Studay of a Radio Galaxy

\bibitem[\protect\astroncite{{Spergel} et~al.}{2007}]{Spergel07}
{Spergel}, D.~N., et~al.\ 2007,
\newblock {\em \apjs} {\bf 170}, 377

\bibitem[\protect\astroncite{{Tadhunter} et~al.}{2003}]{Tadhunter03}
{Tadhunter}, C., et~al. \ 2003,
\newblock {\em \mnras} {\bf 342}, 861

\bibitem[\protect\astroncite{{Tadhunter} et~al.}{1994}]{Tadhunter94}
{Tadhunter}, C.~N., {Metz}, S., \& {Robinson}, A.\ 1994,
\newblock {\em \mnras} {\bf 268}, 989

\bibitem[\protect\astroncite{{Tadhunter} et~al.}{1999}]{Tadhunter99}
{Tadhunter}, C.~N., et~al.\ 1999,
\newblock {\em \apjl} {\bf 512}, L91

\bibitem[\protect\astroncite{{Urry} \& {Padovani}}{1995}]{Urry95}
{Urry}, C.~M. \& {Padovani}, P.\ 1995,
\newblock {\em \pasp} {\bf 107}, 803

\bibitem[\protect\astroncite{{Werner} et~al.}{2004}]{Werner04}
{Werner}, M.~W. et~al.  \ 2004,
\newblock {\em \apjs} {\bf 154}, 1

\bibitem[\protect\astroncite{{Whysong} \& {Antonucci}}{2004}]{Whysong04}
{Whysong}, D. \& {Antonucci}, R.\ 2004,
\newblock {\em \apj} {\bf 602}, 116

\bibitem[\protect\astroncite{{Wright} et~al.}{2004}]{Wright04}
{Wright}, G.~S., et~al.  \ 2004,
\newblock in J.~C. {Mather} (ed.), {\em Society of Photo-Optical
  Instrumentation Engineers (SPIE) Conference Series}, Vol. 5487 of {\em
  Presented at the Society of Photo-Optical Instrumentation Engineers (SPIE)
  Conference}, pp 653--663

\bibitem[\protect\astroncite{{Wright} \& {Birkinshaw}}{1984}]{Wright84}
{Wright}, M. \& {Birkinshaw}, M.\ 1984,
\newblock {\em \apj} {\bf 281}, 135

\bibitem[\protect\astroncite{{Young} et~al.}{2002}]{Young02}
{Young}, A.~J., et~al.\ 2002,
\newblock {\em \apj} {\bf 564}, 176

\bibitem[\protect\astroncite{{Zirbel} \& {Baum}}{1998}]{Zirbel98}
{Zirbel}, E.~L. \& {Baum}, S.~A.\ 1998,
\newblock {\em \apjs} {\bf 114}, 177

\end{thebibliography}

\end{document}